\documentclass[]{aastex631}
\usepackage{amsmath}
\usepackage{appendix}
\usepackage{multirow}
\usepackage{graphicx}
\usepackage{subfigure}
\usepackage{threeparttable}

\newcommand{\kms}{km~s$^{-1}$}
\newcommand{\coa}{$^{12}$CO}
\newcommand{\cob}{$^{13}$CO}
\newcommand{\coc}{C$^{18}$O}
\newcommand{\hco}{HCO$^{+}$}
\newcommand{\hii}{H$_{\rm II}$}

\begin{document}
\title{Molecular Bubble and Outflow in S~Mon Revealed by Multiband Datasets}
\correspondingauthor{Ye Xu}
\email{xuye@pmo.ac.cn}

\author{Dejian Liu}
\affiliation{Purple Mountain Observatory, Chinese Academy of Sciences, Nanjing 210023, People's Republic of China}
\affiliation{School of Astronomy and Space Science, University of Science and Technology of China, Hefei 230026, People's Republic of China}
\author{Ye Xu}
\affiliation{Purple Mountain Observatory, Chinese Academy of Sciences, Nanjing 210023, People's Republic of China}
\affiliation{School of Astronomy and Space Science, University of Science and Technology of China, Hefei 230026, People's Republic of China}
\author{YingJie Li}
\affiliation{Purple Mountain Observatory, Chinese Academy of Sciences, Nanjing 210023, People's Republic of China}
\author{Zehao Lin}
\affiliation{Purple Mountain Observatory, Chinese Academy of Sciences, Nanjing 210023, People's Republic of China}
\affiliation{School of Astronomy and Space Science, University of Science and Technology of China, Hefei 230026, People's Republic of China}
\author{Chaojie Hao}
\affiliation{Purple Mountain Observatory, Chinese Academy of Sciences, Nanjing 210023, People's Republic of China}
\affiliation{School of Astronomy and Space Science, University of Science and Technology of China, Hefei 230026, People's Republic of China}
\author{WenJin Yang}
\affiliation{School of Astronomy \& Space Science, Nanjing University, 163 Xianlin Avenue, Nanjing 210023, People's Republic of China}
\affiliation{Max-Plank-Institut f$\ddot{u}$r Radioastronomie, auf dem H$\ddot{u}$gel 69, 53121 Bonn, Germany
}
\author{Jingjing Li}
\affiliation{Purple Mountain Observatory, Chinese Academy of Sciences, Nanjing 210023, People's Republic of China}
\affiliation{School of Astronomy and Space Science, University of Science and Technology of China, Hefei 230026, People's Republic of China}
\author{Xinrong Liu}
\affiliation{Purple Mountain Observatory, Chinese Academy of Sciences, Nanjing 210023, People's Republic of China}
\affiliation{School of Astronomy and Space Science, University of Science and Technology of China, Hefei 230026, People's Republic of China}
\author{Yiwei Dong}
\affiliation{Purple Mountain Observatory, Chinese Academy of Sciences, Nanjing 210023, People's Republic of China}
\affiliation{School of Astronomy and Space Science, University of Science and Technology of China, Hefei 230026, People's Republic of China}
\author{Shuaibo Bian}
\affiliation{Purple Mountain Observatory, Chinese Academy of Sciences, Nanjing 210023, People's Republic of China}
\affiliation{School of Astronomy and Space Science, University of Science and Technology of China, Hefei 230026, People's Republic of China}
\author{Deyun Kong}
\affiliation{School of Physics, Harbin Institute of Technology, Harbin 150001, People's republic of China}
\begin{abstract}

We identify a molecular bubble, and study the star formation and its feedback in the S~Mon region, using multiple molecular lines, young stellar objects (YSOs), and infrared data. We revisit the distance to S~Mon, $\sim$ 722 $\pm$ 9~pc, using \emph{Gaia} Data Release 3 parallaxes of the associated Class~II YSOs. The bubble may be mainly driven by a massive binary system (namely 15~Mon), the primary of which is an O7V-type star. An outflow is detected in the shell of the bubble, suggesting ongoing star formation activities in the vicinity of the bubble. The total wind energy of the massive binary star is three orders of magnitude higher than the sum of the observed turbulent energy in the molecular gas and the kinetic energy of the bubble, indicating that stellar winds help to maintain the turbulence in the S~Mon region and drive the bubble. We conclude that the stellar winds of massive stars have an impact on their surrounding environment.

\end{abstract}

\section{Introduction}
The mass-loss phase is a very common phenomenon during the early evolutionary stages of a star. Strong winds from young stars inject momentum and energy into the surrounding environment, thus affecting the dynamics and structure of their parent clouds~\citep[e.g.,][]{Lada+1985,Arce+etal+2011,Frank+etal+2014,Bally+2016,Li+etal+2020}. High-mass stars evolve very quickly and they may still be embedded in their parent cloud when they reach the main-sequence stage. When a high-mass star forms in the molecular cloud, its ultraviolet (UV) radiation can ionize and heat the surrounding gas to create an \hii~region. The expanding \hii~region can reshape the surrounding molecular gas and drive a bubble~\citep[e.g.,][]{Churchwell+etal+2006,Zhang+etal+2015}. During the main-sequence stage, high-mass stars can drive spherical winds that blow away the gas around the star and then form bubbles~\citep[e.g.,   ][]{Castor+etal+1975,Arce+etal+2011}. Bubbles and outflows are common stellar-feedback phenomena in the star-formation process and provide information about the physical properties of their surroundings~\citep[e.g.,][]{Lada+1985,Arce+etal+2010,Arce+etal+2011,Li+etal+2018,Li+etal+2020}. 

There have been quite a few studies concerned with bubbles, outflows and their feedback~\citep[e.g., see][for reviews]{Frank+etal+2014,Dale+2015,Bally+2016}. 
Since directly measuring the proper motion of extended structures, like molecular clouds, is very difficult, many previous studies only analyzed the line-of-sight dynamics in a region. Additionally, recent studies started to investigate the 3D structure of molecular clouds in the solar neighborhood~\citep[e.g.,][]{Green+etal+2019,Lallement+etal+2019,Guo+etal+2021}, mainly enabled by the \emph{Gaia} space mission~\citep{Gaia+2016}. 
Young stellar objects (YSOs) that only recently formed in star-forming molecular clouds are still close to their birth sites; hence, they still, on average, share the distance and kinematic properties of their parent clouds~\citep[e.g.,][]{Furesz+etal+2008,Tobin+etal+2009,Hacar+etal+2016,Groschedl+etal+2021}.
On this basis, cloud distances can be estimated by using the distances of YSOs located within the clouds as proxies~\citep[e.g.,][]{Groschedl+etal+2018,Groschedl+etal+2021,Zhang+2023}. The recent release of astrometric data from \emph{Gaia}~Data Release 3~\citep[DR3,][]{gaia2022} allows one to obtain a large number of high-quality parallaxes and proper motions of nearby YSOs, which also makes it possible to explore further the 3D kinematics of molecular clouds that harbor these YSOs~\citep[e.g.,][]{Groschedl+etal+2021,Flaccomio+etal+2023}.

NGC~2264 is a nearby high-mass star-forming complex in Monoceros with a distance of about 700--800~pc~\citep[e.g.,][]{Zucker+etal+2020,Flaccomio+etal+2023}, and it is well studied at multiple wavelengths~\citep[for a review see][]{Dahm+2008}. Studies of the spatial and dynamical structure of NGC~2264 revealed that it is composed of two complexes: S~Mon in the northern part and the Cone Nebula in the southern part~\citep[e.g., ][]{Sung+etal+2008,Tobin+etal+2015,Venuti+etal+2018}. 
The gas structure of S~Mon is dominated by massive stars at its center. For instance, 15 Mon, which consists of an O7V primary star with a mass of 35~$M_{\odot}$ and an O9.5V secondary star with a mass of  24~$M_{\odot}$~\citep{Dahm+2008}. 15~Mon exhibits variations in its UV resonance line profile as well as fluctuations in its soft X-ray flux, which can be interpreted as being induced by variations in its mass-loss rate~\citep{Snow+etal+1981,Grady+etal+1984,Dahm+2008}. 

\citet{Sung+Michael+2010} and~\citet{Venuti+etal+2018} studied the star formation history of S Mon by analyzing the properties of young clusters, and found that stars are still forming actively in the region.
\citet{Buckle+etal+2012} found that the CO emission surrounding S~Mon shows many filaments and arcs, and these structures are unlikely due to protostellar outflow activity. Based on \cob~emission, \citet{Tobin+etal+2015} speculated that there exists a bubble in the S~Mon region which may be driven by 15~Mon. Since studies of the S~Mon region either used only young stars or molecular gas, or did not focus on the details of the bubble, the properties of this region, especially the bubble, outflows, and their feedback remain unclear.

In this work, we carried out molecular line observations and collected YSO information in the S~Mon region, aiming to provide insights into the star formation activities and their feedback. This paper is organized as follows. In Section~\ref{sect:data}, we present the data used in this work. The correlation between YSOs and molecular gas, the distance of the molecular gas, and details of the bubble and outflow are presented in Section~\ref{sect:res}. We discuss the driving source of the bubble and the energy cascade in S~Mon, as well as the kinematic features of the bubble based on the YSOs in Section~\ref{sect:dis}. In Section~\ref{sect:sum}, we summarize the main results.

\section{Data}
\label{sect:data}
\subsection{Molecular Gas}
Molecular line observations of S~Mon were carried out between 2022 April and November using the Purple Mountain Observatory Delingha (PMODLH) 14 m millimeter-wavelength telescope. Four molecular lines were observed: \coa~(J = 1 - 0), \cob~(J = 1 - 0), \coc~(J = 1 - 0), and \hco~(J = 1 - 0). The three CO lines were observed simultaneously. All lines were observed with the nine-beam Superconducting Spectroscopic Array Receiver system in the sideband separation mode~\citep{Shan+etal+2012}, with the \cob~and \coc~lines in the lower sideband and the \coa~and \hco~lines in the upper sideband. The on-the-fly~\citep[OTF; ][]{Sun+etal+2018} mode was used during the observations, and the OTF raw data were gridded in fits cube with a pixel size of $30''$ using the GILDAS software package~\citep{Pety+2005}.\footnote[1]{\url{http://www.iram.fr/IRAMFR/GILDAS}} The typical integration time per position was approximately 3 minutes. Each fast Fourier transform spectrometer with a bandwidth of 1~GHz provides 16\,384 channels, yielding a spectral resolution of 61~kHz.\footnote[2]{\url{http://www.radioast.nsdc.cn/english/zhuangtaibaogao.php}} All results presented in this work are expressed as brightness temperatures, $T_{\rm R}^{*} = T_{A}^{*} / \eta_{\rm mb}$, where $T_{A}^{*}$ is the antenna temperature and $\eta_{\rm mb}$ is the main beam efficiency. Table~\ref{tab:obs} lists the observational parameters of the molecular lines.

\begin{deluxetable}{ccccccc}[htbp]
\label{tab:obs}
\tablecaption{Observational Parameters of the Molecular Lines}
\tablehead{
    Molecular line & Rest frequency & HPBW & $T_{\rm sys}$ & $\eta_{\rm mb}$ & $\delta_{v}$ & rms noise \\
    (J = 1 - 0) & (GHz) & ($''$) & (K) & & (\kms) & (K) \\
   (1) & (2) & (3) & (4) & (5) & (6) & (7) 
}
\startdata
    \coa & 115.271 & 50 & $250 - 300$ & 51.1\% & 0.159 & 0.15 \\
    \cob & 110.201 & 52 & $150 - 200$ & 56.3\% & 0.166 & 0.07 \\
    \coc & 109.782 & 52 & $150 - 200$ & 56.3\% & 0.166 & 0.06 \\
    \hco &  89.189 & 59 & $100 - 150$ & 58.1\% & 0.187 & 0.03 
\enddata
\tablecomments{(1) Molecular line. (2) Frequency of the molecular line. (3) Half-power beamwidth (HPBW). (4) Typical system temperature. (5) Main beam efficiency. (6) Velocity resolution. (7) Main beam rms noise.}
\end{deluxetable}

\subsection{Young Stellar Object Sample}
The classification of YSOs was initially proposed by~\citet{Lada+Wilking+1984} to delineate their evolutionary stages, and is defined as follows. Class~I YSOs are protostars that are still embedded in an envelope and surrounded by a circumstellar disk, causing significant infrared-excess and a rising slope of the spectral energy distribution (SED) in the mid- to far-infrared range; Class~II YSOs are pre-main-sequence (PMS) stars with circumstellar disks, which show a negative SED slope, that is significantly above the slopes of main-sequence stars \citep[see][]{Lada+etal+2006}. Class~II YSOs with inner holes in their accretion disks are named transition disks and Class~III YSOs are PMS stars surrounded by thin disk remnants (anaemic disks, ADs) or they could be already disk free.

To study the young cluster NGC~2264, a variety of multiwavelength datasets and criteria have been employed to identify the evolutionary stages of YSOs. For example, \citet{Sung+etal+2008,Sung+etal+2009} classified YSOs based on SED slopes of infrared data combined with X-ray data from the {\it Chandra X-ray Observatory}~\citep{Weisskopf+etal+2002}. \citet{Broos+etal+2013} adopted a Naive Bayes approach to classify YSOs, using data from the Massive Young star-forming Complex Study in Infrared and X-rays (MYStIX) project~\citep{Feigelson+etal+2013}. \citet{Cody+etal+2014} classified YSOs based on the slope of SEDs at near- and mid-infrared wavelengths. \citet{Rapson+etal+2014} utilized a color-based classification scheme to classify YSOs with \textit{Spitzer}~ \citep{Fazio+etal+2004} data and Two Micron All Sky Survey~\citep[2MASS;][]{Skrutskie+etal+2006} photometry. \citet{Venuti+etal+2018} classified YSOs using spectroscopic and photometric data from the \emph{Gaia}-ESO Survey \citep[GES,][]{Gilmore+etal+2012,Randich+etal+2013} and the Coordinated Synoptic Investigation of NGC~2264 Survey \citep[CSI~2264,][]{Cody+etal+2014}. 

We describe in detail the YSO data collection, classification criteria, and astrometric parameters acquisition in Appendix~\ref{app:data}, and briefly introduce these procedures here. We collected YSO candidates in the S~Mon region from~\citet{Sung+etal+2009}, \citet{Broos+etal+2013}, \citet{Cody+etal+2014} and \citet{Rapson+etal+2014}, within an area of $202.8^{\circ} < l < 203.1^{\circ}$, and $2.0^{\circ} < b < 2.3^{\circ}$. Duplicates were eliminated through coordinate cross-matching with a radius of 1$\arcsec$, yielding in an initial sample of 1\,023 YSO candidates. As these candidates were classified by different methods, we reclassified them through a homogeneous classification scheme based on infrared data. 
We adopted the SED slope \citep[$\alpha = {\rm d} \log \lambda F_{\lambda} / {\rm d} \log \lambda$,][]{Lada+1987} and the classification scheme of \citet{Greene+etal+1994} to classify the evolutionary stage of the YSOs. Flat-spectrum sources are classified as Class II YSOs in this study. In line with \citet{Sung+etal+2009}, we use $\alpha \geq 0.3$ to classify objects as Class~I and $-1.8$ to 0.3 for Class~II. 
When using selection criteria based on infrared photometry, only the part of Class III YSOs with infrared-excess (i.e., ADs) can be classified. Therefore, X-ray data collected by~\citet{Flaccomio+etal+2023} were also employed to classify additional Class~III YSOs. The Class~III YSOs in this work are comprised of sources with ADs and disk-less PMS stars. For the former, the lower limit for a source with infrared-excess is given by~\citet{Lada+etal+2006} with $\alpha >$ -2.56. This slope is typical for stellar photospheres when there is no additional infrared-excess from a disk present, and when using the \textit{Spitzer} IRAC bands to calculate the infrared SED~\citep[see e.g.,][]{Lada+etal+2006,Teixeira+etal+2012}. Given the uncertainties in photometry, however, we took -2.3 as a rough limit to exclude possible contaminants from field stars. For disk-less PMSs, they have a SED slope of below -2.3 and X-ray emission. 
The SED fitting was performed without reddening corrections following~\citet{Cody+etal+2014}. In this case, we gain a sample of seven Class~I, 160 Class~II, and 245 Class~III YSOs (101 ADs + 144 disk-less PMSs). The parallaxes and proper motions of these YSOs were taken from \emph{Gaia}~DR3~\citep{gaia2022}. The radial velocities (RVs) of the YSOs were collected from the Sloan Digital Sky Survey (SDSS) DR17 APOGEE-2~\citep{Abdurrouf+etal+2022} and the catalog of \citet{Tobin+etal+2015}. 

To summarize, we end up with a sample of one Class I, 145 Class II, and 231 Class III YSOs that had matched with the Gaia DR3 catalog, and 47 Class II and 91 Class III YSOs with available RV information.

To compare the gas and stellar RVs, the heliocentric RVs ($V_{\rm Helio}$) of the YSOs were converted to the local standard of rest velocity ($V_{\rm LSR}$). We followed the conversion method described by~\citet{Reid+etal+2009} via:
\begin{equation}
    V_{\rm LSR} = V_{\rm Helio} + (V_{\odot} \sin l + U_{\odot} \cos l) \cos b + W_{\odot} \sin b,
\end{equation}
where $l$ and $b$ are the Galactic longitude and latitude, respectively, and $U_{\odot}, V_{\odot}$, and $W_{\odot}$ denote the solar motion components in the direction of the Galactic Center, along Galactic rotation, and toward the north Galactic pole, respectively. The used values for $U_{\odot}, V_{\odot},$ and $W_{\odot}$ are 10.3, 15.3, and 7.7~\kms, respectively~\citep{Kerr+Bell+1986}.

\section{Results}
\label{sect:res}
In the following we present the relationship between the molecular gas and YSOs in Section~\ref{subsect:gas_ysos}. Then, the distance to the S~Mon region is estimated in Section~\ref{subsect:dist} using YSOs that are closely correlated to the gas. In Section~\ref{subsect:bubble}, we present the bubble observed in the S Mon region. An outflow is detected in one of clumps, as described in Section~\ref{subsect:outflow}.

\subsection{Relationship between the Molecular Gas and YSOs}
\label{subsect:gas_ysos}
\begin{figure}[htbp]
    \centering
    \subfigure[]{\includegraphics[width = 0.49\textwidth]{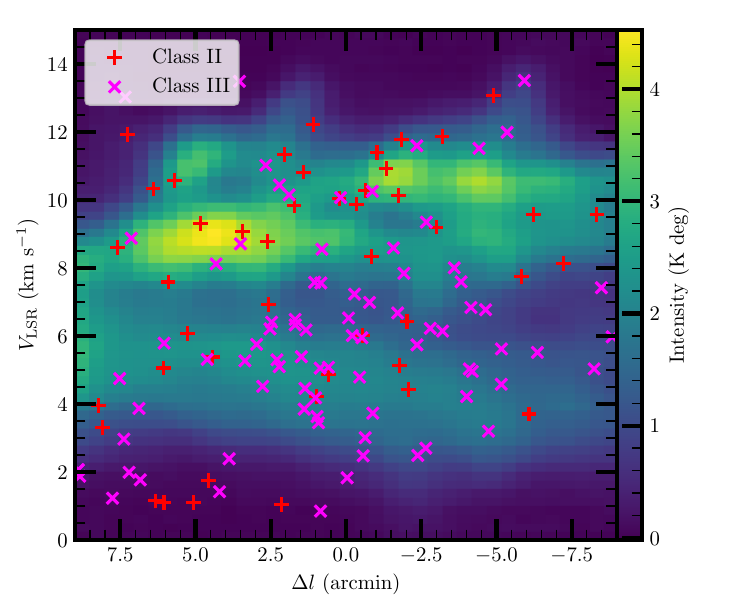}}
    \subfigure[]{\includegraphics[width = 0.49\textwidth]{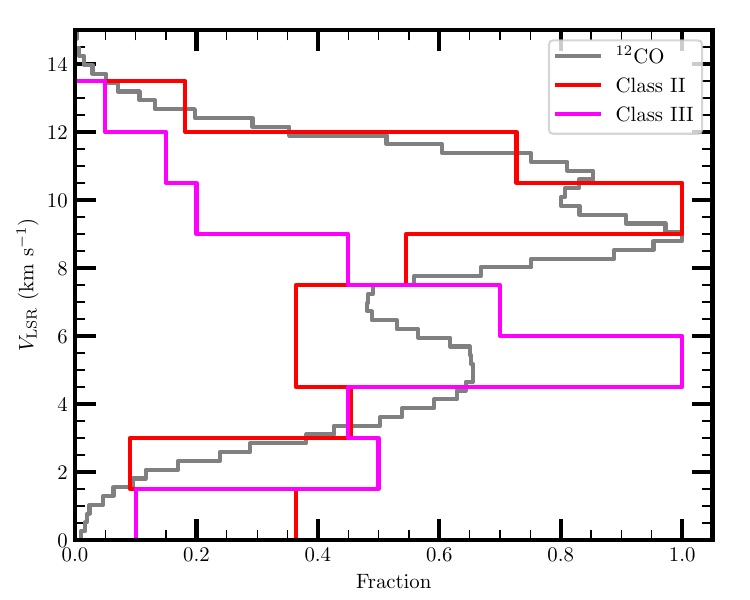}}
    \caption{(a) P-V diagram showing \coa~(background), Class~II (red pluses), and Class~III (magenta crosses) YSOs in S~Mon. (b) Normalized histogram of $V_{\rm LSR}$ for \coa~(gray line), Class~II (magenta line), and Class III (red line) YSOs.}
    \label{fig:lv}
\end{figure}

Figure~\ref{fig:lv} shows a position--velocity (P-V) diagram (a) and the normalized $V_{\rm LSR}$ distribution (b) for YSOs and molecular gas traced by \coa~in S~Mon. The \coa~emission shows two velocity components in this region, with one ranging from 0 to 7~\kms\, and the other from 7 to 15~\kms. The presence of a gap or minimum in the CO emission at $V_{\rm LSR} \sim 7$~\kms~is in agreement with the results of~\citet{Tobin+etal+2015}. In panel (b) of Figure~\ref{fig:lv}, the $V_{\rm LSR}$ of the majority (62\%) of the Class II YSOs peak at about 10~\kms, which is consistent with the \coa~emission component ranging from 7 to 15~\kms. Most (71\%) Class~III YSOs have LSR velocities between 0 and 7~\kms, with a peak at 5~\kms.

\begin{figure}[htbp]
    \centering
    \subfigure[\coa, 0--7~\kms]{\includegraphics[width = 0.49\textwidth]{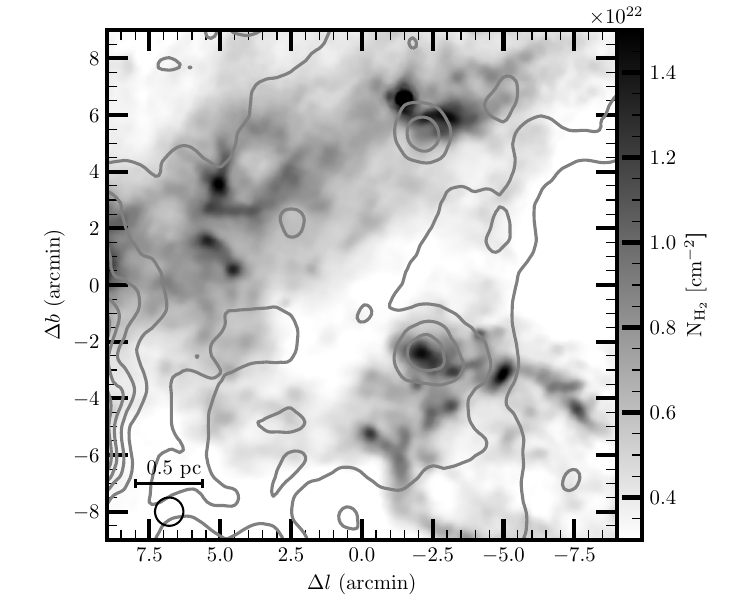}}
    \subfigure[\hco, 0--7~\kms]{\includegraphics[width = 0.49\textwidth]{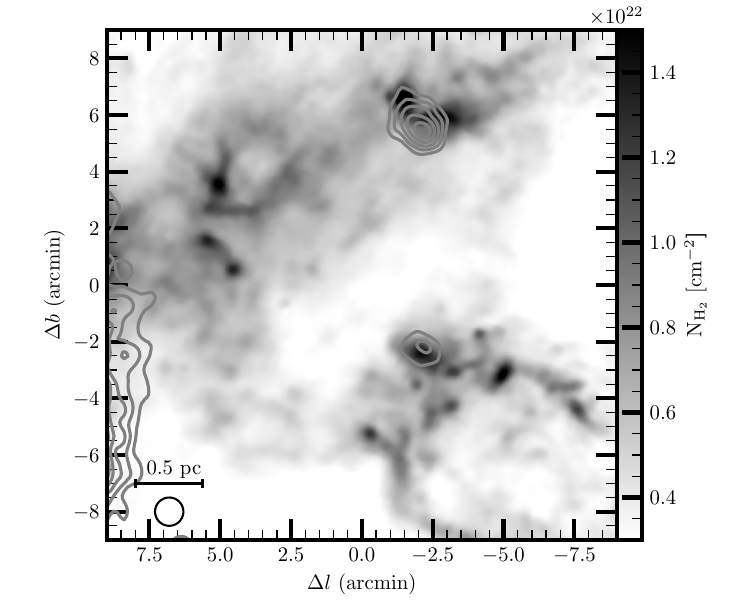}}
    \subfigure[\coa, 7--15~\kms]{\includegraphics[width = 0.49\textwidth]{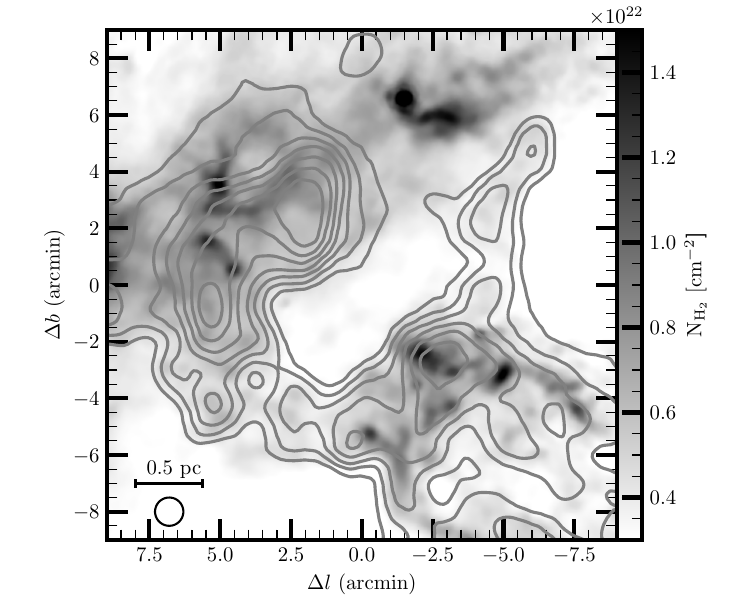}}
    \subfigure[\hco, 7--15~\kms]{\includegraphics[width = 0.49\textwidth]{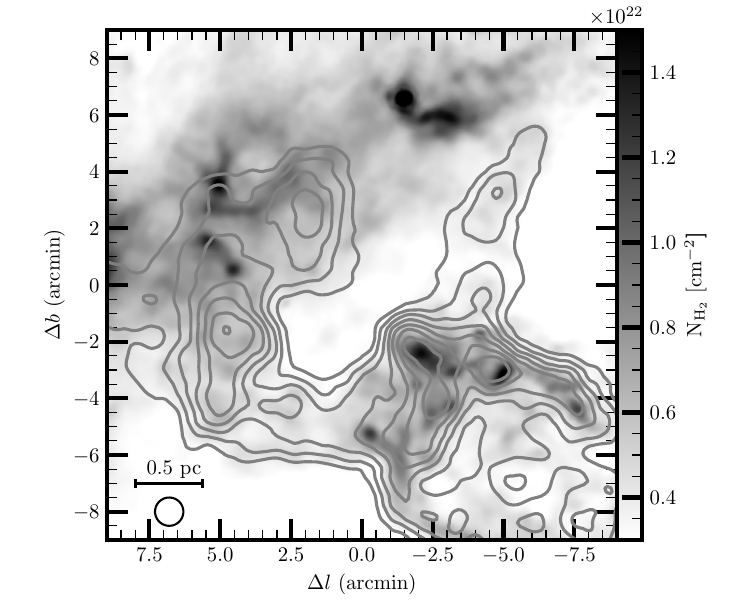}}
    \caption{Integrated maps of \coa~and \hco~over the velocity ranges of 0 to 7~\kms~and 7 to 15~\kms. The contour levels are shown from 30\% to 90\% with steps of 10\% of the peak integrated intensity, and the black open circle in the lower left corner of each panel represents the beam size of PMODLH. The gray background shows the {\it Herschel} H$_{2}$ column density map~\citep{Nony+etal+2021}. In all panels, the (0, 0) offsets corresponds to $(l, b) = (202.95^{\circ}, 2.15^{\circ})$.}
    \label{fig:channel_0_7}
\end{figure}

We investigate the two velocity components of molecular gas in S~Mon separately. Figure~\ref{fig:channel_0_7} shows integrated maps of \coa~and \hco~within the velocity range from 0 to 7~\kms~(panels (a) and (b)) and from 7 to 15~\kms~(panels (c) and (d)). The integrated map of molecular gas is superposed on a {\it Herschel}~\citep{Pilbratt+etal+2010} ${\rm H_{2}}$ column density map of this region, with a resolution of 18$\arcsec$~\citep{Nony+etal+2021}. The \coa~and \hco~emission between 0 and 7~\kms~appears to be diffuse and is not consistent with the {\it Herschel} ${\rm H_{2}}$ column density map. This component seems to dominate in the easternmost area of the region. In particular in the \coa~map (Figure~\ref{fig:channel_0_7} (a)) there appears to be a somewhat stronger component starting from the left of the image, beyond the shown sky cutout. Our investigation does not include the easternmost region. The \coa~and \hco~emission between 7 and 15~\kms~shows an arc-like structure, possibly indicative of a ``bubble" (centered at (0, 0) offset, see more details in Section~\ref{subsect:bubble}). We note that the distribution of the molecular gas within this velocity range is in agreement with the distribution of the {\it Herschel} H$_{2}$ column density map.

\begin{figure}[htbp]
    \centering
    \subfigure[Class~II]{\includegraphics[width = 0.49\textwidth]{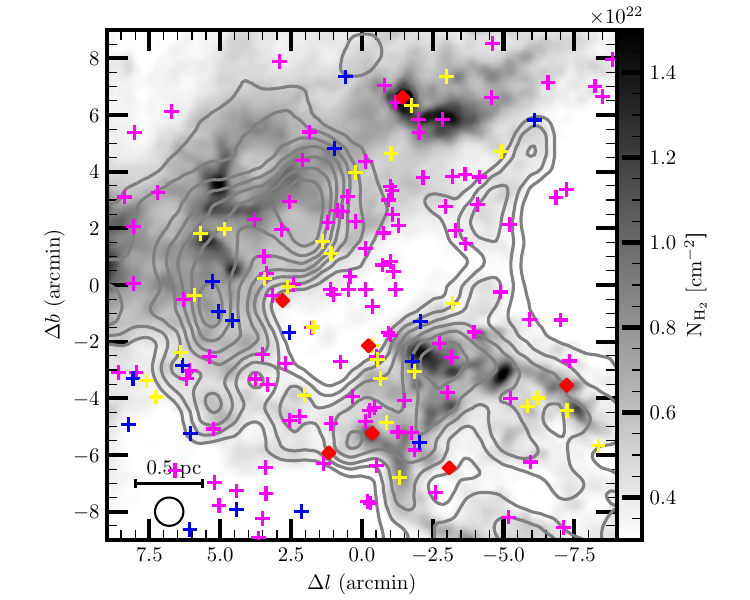}}
    \subfigure[Class~III]{\includegraphics[width = 0.49\textwidth]{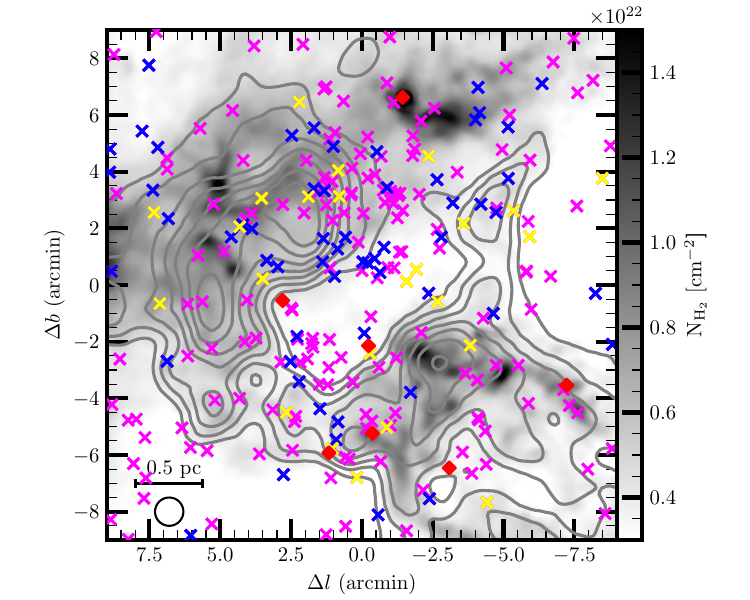}}
    \caption{Projected distribution of Class~II (left) and Class~III (right) YSOs with different velocity ranges. The blue pluses (crosses) indicate Class~II (III) YSOs with $V_{\rm LSR}$ ranging from 0 to 7~\kms. The yellow pluses (crosses) indicate Class~II (III) YSOs with $V_{\rm LSR}$ ranging from 7 to 15~\kms. The magenta pluses (crosses) represent Class~II (III) YSOs with unknown $V_{\rm LSR}$. The red diamonds represent Class~I YSOs. The gray contours and background are the same as in Figure~\ref{fig:channel_0_7}.}
    \label{fig:map_yso}
\end{figure}

Figure~\ref{fig:map_yso} depicts the projected distribution of the YSOs in combination with \coa~emission and {\it Herschel} H$_{2}$ column density map. Both the Class II and Class III YSOs, which have LSR velocities between 7 and 15~\kms~ (yellow pluses), generally coincide with the denser regions of the molecular cloud. More members associated with the molecular gas are found belong to the Class~II YSOs.

Compared with the Class~III YSOs, the Class~II YSOs are better correlated with the molecular gas in both spatial distribution and kinematic properties.
Considering the association between the molecular gas and the Class~II YSOs, we are able to use the Class~II YSOs to characterize the properties of the molecular gas, such as its distance and kinematics. The proper motions of the YSOs are used to derive the kinematics of the molecular gas in the plane of the sky, since they are the missing part, and cannot be measured directly for clouds.

\clearpage
\subsection{Distance to S~Mon}
\label{subsect:dist}
\begin{figure}[htbp]
    \centering
    \subfigure[Distribution of Class~II YSOs]{\includegraphics[width = 0.49\textwidth]{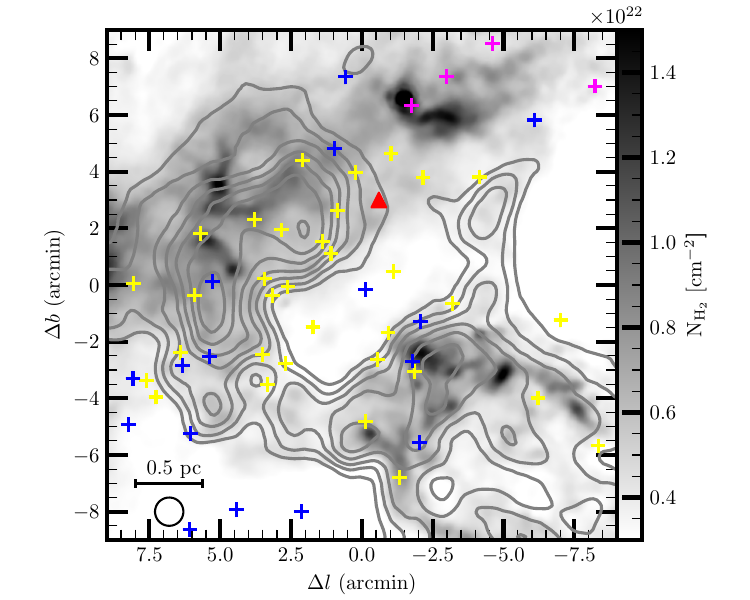}}
    \subfigure[CDF of \emph{Gaia} distance]{\includegraphics[width = 0.48\textwidth]{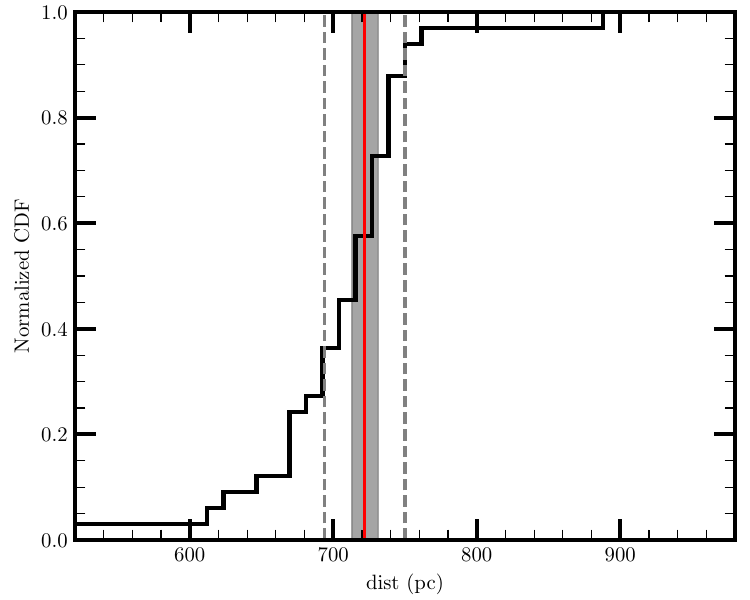}}
    \caption{(a) Projected distribution of Class~II YSOs with parallax uncertainties smaller than 10\%. The yellow pluses are the Class~II YSOs used for the distance estimation. The blue pluses show YSOs with LSR velocities smaller than 7~\kms. The magenta pluses show YSOs which do not coincide with the gas distribution. The red triangle represents the projected position of 15~Mon. (b) CDF of \emph{Gaia} distances. The red vertical line indicates the median value of 722~pc, with a standard error of 9~pc denoted by the gray shadow. The upper and lower percentiles within 1$\sigma$ around the median are labeled by the dashed gray lines.}
    \label{fig:dist}
\end{figure}

Accurate distances to molecular clouds are essential to study their physical properties. We employ the \emph{Gaia}~DR3 parallaxes of the Class~II YSOs to estimate the distance to S~Mon. In this work, only Class~II YSOs with parallax accuracies better than 10\% are selected for distance estimation. First, outliers of parallaxes beyond $3 \sigma$ from the median value are excluded.
Class II YSOs, which do not coincide with the gas distribution (CO at 7--15~\kms, magenta pluses in Figure~\ref{fig:dist} (a)), are also excluded from the distance estimation.
Finally, a total 35 YSOs that meet these criteria are selected to estimate the distances shown (yellow pluses in Figure~\ref{fig:dist} (a)).

The distance is determined from \emph{Gaia} parallaxes with $d({\rm pc}) = 1\,000 / \varpi$, where $\varpi$ is the parallax in units of mas. \citet{Lindegren+etal+2021} provided a method to correct the parallax zero points, however for red sources with $m_{G_{\rm BP}} - m_{G_{\rm RP}} > 1.6$, the correction could lead to a large parallax bias of about 40~$\mu$as. Since the photometric data of the used Class II YSOs have generally redder colors, we do not apply the parallax zero-point correction in this work.

We estimate the distance to the cloud directly using the YSOs associated with the cloud and their \textit{Gaia}~DR3 parallaxes. The cumulative distribution function (CDF) of \emph{Gaia} distances is shown in panel (b) of Figure~\ref{fig:dist}, where the median distance estimate is of 722 $\pm$ 9~pc. The uncertainty is derived by $\sigma / \sqrt{N}$, where $\sigma$ represents the dispersion of the distance, and $N$ represents the number of stars.

\citet{Flaccomio+etal+2023} conducted a study on the NGC~2264 star-forming region using X-ray data obtained with {\it XMM-Newton} telescope. The S~Mon core region in their study appears to be close to the region we studied, although we note that they did not give the specific coordinates of the region they investigated. Although only half of the samples we used for distance estimation aligned with their sample counterparts, our distance estimation is similar to their results (719 $\pm$ 9~pc). However, the distance we derive is smaller than the distance estimated ($759_{-26}^{+10} \pm 37$~pc) by~\citet{Zucker+etal+2020}, who inferred the distance by fitting a line-of-sight dust model with \textit{Gaia}~DR2 data. We conducted a parallax difference test between Gaia DR2 and DR3 using a selected set of Class II YSOs in our region. These YSOs have LSR velocities ranging from 0 to 15~\kms, and both DR2 and DR3 exhibit parallax accuracies better than 10\%. The difference between these two groups is approximately 25~pc, meaning the two estimates agree within their given uncertainties. Although we did not test the method of \citet{Zucker+etal+2020} explicitly, variations in methods could also introduce some differences in distance estimations. Therefore, the difference in distance between this work and that of \citet{Zucker+etal+2020} could be caused by the difference in both method and data. In the subsequent analysis, we adopt 722 $\pm$ 9~pc as the distance to the S Mon cloud, which has a velocity ranging between 7 and 15~\kms.

\clearpage
\subsection{Bubble}
\label{subsect:bubble}
We identified the bubble in S~Mon based on four criteria summarized from previous studies~\citep[e.g.,][]{Arce+etal+2011,Li+etal+2015,Feddersen+etal+2018}: (1) the shell shows a circular or arc-like structure in integrated CO or \hco~maps; (2) the P-V diagram of the shell shows an expansion signature (i.e., a ``U" or ``V" shape); (3) the CO or \hco~shell matches the morphology of infrared data in at least one band; and (4) the shell contains a candidate driving source. An expanding bubble model requires three parameters: bubble radius ($R_{\rm bub}$), expansion velocity ($V_{\rm exp}$), and central velocity ($V_{\rm cnt}$). We followed the method of~\citet{Arce+etal+2011} to estimate these parameters. The radius is determined by a Gaussian fit to the azimuthally averaged profile of the integrated intensity map. The central and expansion velocities can be roughly read from a P-V diagram, where the upper part of the ``U" or ``V" structure corresponds to the central velocity and the difference between the central velocity and the lowest point of ``U" or ``V" structure indicates the expansion velocity. Then, the central and expansion velocities are corrected by visual fitting based on the channel maps. As indicated by~\citet{Arce+etal+2011}, the uncertainty in the central velocity of a bubble is about $\pm 0.5$~\kms, and since the bubble is not detected over its entire velocity range, the expansion velocity is obtained as a lower limit.

\begin{figure}[htbp]
    \centering
    \includegraphics[width = 5.5cm]{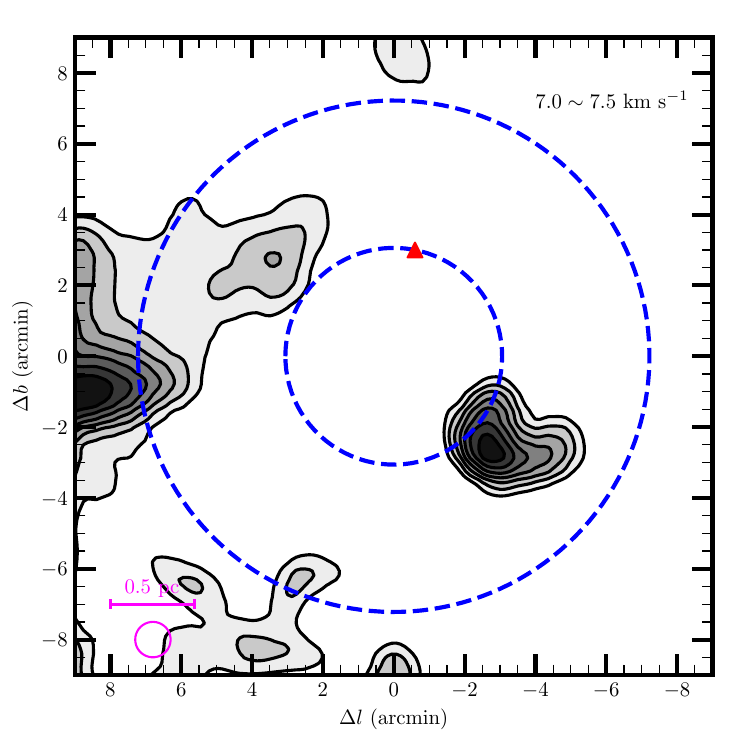}
    \includegraphics[width = 5.5cm]{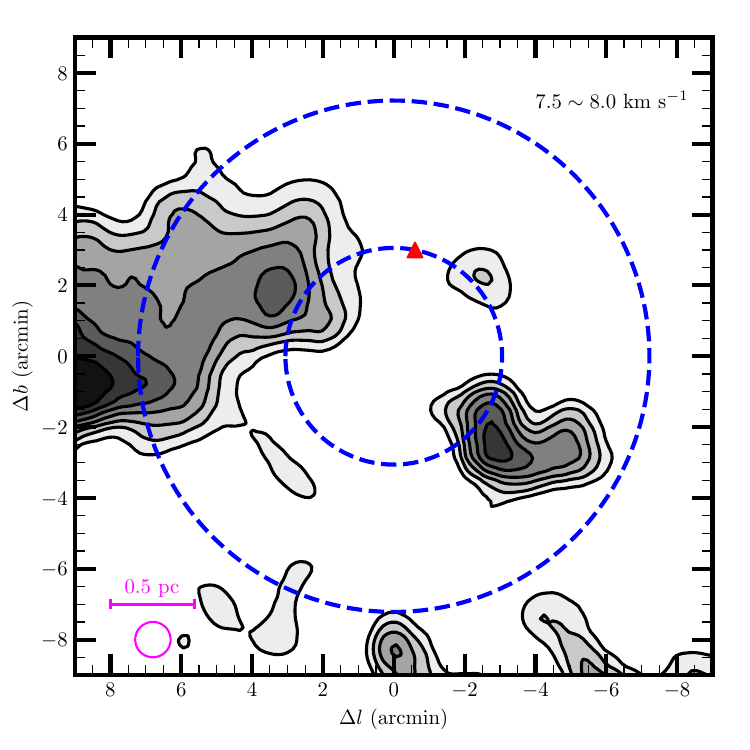}
    \includegraphics[width = 5.5cm]{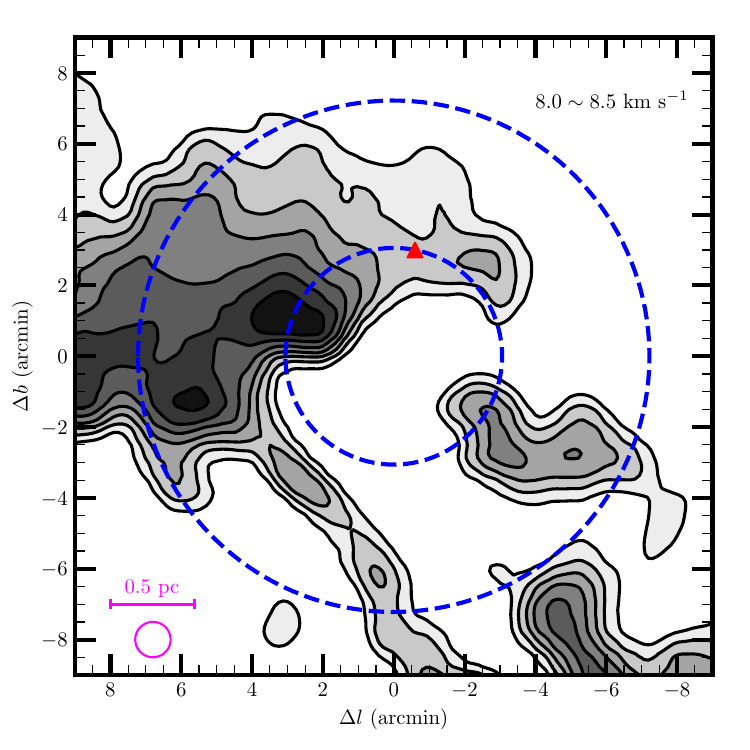}
    \includegraphics[width = 5.5cm]{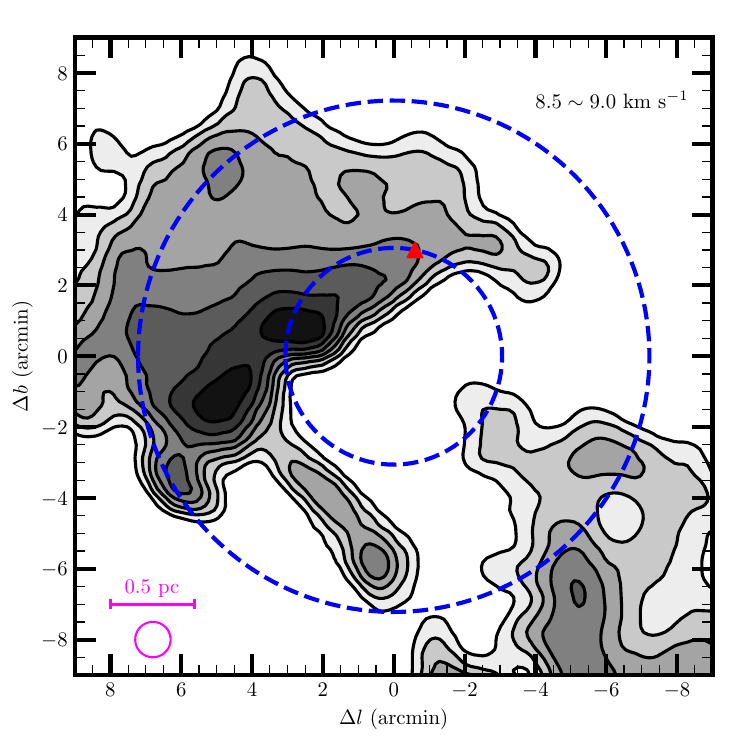}
    \includegraphics[width = 5.5cm]{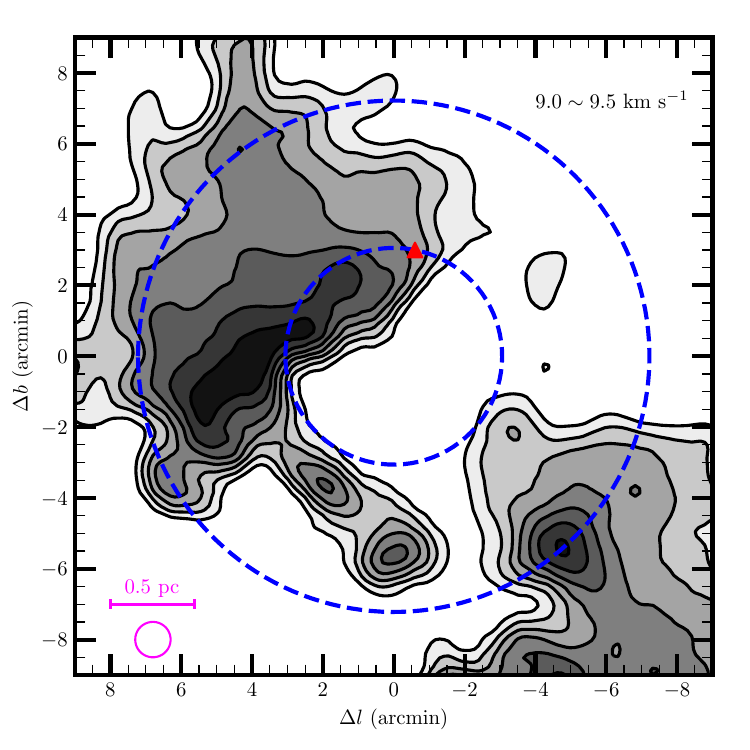}
    \includegraphics[width = 5.5cm]{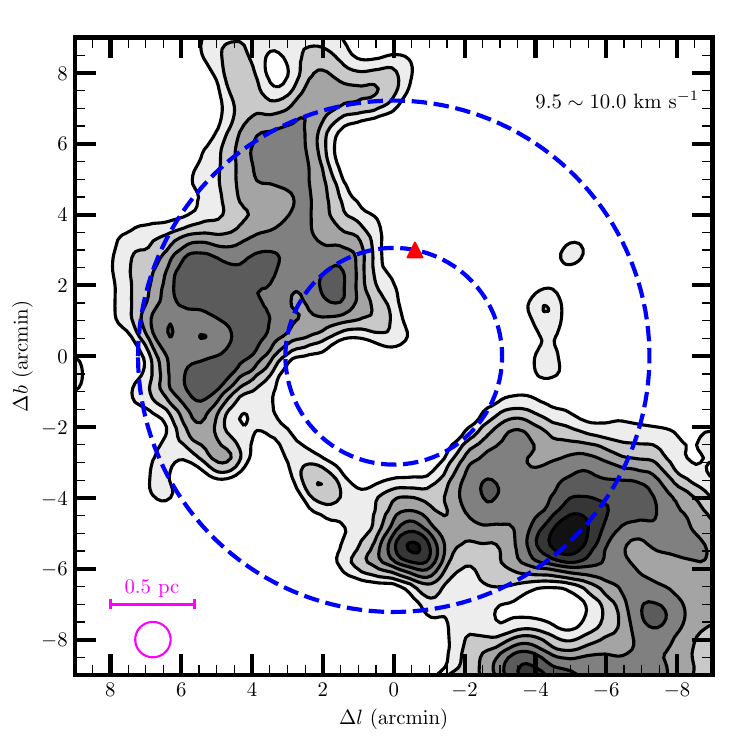}
    \includegraphics[width = 5.5cm]{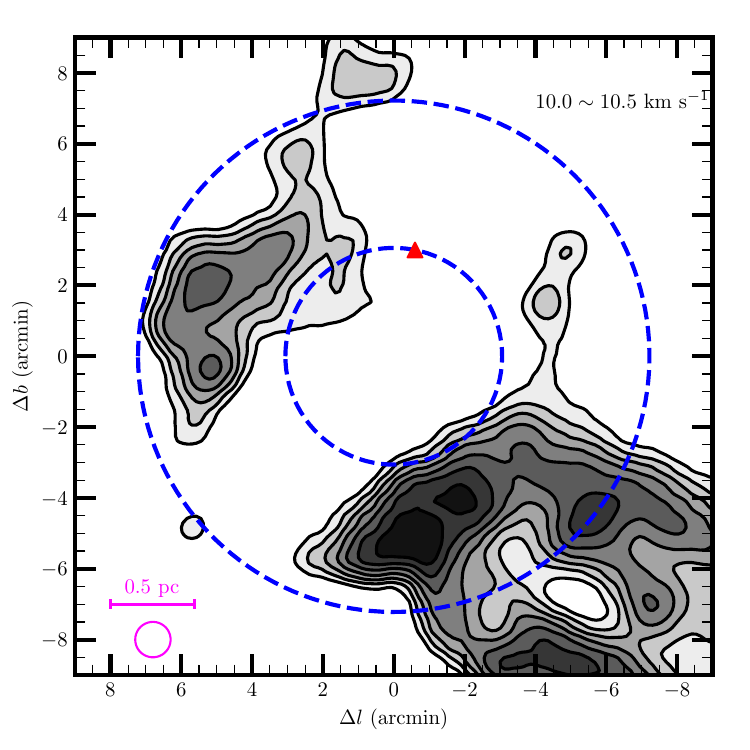}
    \includegraphics[width = 5.5cm]{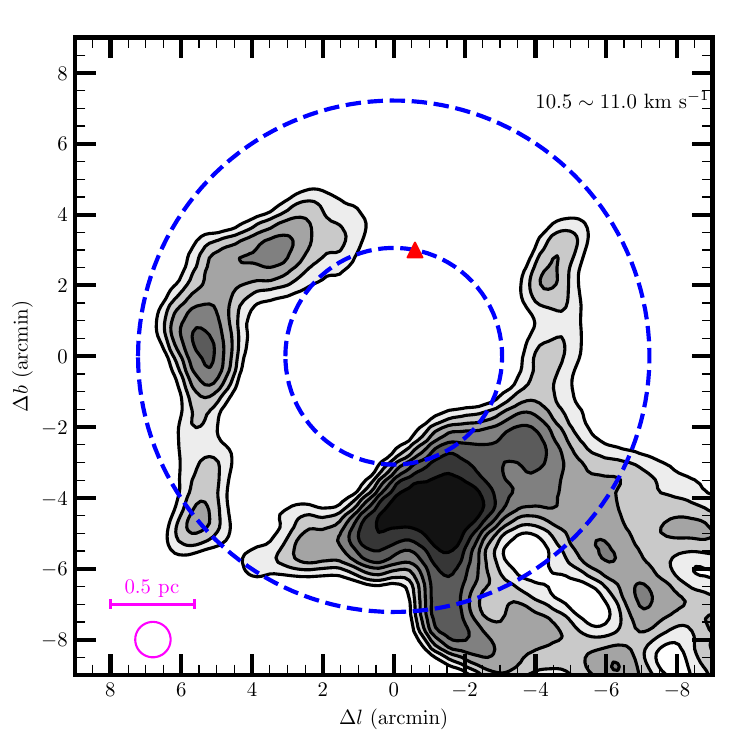}
    \includegraphics[width = 5.5cm]{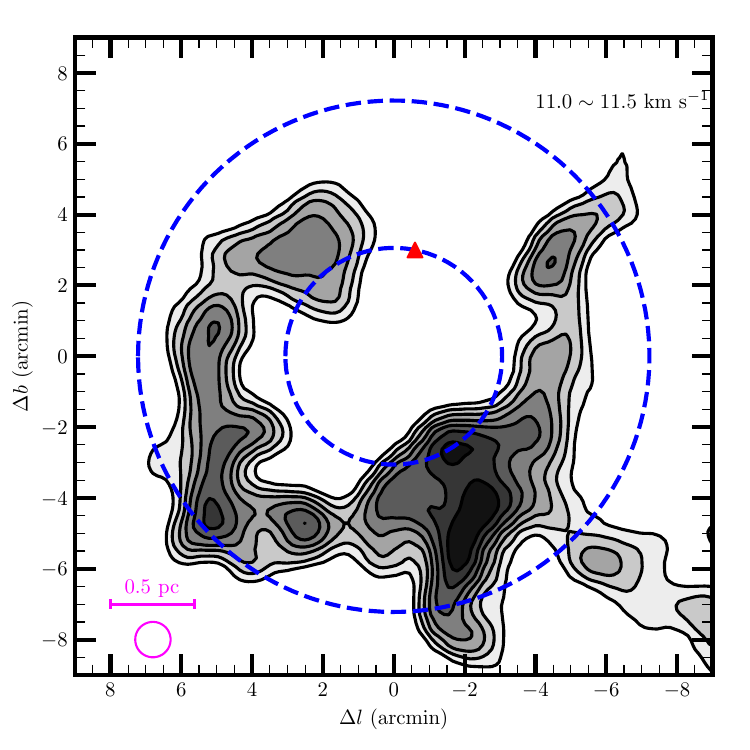}
    \includegraphics[width = 5.5cm]{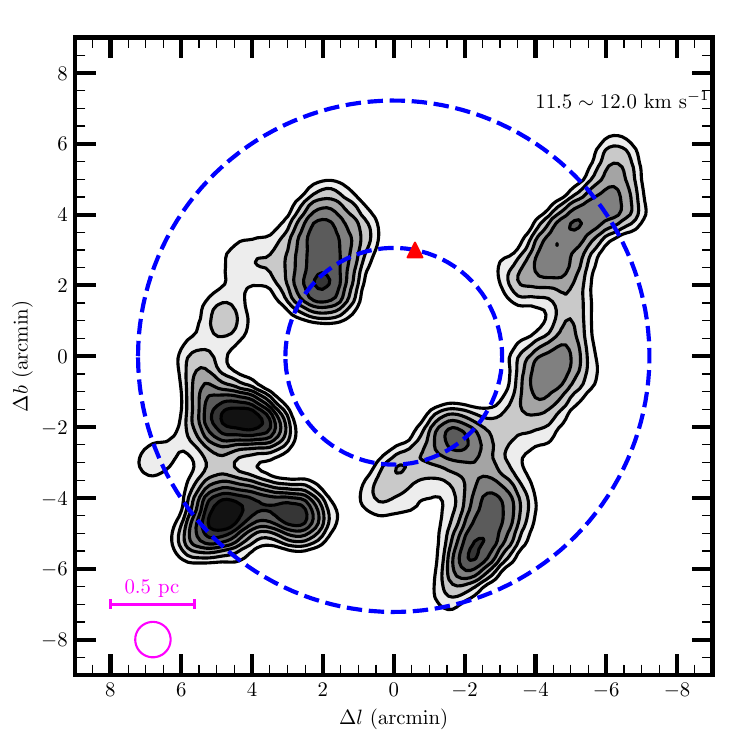}
    \includegraphics[width = 5.5cm]{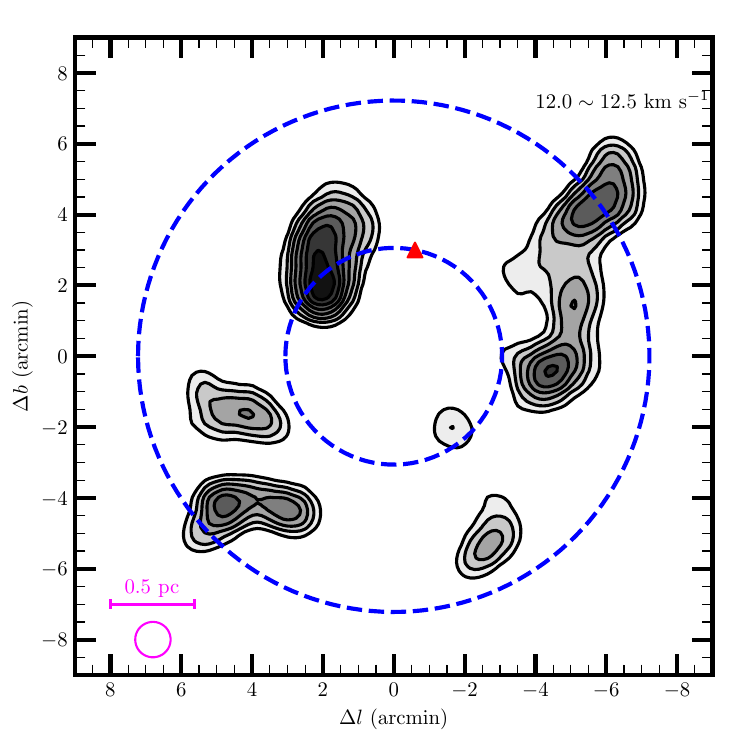}
    \includegraphics[width = 5.5cm]{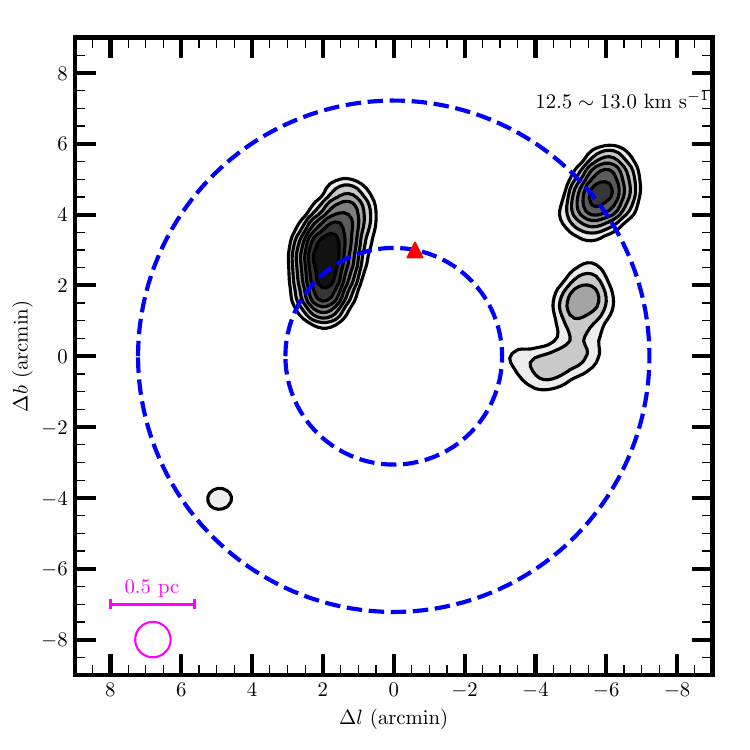}
    \caption{Channel maps of \coa~emission. The red triangle represents the star 15~Mon. The blue dashed circle shows the bubble obtained by fitting the gas, which has a radius and thickness of $\sim 5.4$ and $\sim 4.2$~arcmin, respectively. The magenta circle indicates the beam size of PMODLH. The velocity range for each map is labeled in the upper right of the map. }
    \label{fig:channel_map_12co}
\end{figure}

\begin{figure}[htbp]
    \centering
    \includegraphics[width = 5.5cm]{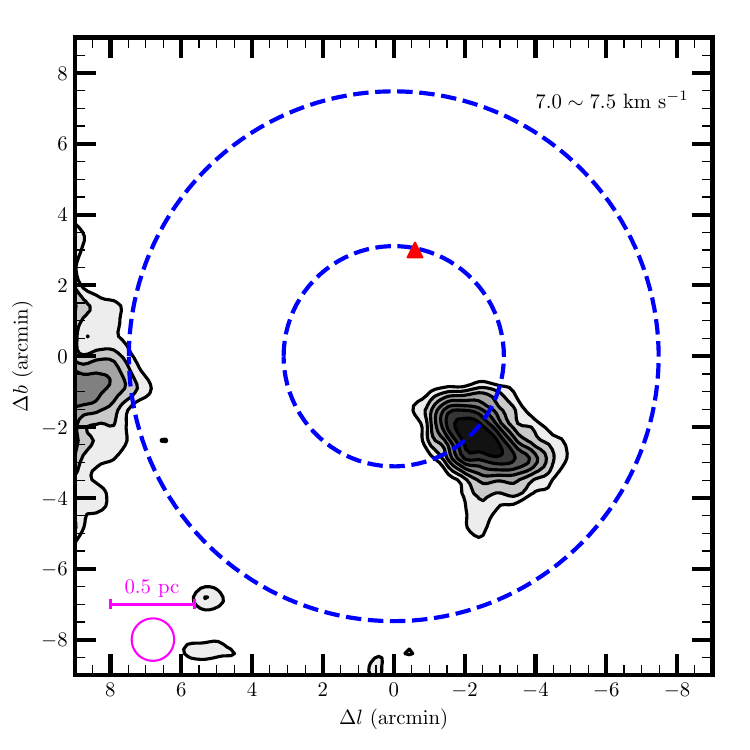}
    \includegraphics[width = 5.5cm]{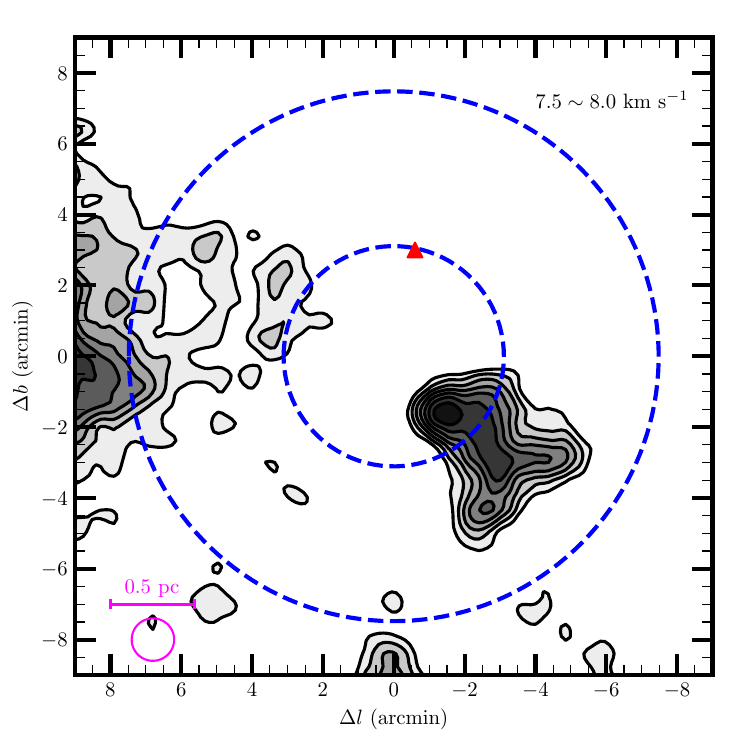}
    \includegraphics[width = 5.5cm]{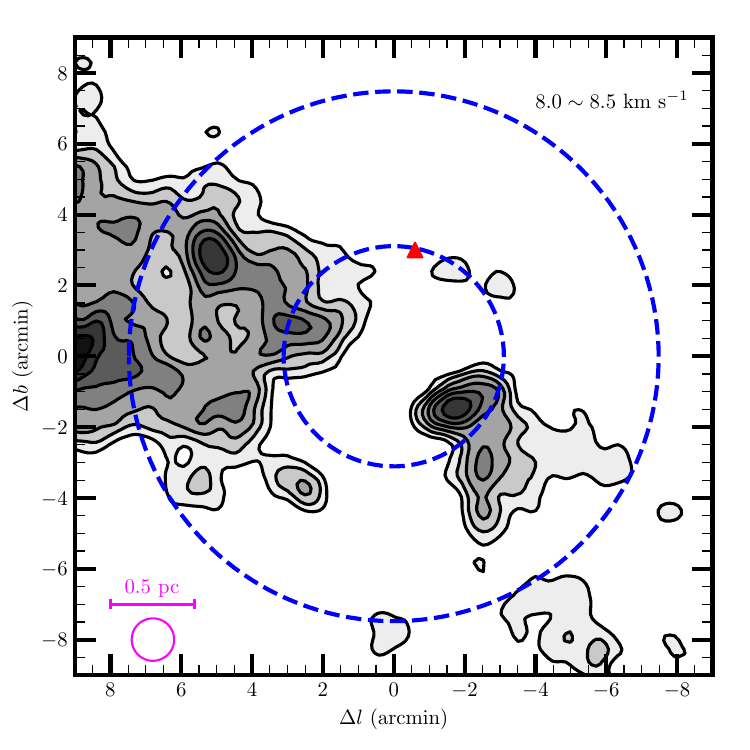}
    \includegraphics[width = 5.5cm]{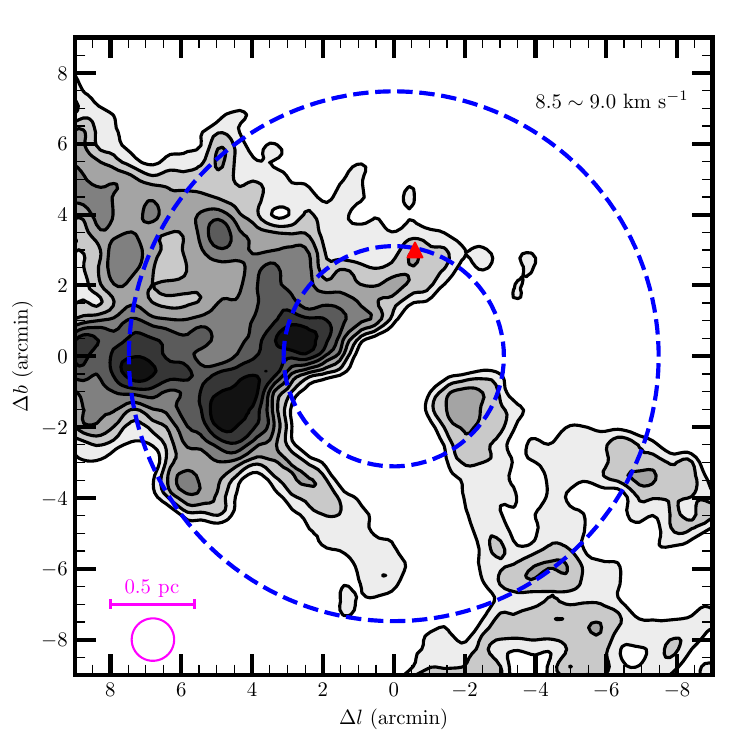}
    \includegraphics[width = 5.5cm]{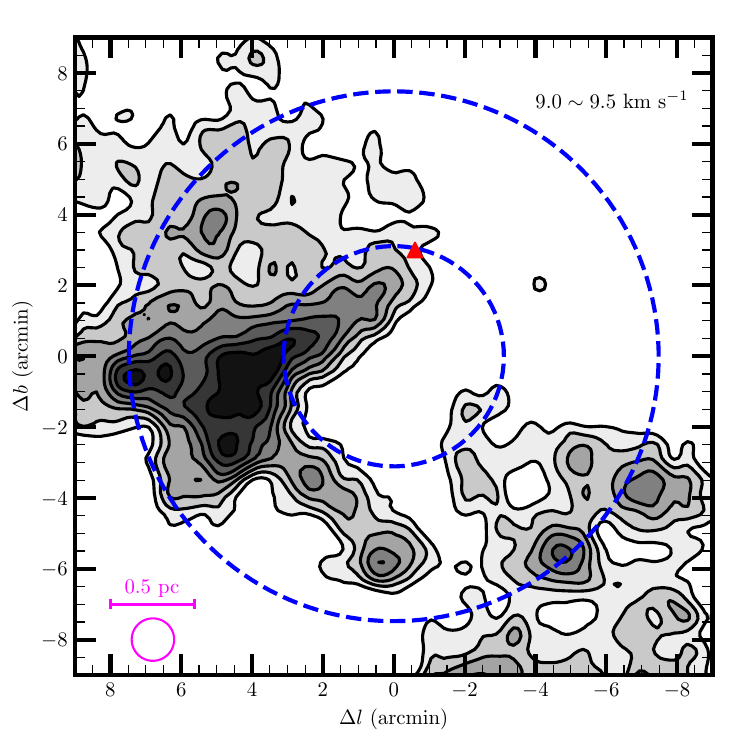}
    \includegraphics[width = 5.5cm]{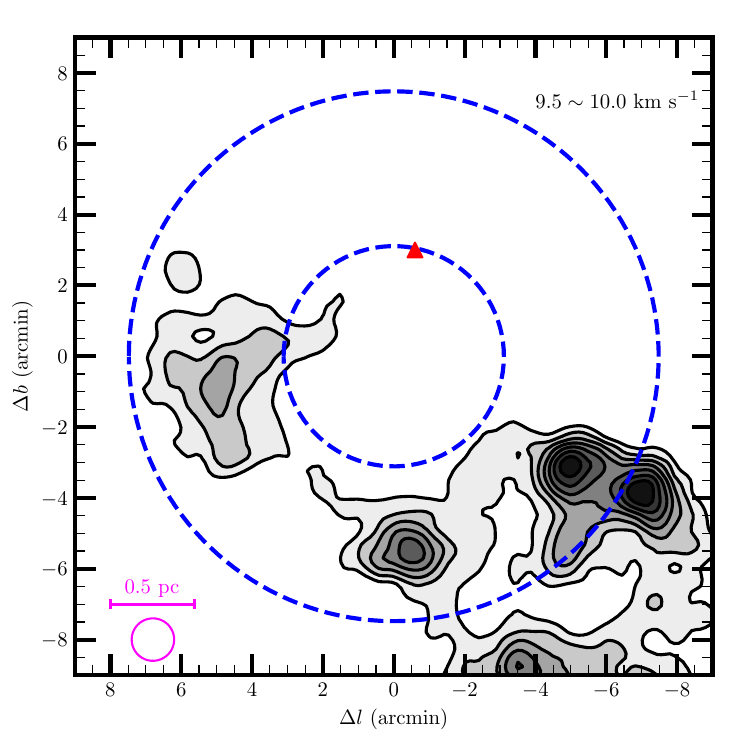}
    \includegraphics[width = 5.5cm]{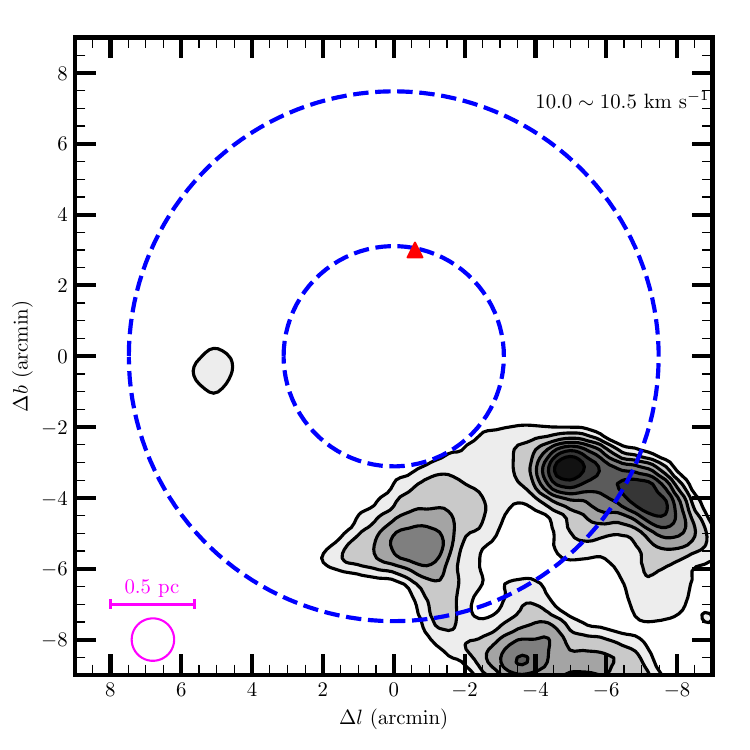}
    \includegraphics[width = 5.5cm]{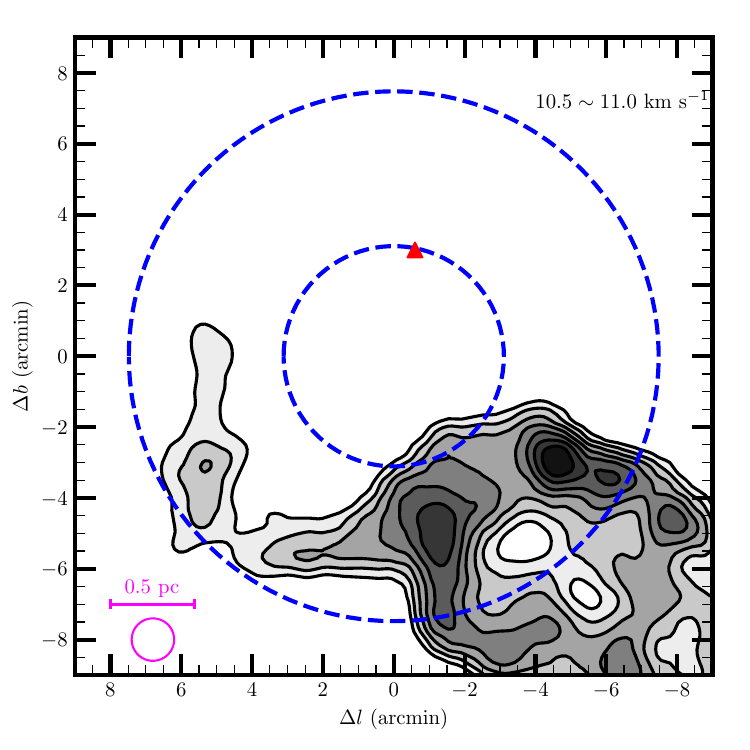}
    \includegraphics[width = 5.5cm]{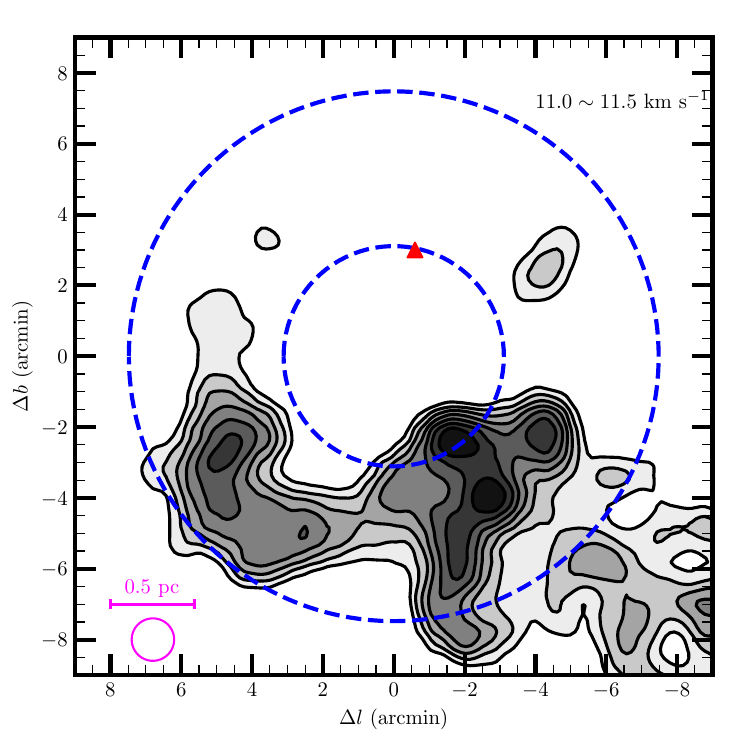}
    \includegraphics[width = 5.5cm]{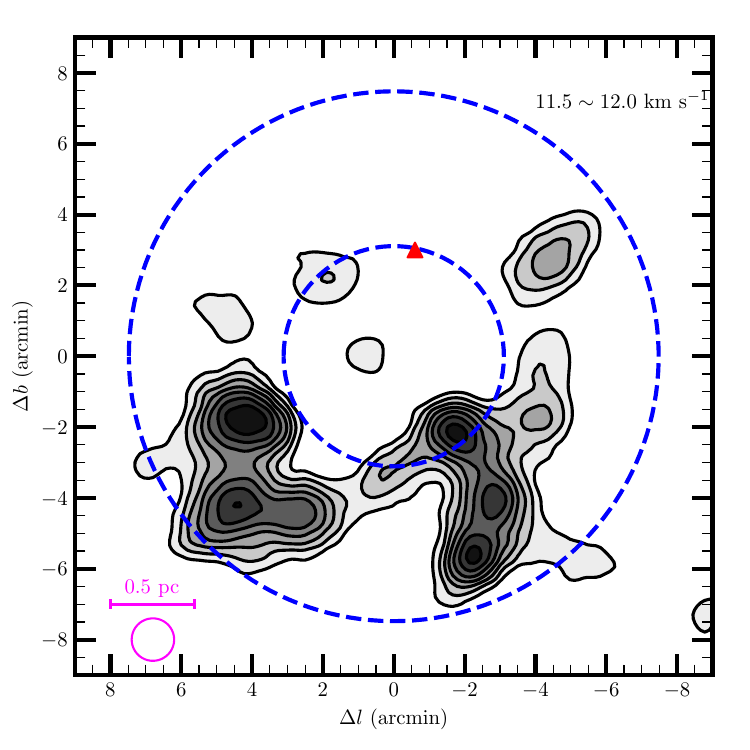}
    \includegraphics[width = 5.5cm]{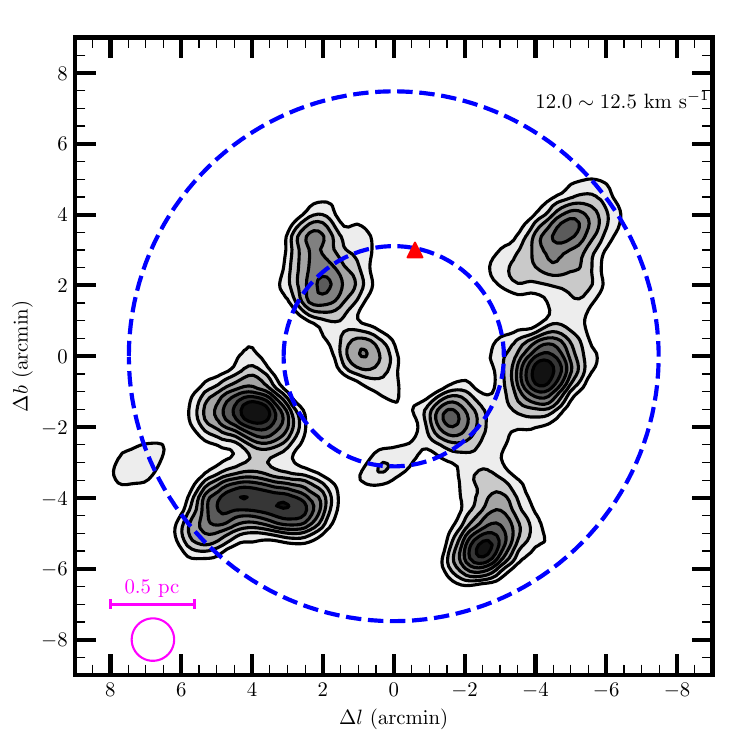}
    \includegraphics[width = 5.5cm]{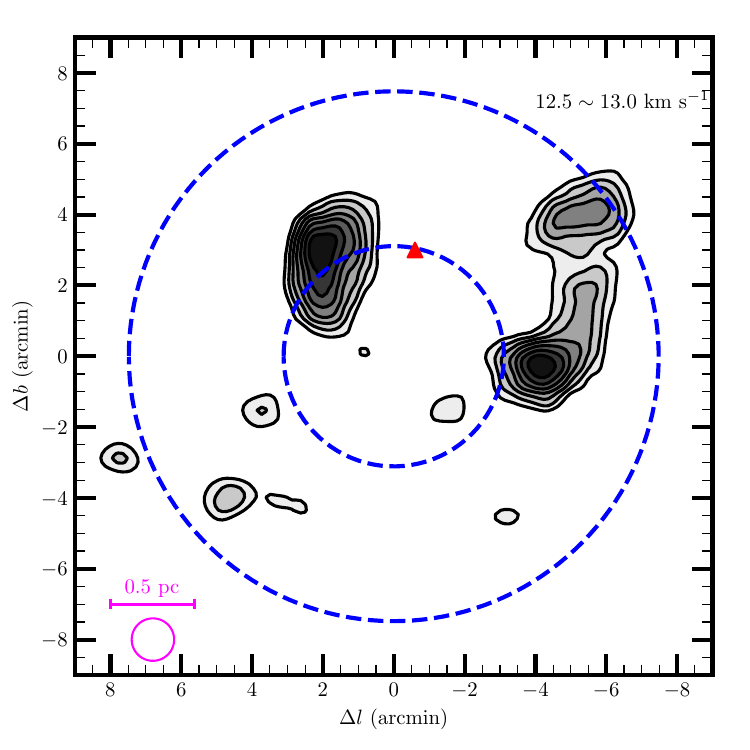}
    \caption{Channel maps of \hco. The blue dashed circle shows the bubble obtained by fitting the gas, which has a radius and thickness of $\sim 5.3$ and $\sim 4.4$~arcmin, respectively. The other labels and colors are the same as in Figure~\ref{fig:channel_map_12co}.}
    \label{fig:channel_map_hco}
\end{figure}

\begin{figure}[htbp]
    \centering
    \subfigure[Intensity map of  \coa]{\includegraphics[width = 0.35\textwidth]{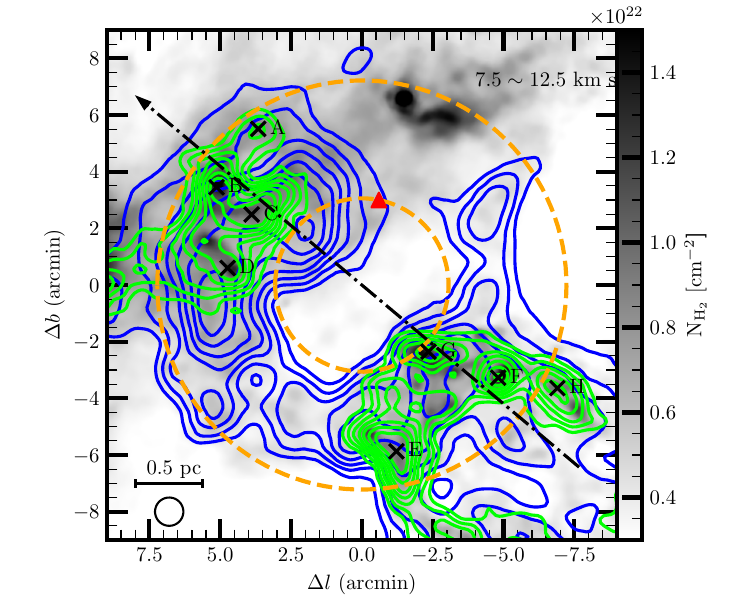}}
    \subfigure[P-V diagram of \coa]{\includegraphics[width = 0.35\textwidth]{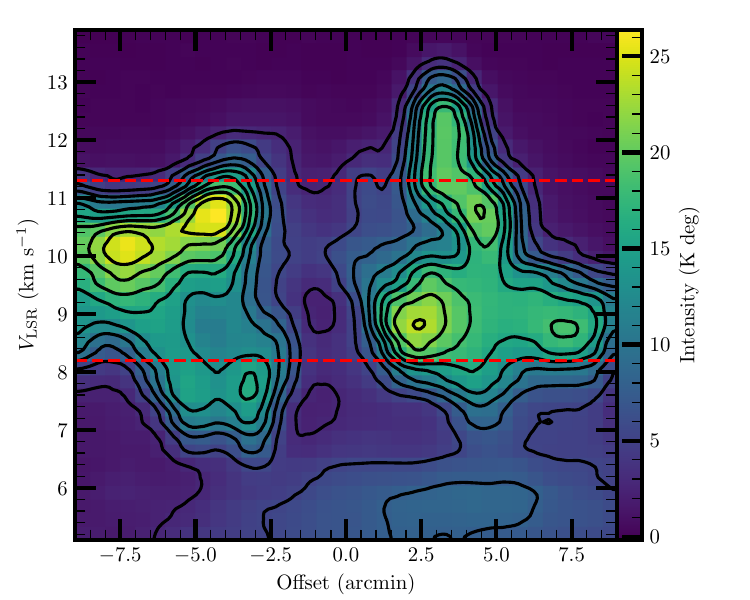}}
    \subfigure[Radial intensity profile of \coa]{\includegraphics[width = 0.28\textwidth]{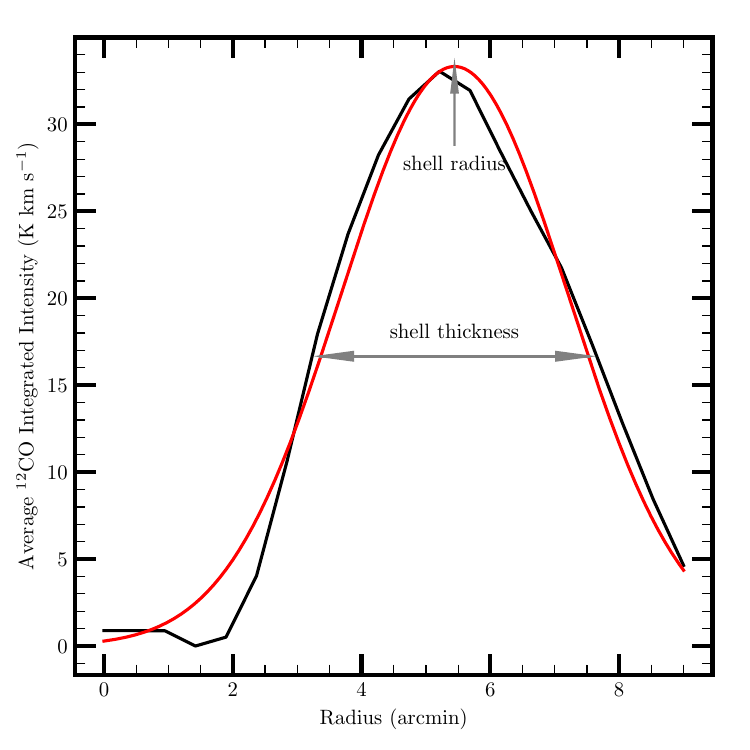}}
    \subfigure[Intensity map of \hco]{\includegraphics[width = 0.35\textwidth]{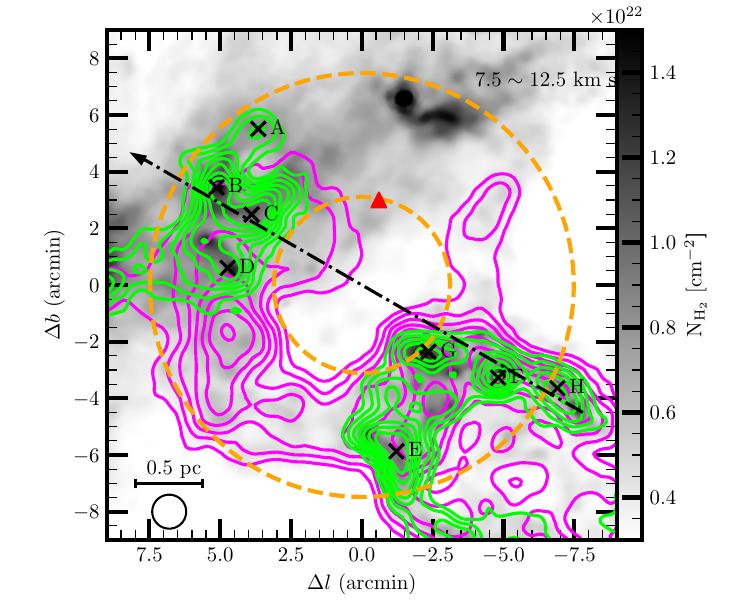}} 
    \subfigure[P-V diagram of \hco]{\includegraphics[width = 0.35\textwidth]{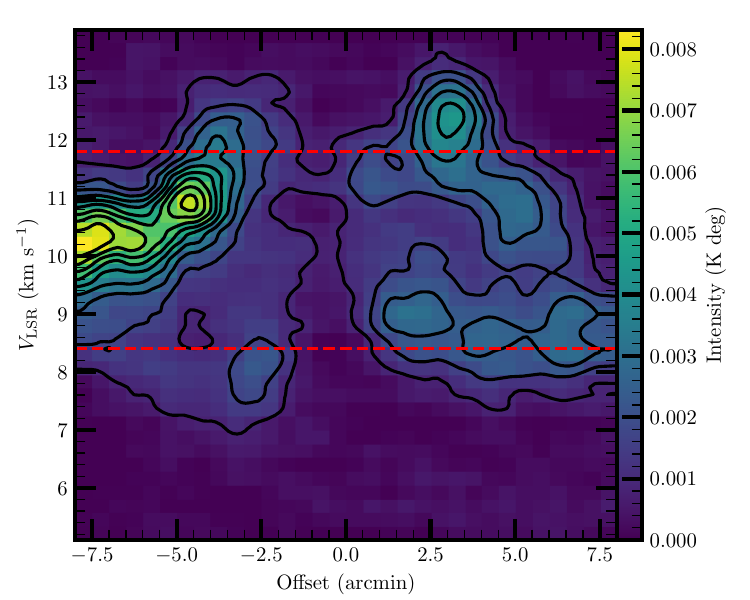}}
    \subfigure[Radial intensity profile of \hco]{\includegraphics[width = 0.28\textwidth]{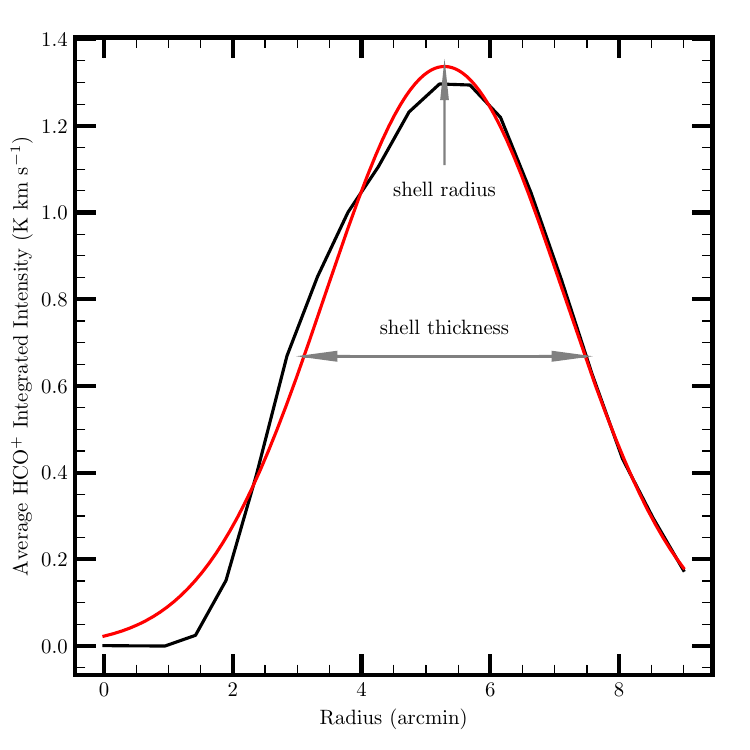}}
    \caption{(a) Integrated map of \coa~emission(blue contours) superposed on the \textit{Herschel} H$_{2}$ column density map~\citep[gray background,][]{Nony+etal+2021}, and the green contour lines are the emission of \coc. The integrated velocity range is $7.5 - 12.5$~\kms~ for both \coa~and \coc, and the contour levels are shown from 30\% to 90\% with steps of 10\% of the peak integrated intensity of the corresponding emission. The orange dashed circle shows the bubble. Clumps identified based on the \coc~integrated map are labeled A to H. (b) P-V diagram of \coa~along the arrow in panel (a), where the contour levels are shown from 10\% to 90\% with steps of 10\% of the peak value. The red dashed lines show the expansion velocity range from the visual inspection, i.e., $8.2 - 11.3$~\kms. (c) Azimuthally averaged radial intensity profile of the \coa~bubble (black curve), where a Gaussian fit (red curve) to the intensity profile was performed to estimate the radius (peak) and thickness (Full Width at Half Maximum, FWHM). (d) Integrated map of \hco~(magenta contours). The other labels are the same as in panel (a). (e) P-V diagram of \hco~along the arrow in panel (d), and the expansion velocity range is $8.4 - 11.8$~\kms. The other features are the same as in panel (b). (f) Azimuthally averaged radial intensity profile of the \hco~bubble, where the other labels and symbols are the same as in panel (c).}
    \label{fig:bubble}
\end{figure}

An arc-like structure is clearly visible in the channel maps of \coa~and \hco~(see the details in Figures~\ref{fig:channel_map_12co}--\ref{fig:channel_map_hco}), spanning a broad $V_{\rm LSR}$ range from 7.5 to 12.5~\kms. This structure is also evident in the integrated maps of \coa~and \hco~from 7.5 to 12.5~\kms~and the  \textit{Herschel} H$_{2}$ column density map~\citep[see panels (a) and (d) of Figure~\ref{fig:bubble},][]{Nony+etal+2021}. The P-V diagrams of \coa~and \hco~(panels (b) and (e) of Figure~\ref{fig:bubble}) both show an upside down ``U'' or ``V" shape. Based on the P-V diagram of \coa~(\hco), $V_{\rm exp}$  and $V_{\rm cnt}$ are derived as $\sim 3.1$~\kms ($\sim 3.4$~\kms) and $\sim 8.2$~\kms ($\sim 8.4$~\kms), respectively. The $V_{\rm exp}$ values estimated from \coa~and \hco~are similar, thus an average of 3.3~\kms~is adopted as a lower limit for the expansion velocity. The center of the bubble is set to be (0, 0). The radius and thickness of the bubble traced by the \coa~emission (see panel (c) of Figure~\ref{fig:bubble}) are $\sim 5.4$ and $\sim 4.2$~arcmin, respectively, corresponding to 1.13 and 0.88~pc at a distance of 722~pc. These values are compared with the bubble traced by \hco~emission (see panel (f) of Figure~\ref{fig:bubble}), i.e., a radius of $\sim 5.3$ arcmin ($\sim 1.11$~pc) and a thickness of $\sim 4.4$~arcmin ($\sim 0.92$~pc). Based on both the \coa~and \hco~emission and considering the uncertainties from the distance and the Gaussian fits, we derive the radius and thickness of the bubble to be $1.1 \pm 0.1$~pc and $0.9 \pm 0.1$~pc, respectively. The mean column density of the bubble is $\sim 2.5 \times 10^{20}~{\rm cm}^{-2}$ and the mass ($M_{\rm shell}$) is $\sim 1\,400 \pm 400~M_{\odot}$ (see the calculation and estimation of uncertainties in Appendix~\ref{sect:phy_para}).

\begin{figure}
    \centering
    \subfigure[\textit{Herschel} H$_{2}$ column density map]{\includegraphics[width = 0.32\textwidth]{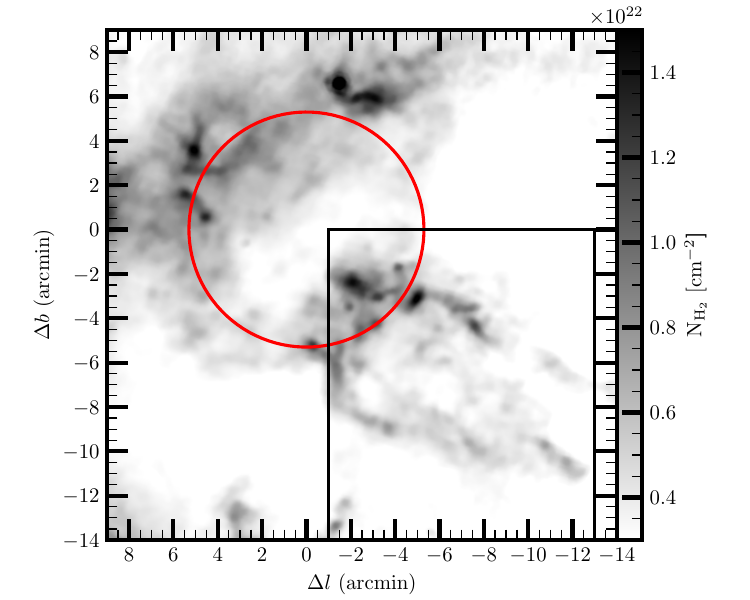}}
    \subfigure[Intensity map of  \cob]{\includegraphics[width = 0.32\textwidth]{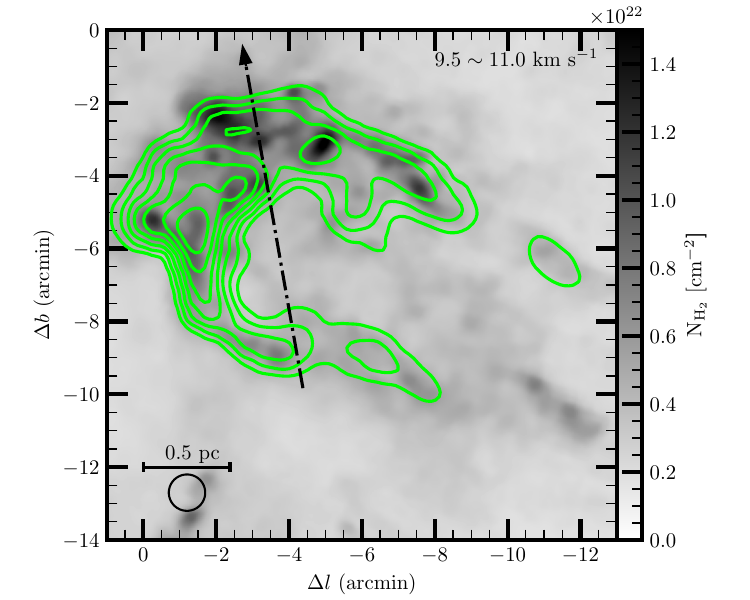}}
    \subfigure[P-V diagram of \cob]{\includegraphics[width = 0.32\textwidth]{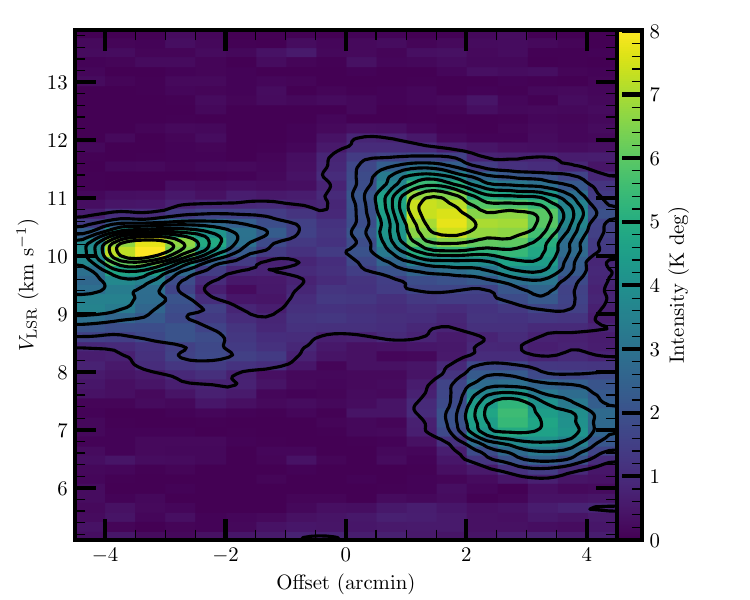}}
    \caption{(a) \textit{Herschel} H$_{2}$ column density map, showing a larger area to the southwest of the investigated region~\citep{Nony+etal+2021}. The red circle depicts the bubble we identified, and the black box shows the region housing the small arc-like structure. (b) Integrated map of \cob~emission (green contours) in the boxed region in panel (a). The integrated velocity range is $9.5 - 11.0$~\kms, and the contour levels are shown from 30\% to 90\% with steps of 10\% of the peak integrated intensity of \cob~emission. (c) P-V diagram of \cob~along the arrow in panel (b), where the contour levels are shown from 10\% to 90\% with steps of 10\% of the peak integrated intensity of the corresponding emission.}
    \label{fig:new_bubble}
\end{figure}

Another arc-like structure appears to be located in the southwest region of the {\it Herschel} H$_{2}$ column density map (see the boxed region of Figure~\ref{fig:new_bubble} (a)). The low-sensitivity CO data, observed using PMODLH, were employed to investigate the structure. The rms noise of the \coa, \cob~and \coc~spectra is 0.37, 0.20, and 0.20~K, respectively. A small arc-like structure is only evident in the \cob~integrated map (see panel (b) of Figure~\ref{fig:new_bubble}), so we use the \cob~data to produce a P-V diagram for the arc-like structure. However, there is no clear ``U" or ``V" shape in the P-V diagram (see panel (c) of Figure~\ref{fig:new_bubble}). In addition, we do not find a potential driving source for this arc-like structure. Moreover, using high-spatial-resolution CO maps (20$\arcsec$), \citet{Tauber+etal+1993} found that the properties of the northern region of the arc-like structure ($\Delta b > -7 \arcmin$) are consistent with the ionizing radiation from 15~Mon, while those of the southern region are not. In conclusion, the arc-like structure shown is unlikely to be a bubble. Given the wind-blown appearance, this feature might be caused by feedback from the northeast side, coincident with the general location of the massive O- and B-type stars in the region.

\begin{figure}
    \centering
    \subfigure[clump~A]{\includegraphics[width = 4.3cm]{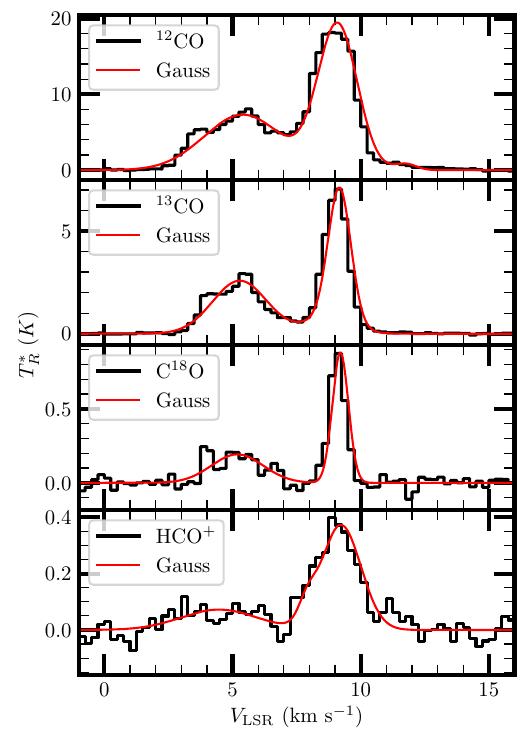}}
    \subfigure[clump~B]{\includegraphics[width = 4.3cm]{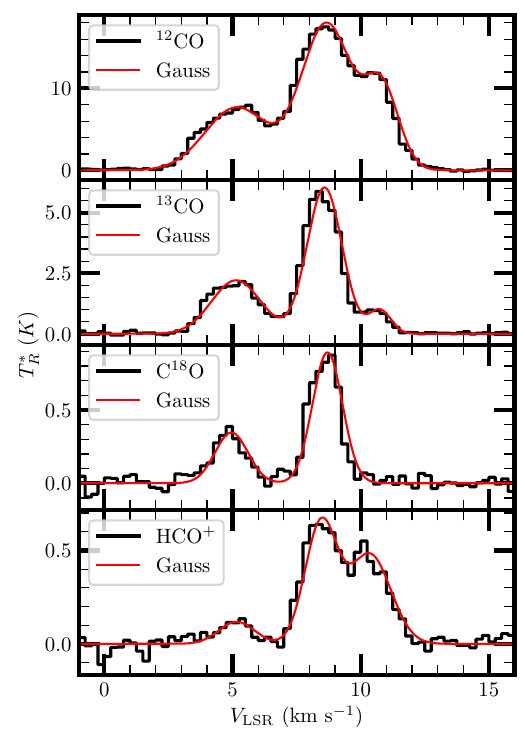}}
    \subfigure[clump~C]{\includegraphics[width = 4.3cm]{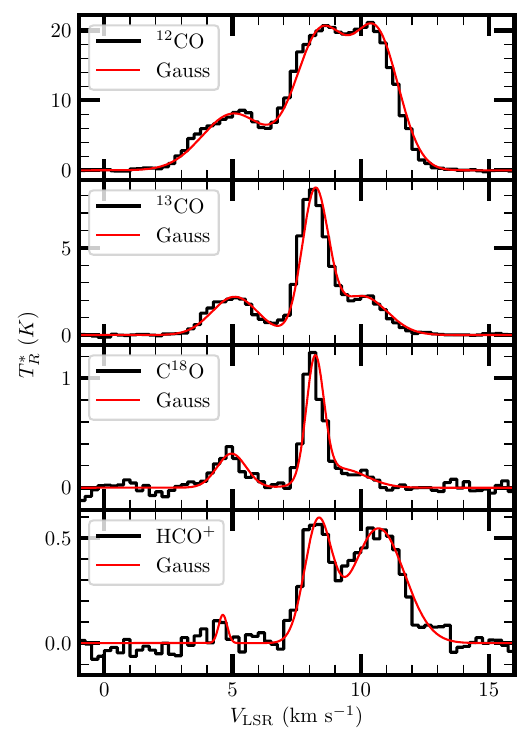}}
    \subfigure[clump~D]{\includegraphics[width = 4.3cm]{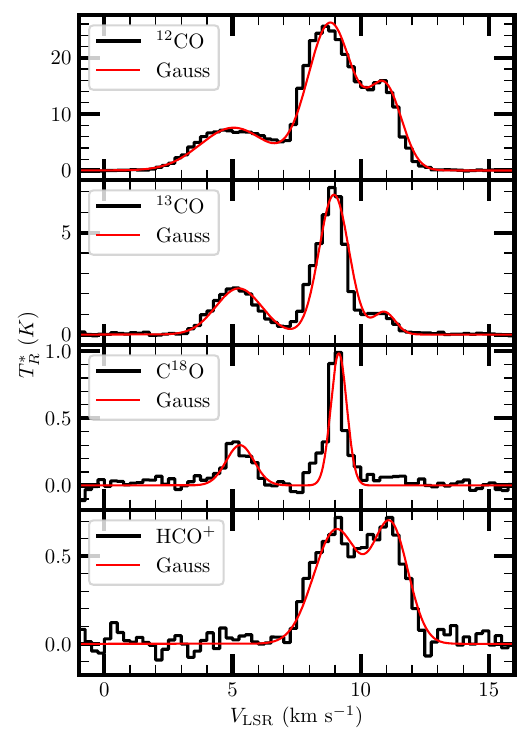}}
    \subfigure[clump~E]{\includegraphics[width = 4.3cm]{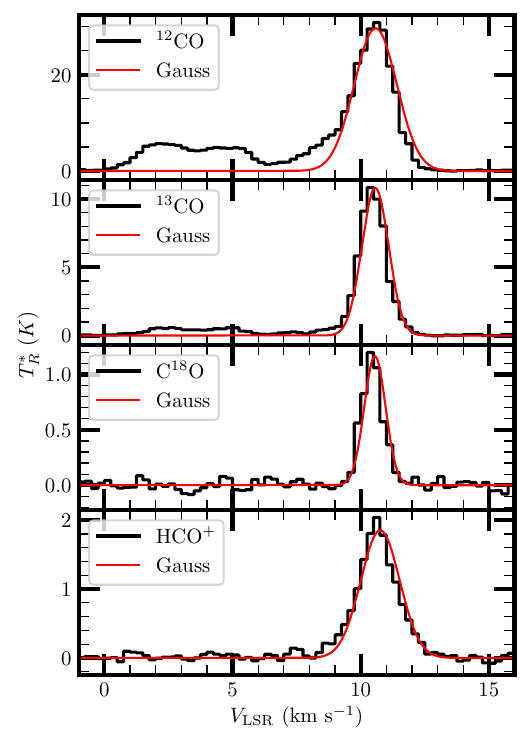}}
    \subfigure[clump~F]{\includegraphics[width = 4.3cm]{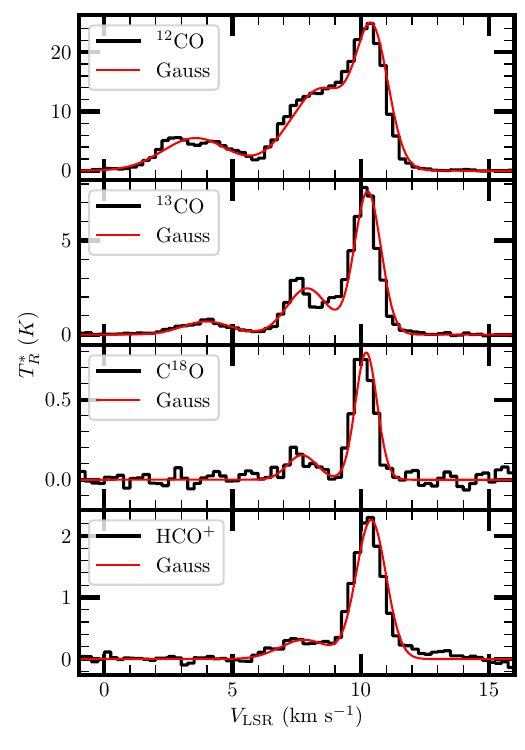}}
    \subfigure[clump~G]{\includegraphics[width = 4.3cm]{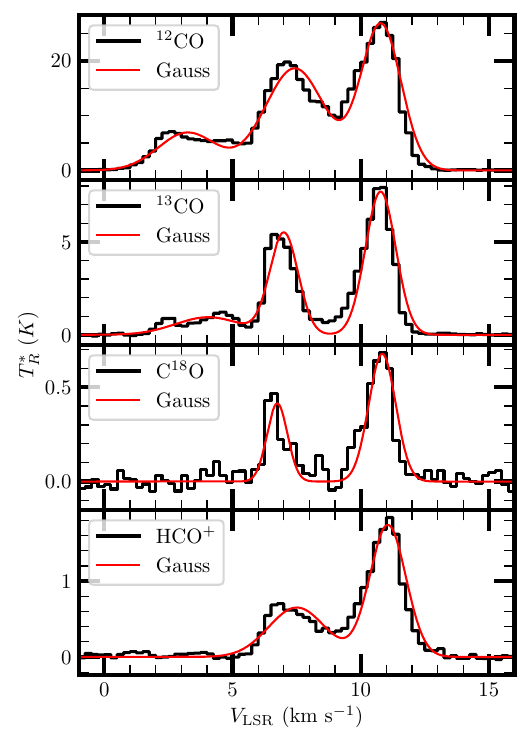}}
    \subfigure[clump~H]{\includegraphics[width = 4.3cm]{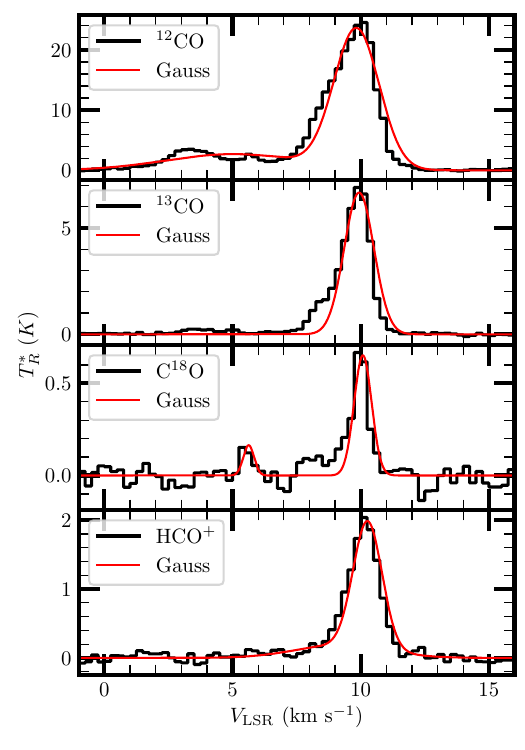}}
    \caption{Spectra of \coa, \cob, \coc, and \hco~of the \coc~peak position of clumps A to H. The red lines represent the sum of the multiple Gaussian fitting components.}
    \label{fig:spec}
\end{figure}

\begin{figure}
    \centering
    \includegraphics[width = 0.45\textwidth]{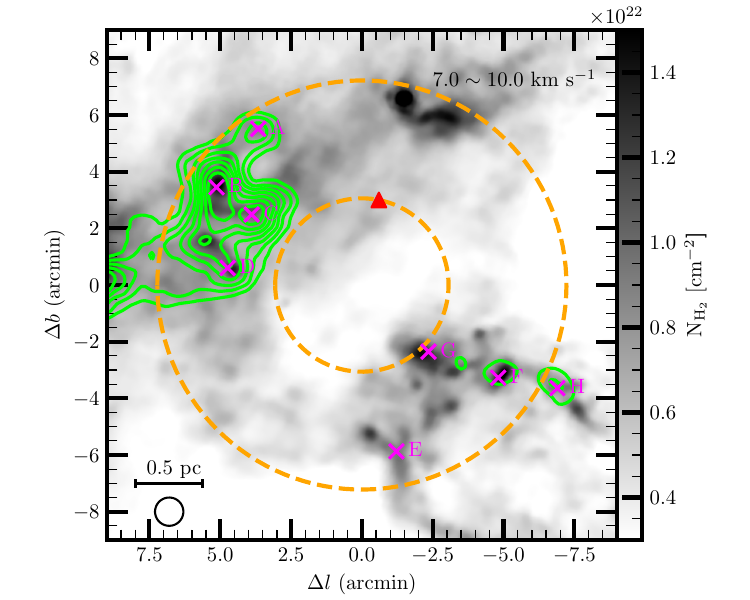}
    \includegraphics[width = 0.45\textwidth]{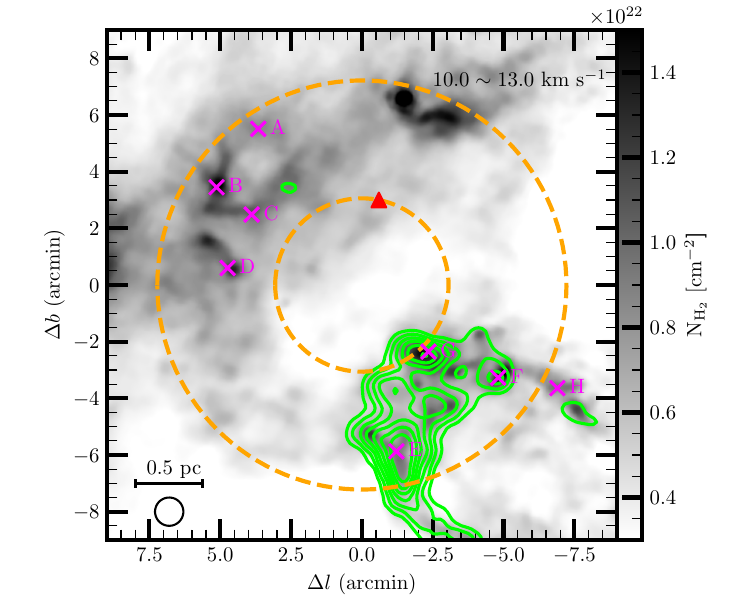}
    \caption{Integrated map of \coc~superposed on the {\it Herschel} column density map. The velocity range for the left panel is 7--10~\kms, and for the right panel is 10--13~\kms. The other features are the same as in Figure~\ref{fig:bubble} (a).}
    \label{fig:two_lobes}
\end{figure}

\begin{figure}
    \centering
    \includegraphics{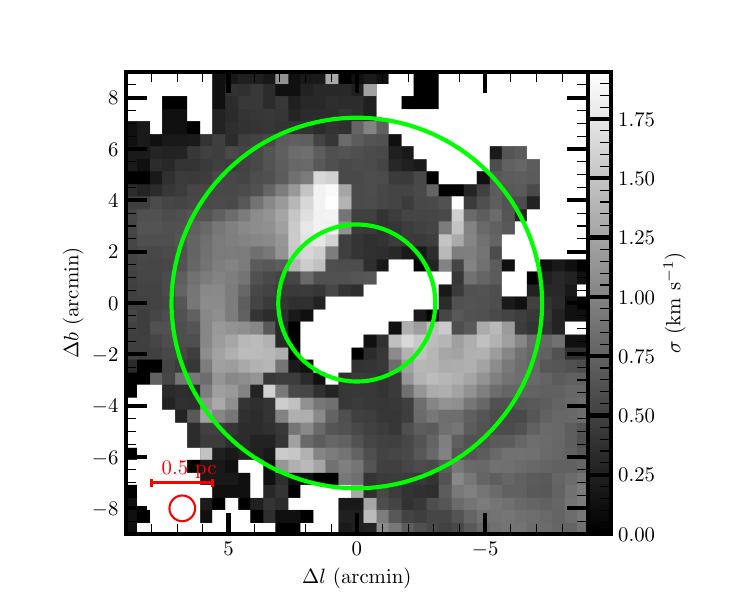}
    \caption{Distribution of the velocity dispersion ($\sigma$). The green circles indicate the bubble region.}
    \label{fig:sigma}
\end{figure}

There are eight clumps, labeled as A--H in Figure~\ref{fig:bubble}, identified from the integrated map of \coc~emission. Figure~\ref{fig:spec} plots the four molecular lines we observed toward the \coc~peak position of each clump. Table~\ref{tab:clump} lists the fits for the \coc~peak emission for each clump. There are three main components among these eight clumps, i.e., $3~{\rm km~s^{-1}} < V_{\rm LSR} < 7~{\rm km~s^{-1}}$, $7~{\rm km~s^{-1}} < V_{\rm LSR} < 10~{\rm km~s^{-1}}$, and $10~{\rm km~s^{-1}} < V_{\rm LSR} < 13~{\rm km~s^{-1}}$. As discussed in Section~\ref{subsect:gas_ysos}, the bubble is not associated with the component of $3~{\rm km~s^{-1}} < V_{\rm LSR} < 7~{\rm km~s^{-1}}$, but instead with the other two components. Clumps A to D correlate with the velocity component of $7~{\rm km~s^{-1}} < V_{\rm LSR} < 10~{\rm km~s^{-1}}$ and centered at $\sim$ 8.5~\kms, and clumps E to H correlate with the velocity component of $10~{\rm km~s^{-1}} < V_{\rm LSR} < 13~{\rm km~s^{-1}}$ and centered at $\sim$10.5~\kms. Figure~\ref{fig:two_lobes} shows integrated maps of \coc~with different velocity components, which are in agreement with the parameters given in Table~\ref{tab:clump}. These two groups of clumps correspond to different lobes of the bubble, so the different kinematic characteristics may be caused by the expansion of the bubble. Furthermore, the velocity dispersion around the bubble is larger (see Figure~\ref{fig:sigma}), indicating possible interaction between the bubble and surrounding gas. Table~\ref{tab:phy_clump} lists the H$_{2}$ column density, mass, and turbulent energy of each clump dervied using the optically thin \cob~line at the peak of the clump (see the calculations in Appendix~\ref{sect:phy_para}).

\begin{deluxetable}{cccccccccccccccc}[htbp]
\tablecaption{Observed Parameters of Each Clump}
\tabletypesize{\footnotesize}
\tablehead{
    \multirow{3}[0]{*}{Clump} & \multicolumn{3}{c}{\coa ($J$ = 1-0)} & & \multicolumn{3}{c}{\cob ($J$ = 1-0)} & & \multicolumn{3}{c}{\coc ($J$ = 1-0)} & & 
    \multicolumn{3}{c}{\hco ($J$ = 1-0)} \\
    \cline{2-4} \cline{6-8} \cline{10-12} \cline{14-16}
    & $T_{\rm R}^{*}$ & FWHM & $V_{\rm LSR}$ & & $T_{\rm R}^{*}$ & FWHM & $V_{\rm LSR}$ & & $T_{\rm R}^{*}$ & FWHM & $V_{\rm LSR}$ & & $T_{\rm R}^{*}$ & FWHM & $V_{\rm LSR}$\\ 
    & (K) & (\kms) & (\kms) & & (K) & (\kms) & (\kms) & & (K) & (\kms) & (\kms) & & (K) & (\kms) & (\kms) \\
    (1) & (2) & (3) & (4) & & (5) & (6) & (7) & & (8) & (9) & (10) & & (11) & (12) & (13)
}
\startdata
    A & 19.1 & 2.5 &  9.1 & &  6.5 & 1.5 &  9.2 & & 2.1 & 0.4 &  9.2 & & 0.4 & 2.6 &  9.2 \\
    B & 17.9 & 3.3 &  8.7 & &  6.0 & 2.3 &  8.6 & & 2.1 & 0.8 &  8.7 & & 0.6 & 2.1 &  8.5 \\
    C & 19.5 & 3.3 &  8.5 & &  8.1 & 1.7 &  8.2 & & 2.7 & 0.6 &  8.3 & & 0.6 & 1.8 &  8.3 \\
    D & 26.1 & 3.0 &  8.8 & &  6.9 & 1.9 &  9.0 & & 2.3 & 0.4 &  9.2 & & 0.7 & 2.9 &  9.0 \\
    E & 29.7 & 2.8 & 10.6 & & 10.8 & 1.7 & 10.6 & & 2.7 & 0.6 & 10.6 & & 1.9 & 2.5 & 10.8 \\
    F & 25.3 & 2.0 & 10.5 & &  7.6 & 1.7 & 10.3 & & 1.9 & 0.6 & 10.2 & & 2.3 & 1.9 & 10.4 \\ 
    G & 26.5 & 2.6 & 10.8 & &  7.7 & 2.0 & 10.8 & & 1.6 & 0.7 & 10.8 & & 1.7 & 2.4 & 11.1 \\
    H & 23.1 & 2.9 &  9.8 & &  6.7 & 1.9 &  9.9 & & 1.5 & 0.5 & 10.1 & & 1.8 & 1.9 & 10.3
\enddata
\tabletypesize{\normalsize}
\tablecomments{(1) Source name. (2), (5), (8), (11) Brightness temperature(s) of \coa, \cob, \coc, and \hco, respectively, at the emission peak of \coc. (3), (6), (9), (12) FWHM of \coa, \cob, \coc, and \hco, respectively. (4), (7), (10), (13) Central velocity of \coa, \cob, \coc, and \hco, respectively.}
\label{tab:clump}
\end{deluxetable}

\begin{deluxetable}{cccccccc}[htbp]
\label{tab:phy_clump}
\tablecaption{Physical Parameters of the Clumps}
\tablehead{
    \multirow{2}[0]{*}{Clump} & $T_{\rm ex}$ & $d_{\rm clump}$ & $\int T_{\rm mb, 13} dv$ & $N_{\rm H_{2}, clump}$  & $M_{\rm clump}$ & $E_{\rm turb, clump}$ & $E_{\rm grav, clump}$ \\
    & (K) & (pc) & (K \kms) & ($10^{20}$ cm$^{-2}$) & ($M_{\odot}$)  & (10$^{44}$~erg) & (10$^{43}$~erg) \\
    (1) & (2) & (3) & (4) & (5) & (6) & (7) & (8)
}
\startdata
    A & 22.6 & $0.2 \pm 0.1$ & 15.4 &  2.7 & $ 8.1 \pm 2.1$ &  $1.3 \pm 0.3$ & $ 5.6 \pm 0.8$ \\
    B & 21.3 & $0.2 \pm 0.1$ & 16.5 &  2.4 & $ 5.8 \pm 2.0$ &  $2.2 \pm 0.8$ & $ 2.9 \pm 0.7$ \\
    C & 23.0 & $0.3 \pm 0.1$ & 20.5 &  3.7 & $14.3 \pm 3.6$ &  $3.0 \pm 0.8$ & $ 11.7 \pm 2.2$ \\
    D & 29.6 & $0.3 \pm 0.1$ & 16.2 &  3.4 & $10.4 \pm 2.7$ &  $2.7 \pm 0.7$ & $ 6.2 \pm 1.2$ \\
    E & 33.2 & $0.5 \pm 0.1$ & 16.8 &  6.1 & $24.8 \pm 5.1$ &  $5.2 \pm 1.1$ & $21.0 \pm 4.4$\\
    F & 28.8 & $0.4 \pm 0.1$ & 16.1 &  3.7 & $11.5 \pm 3.0$ &  $2.4 \pm 0.6$ & $ 5.7 \pm 1.5$ \\ 
    G & 30.0 & $0.5 \pm 0.1$ & 22.6 &  3.9 & $16.7 \pm 4.2$ &  $4.9 \pm 1.2$ & $ 9.5 \pm 3.0$ \\ 
    H & 26.6 & $0.6 \pm 0.1$ & 10.5 &  3.1 & $ 7.7 \pm 1.7$ &  $2.0 \pm 0.4$ & $ 1.7 \pm 0.5$
\enddata  
\tablecomments{(1) Source name. (2) Excitation temperature of the clump, which is obtained by the \coa~spectral line from the \coc~peak emission of each clump. (3) Diameter of the clump. (4) Integrated intensity of \cob~emission. (5) H$_{2}$ density of the clump. (6) Mass of the clump. (7) Turbulent energy of the clump. (8) Gravitational energy of the clump.}
\end{deluxetable}

\clearpage
\subsection{The outflow of clump E}
\label{subsect:outflow}
In this section we investigate clump E in more detail, since the velocity profile of this clump alone displays typical outflow characteristics and because the spectra of the other clumps are too complex to identify reliable outflows. Outflows are a direct signature of ongoing star formation, and they can be identified by examining their line profiles, integrated intensity maps, and P-V diagrams. The central velocity and position of the driving source are estimated through the \coc~line. The emission peaks of \coc~can be seen in Figure~\ref{fig:bubble}. The initial velocity ranges of the blue and red wings are determined from the line wings of the spectral line where the \coc~emission reaches the 1$\sigma$ noise level. According to morphologies of the P-V diagram and the integrated map of the line wings of the \cob~and \hco~emission, then visually adjusted the velocity range of the line wings. To improve the signal-to-noise ratio, we smooth the CO and \hco~lines to a velocity resolution of 0.25~\kms.

\begin{figure}[htbp]
    \centering
    \subfigure[Distribution of the outflow]{\includegraphics[width = 8cm]{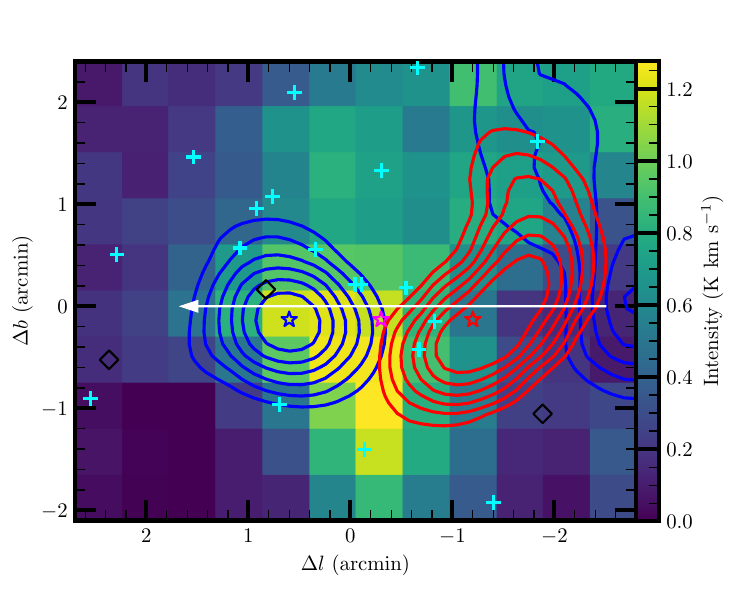}}
    \subfigure[P-V diagram]{\includegraphics[width = 8cm]{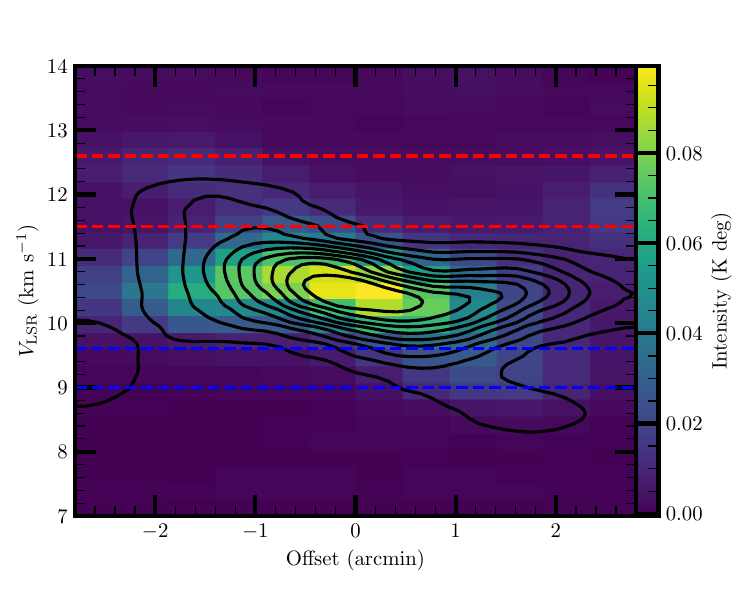}}
    \subfigure[Spectra of the blue side of the lobe]{\includegraphics[width = 8cm]{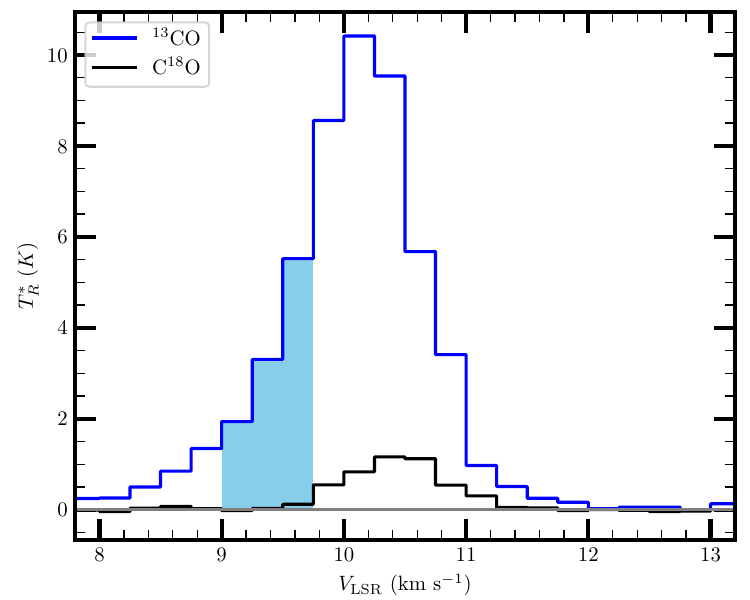}}
    \subfigure[Spectra of the red side of the lobe]{\includegraphics[width = 8cm]{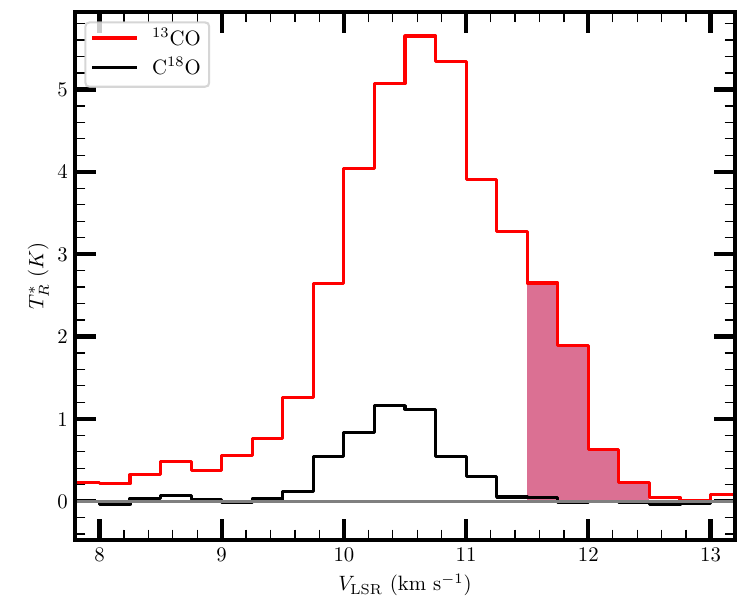}}
    \caption{Outflow in clump E identified by \cob~emission. (a) Integrated \coc~map with \cob~blue and red lobe contours. The contour levels are shown from 40\% to 90\% with steps of 10\% of the peak intensity of each lobe. The black diamonds and cyan pluses represent Class~I and Class~II YSOs, respectively. The magenta, blue, and red stars are the emission peaks of \coc, and the blue lobe and red lobe, respectively. (b) P-V diagram along the white arrow in panel (a). The black contour levels are shown from 10\% to 90\% with steps of 10\% of the peak value. The blue and red dashed lines are the velocity ranges of the blue and red lobes, respectively. (c) The blue spectrum presents \cob~at the peak position of the blue emission of the \cob~outflow. The black spectrum is \coc~at the emission peak of \coc. The blue shading of the spectrum indicates the blue line wing velocity of \cob. The feature at 8.6~\kms~is likely unrelated to the outflow, so emission with a velocity below 8.6~\kms~is not included in the blue wing of the outflow profile. (d) The red spectrum is \cob~at the peak position of the red emission of the \cob~outflow. The black spectrum is \coc~at the emission peak of \coc. The red shading of the spectrum indicates the red line wing velocity of \cob.}
    \label{fig:outflow_13co}
\end{figure}

\begin{figure}[htbp]
    \centering
    \subfigure[Distribution of the outflow]{\includegraphics[width = 8cm]{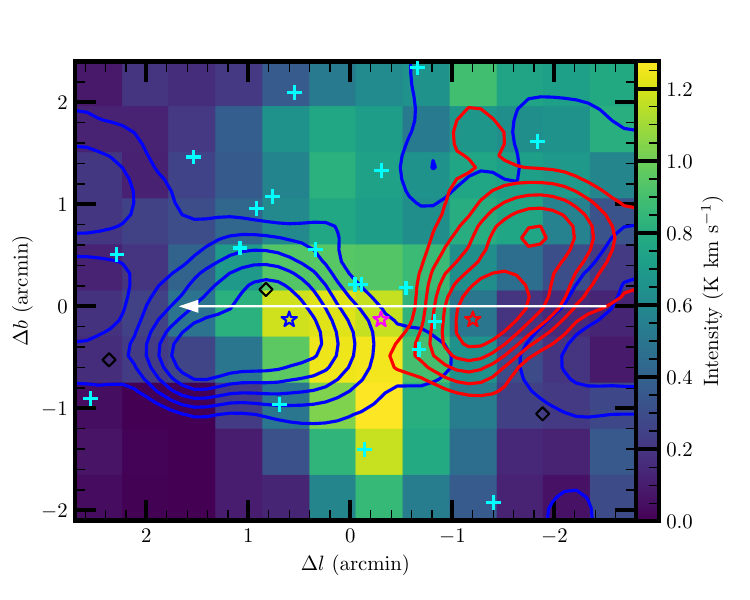}}
    \subfigure[P-V diagram]{\includegraphics[width = 8cm]{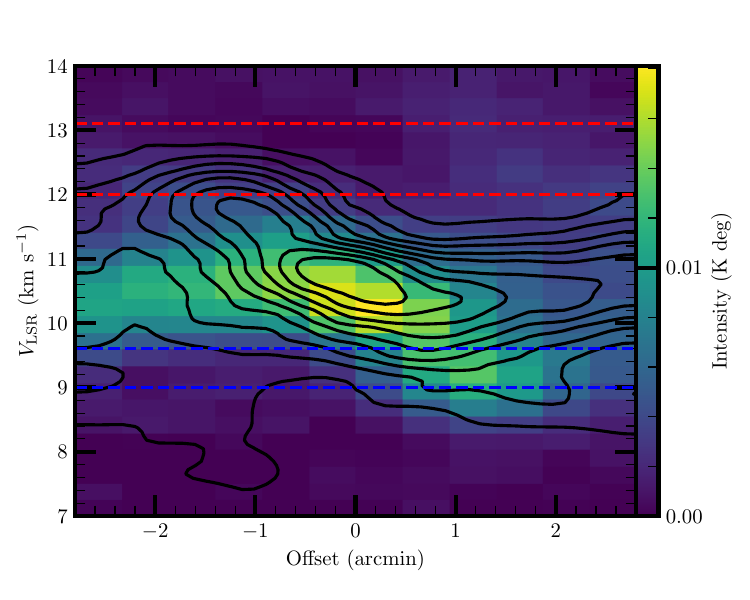}}
    \subfigure[Blue spectra]{\includegraphics[width = 8cm]{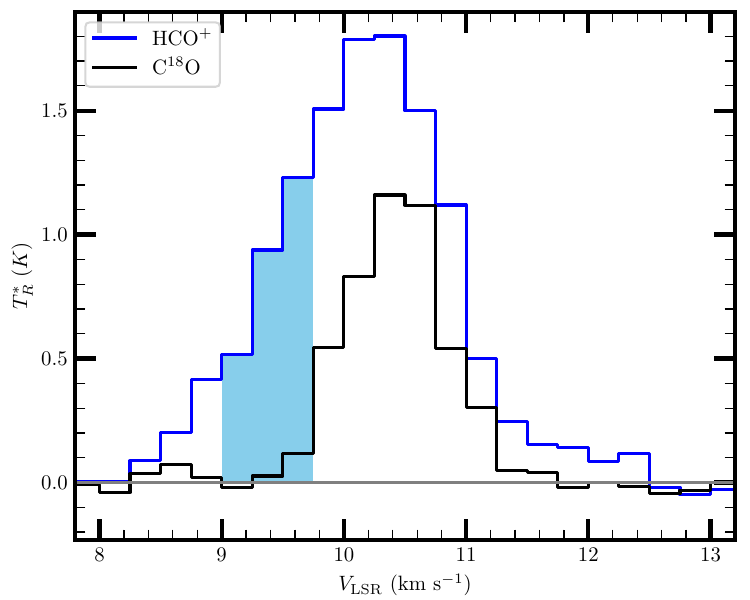}}
    \subfigure[Red spectra]{\includegraphics[width = 8cm]{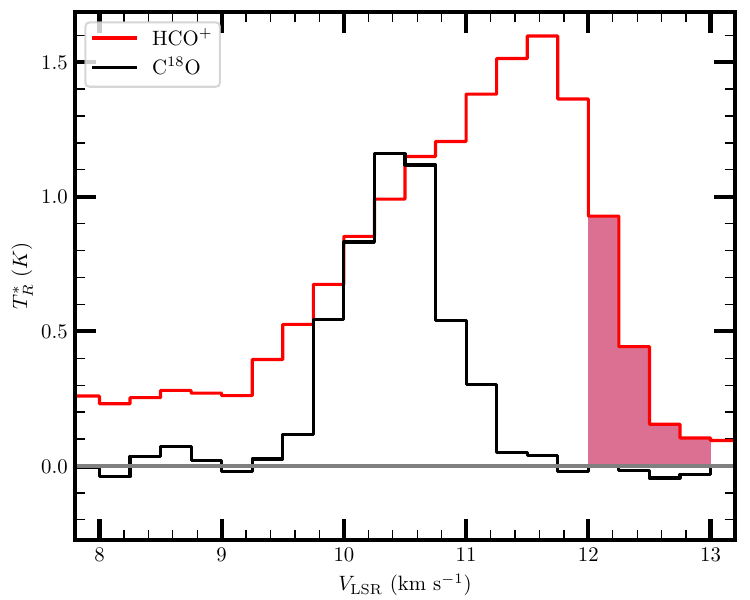}}
    \caption{Outflow in clump E identified by \hco~emission. The description of each panel is the same as that in Figure~\ref{fig:outflow_13co}, except the contour levels in panel (a) are shown from 50\% to 90\% with steps of 10\% of the peak intensity of each lobe. }
    \label{fig:outflow_hco}
\end{figure}

Figures~\ref{fig:outflow_13co} and~\ref{fig:outflow_hco} show the identified outflow in clump E based on \cob~and \hco~emission, respectively. The spatial distributions of the \cob~and \hco~outflows are consistent with each other and are aligned along the east--west direction. The blue and red lobes are symmetrically distributed with respect to the \coc~emission peak, which are similar in size. We note that the \coc~and \hco~emission at $V_{\rm LSR} \sim 8.6$~\kms, and the offset of $\sim 1.5$~arcmin in the P-V diagram, are not associated with the blue lobe of the outflow, thus we did not cut the edge of the blue wing to the 1$\sigma$ noise level.

The method presented in~\citet{Liu+etal+2021} was employed to estimate various properties of the outflows, such as the mass, momentum, and energy. Therein, they derived the H$_{2}$ column density of the outflow lobes assuming that the gas is in local thermodynamic equilibrium (LTE) and the excitation temperature is 30~K (15~K) for \cob~(\hco). The mass of an outflow lobe can be then estimated based on its size and the H$_{2}$ column density. The other physical properties of the outflows are obtained based on the mass, velocity, and size of the lobes~\citep[see Appendix A of][]{Liu+etal+2021}. In our work, we use the ratios of column densities of $N_{\rm H_{2}} / N_{\rm ^{13}CO} \approx 5 \times 10^{5}$ and $N_{\rm H_{2}} / N_{\rm HCO^{+}} \approx 10^{8}$~\citep{Turner+etal+1997,Simon+etal+2001}. We also use the inclination of the outflow, i.e., the angle between the long axis of the outflow and the line of sight, to correct the parameters of the outflow. Since the inclination of an outflow cannot be reliably determined, we adopt 57\fdg3 as a proxy in this work, which is the mean value for a random distribution~\citep[see details in][]{Bontemps+etal+1996,Li+etal+2018}. We note that the calculation of the outflow's physical parameters (see Table~\ref{tab:phy_outflow}), involve many assumptions, such as the excitation temperature and the inclination. Therefore, we only consider the uncertainties in the outflow lobe region and distance when estimating the errors of the physical parameters. The mass, momentum, and dynamical timescale of the outflow are a few $M_{\odot}$, a few tens of $M_{\odot}$~\kms, and $\sim 10^{5}$~yr, respectively. Based on \citet{Yang+etal+2018}, the typical mass of a low-mass outflow is about $0.1 \sim 1~M_{\odot}$ and the typical mass loss rate is about $10^{-7} \sim 10^{-6}~M_{\odot}~{\rm yr^{-1}}$.
These values are much smaller than those found for the outflow in clump E. Moreover, all physical parameters of the outflow in clump E align with the typical values of high-mass outflows as summarized in table~1 of~\citet{Yang+etal+2018}. Therefore, it is unlikely that the outflow in clump E is driven by a single low-mass YSO. We do not find any high-mass YSOs in this region based on the current data, yet there are several low-mass YSOs embedded in the vicinity of the outflow region (see panel (a) of Figures~\ref{fig:outflow_13co} and \ref{fig:outflow_hco}). Therefore, the outflow might be driven by a deeply embedded protostar that has not been detected yet.

\begin{deluxetable}{ccccccccccc}[htbp]
\label{tab:phy_outflow}
\tablecaption{Physical Properties of the Outflow of Clump E}
\tablehead{
    \multirow{2}[0]{*}{Line} & \multirow{2}[0]{*}{Lobe} & $\Delta v$ & $\left\langle \Delta v_{\rm lobe} \right\rangle$ & $N_{\rm H_{2}}$ & $M_{\rm lobe}$  & $l_{\rm lobe}$ & $P_{\rm lobe}$ & $E_{\rm lobe}$ & $t_{\rm lobe}$ & $L_{\rm m~(lobe)}$ \\
    & & (\kms) & (\kms) & ($10^{21}$ cm$^{-2}$) & ($M_{\odot}$) & (pc) & ($M_{\odot}$~\kms) & ($10^{44}$ erg) & ($10^{5}$ yr) & ($10^{-1}~L_{\odot}$) \\
    (1) & (2) & (3) & (4) & (5) & (6) & (7) & (8) & (9) & (10) & (11) 
}
\startdata
    \cob& blue &      (9, 9.6) & 2.1 & 1.1 & $6.2 \pm 1.2$ & $0.8 \pm 0.2$ &  $24.6 \pm 4.7$ & $9.1 \pm 1.8$ & $2.2 \pm 0.3$ & $1.8 \pm 0.2$ \\
    \cob &  red & (11.5, 12.6) & 2.7 & 1.1 & $9.6 \pm 2.8$ & $0.9 \pm 0.2$ & $49.8 \pm 13.9$ & $24.3 \pm 6.4$ & $2.0 \pm 0.3$ & $5.3 \pm 0.6$ \\
    \hco & blue &     (9, 9.6) & 2.1 & 0.4 & $2.5 \pm 0.9$ & $0.8 \pm 0.1$ & $10.1 \pm 3.4$ & $3.7 \pm 1.3$ & $2.3 \pm 0.4$ & $0.7 \pm 0.1$ \\
    \hco &  red &     (12, 13) & 2.7 & 0.5 & $2.9 \pm 0.6$ & $0.8 \pm 0.1$ & $14.7 \pm 3.3$ & $7.1 \pm 1.6$ & $1.8 \pm 0.2$ & $1.8 \pm 0.2$
\enddata
\tablecomments{(1) Molecular line. (2) Blue/red lobe. (3) Outflow velocity range for the blue/red lobe. (4) Velocity of the blue/red lobe. (5) H$_{2}$ density of the blue/red lobe. (6) Mass of the blue/red lobe. (7) Length of the blue/red lobe. (8) Momentum of the blue/red lobe. (9) Kinetic energy of the blue/red lobe. (10) Dynamical timescale of the blue/red lobe. (11) Mechanical luminosity of the blue/red lobe.}
\end{deluxetable}

\clearpage
\section{Discussion}
\label{sect:dis}
In this section, we discuss the physical properties of the bubble (Section~\ref{subsect:phy}), and the energy cascade of S~Mon region (Section~\ref{subsect:energy}). In Section~\ref{subsect:expansion}, we use the Class~II YSOs to investigate the kinematic characteristics of the bubble.
\subsection{Physical Properties of the Bubble}
\label{subsect:phy}

In the following, we discuss the physical properties of the bubble. Using the shell mass ($M_{\rm shell}$) and expansion velocity ($V_{\rm exp}$) of the bubble, we can estimate the shell's momentum ($P_{\rm shell}$), kinetic energy ($E_{\rm shell}$), and kinetic timescale ($t_{\rm kinetic}$). These quantities are determined as $P_{\rm shell} = M_{\rm shell} V_{\rm exp}$, $E_{\rm shell} = 0.5 M_{\rm shell} V^{2}_{\rm exp}$, and $t_{\rm kinetic} = R_{\rm bub} / V_{\rm exp}$, respectively. For the bubble we find that $P_{\rm shell}$ is 4\,600 $\pm$ 1\,300~$M_{\odot}$~\kms, $E_{\rm shell}$ is $(1.5 \pm 0.4) \times 10^{47}$~erg, and $t_{\rm kinetic}$ is $(3.3 \pm 0.3) \times 10^{5}$~yr. Considering that the expansion velocity serves as a lower limit, the uncertainties of these parameters only come from the mass.

We find 13 B-type stars and a massive binary system, 15~Mon, located in the S~Mon region using SIMBAD\footnote{\url{https://simbad.u-strasbg.fr/simbad}}. Most B-type stars are located in the shell of the bubble rather than at its center (see Figure~\ref{fig:OBstars} in Appendix~\ref{sect:OBstars}), and only two B-type stars are located at the center of the bubble. 
We employ Equation (2) from~\citet{Arce+etal+2011} to estimate the mass-loss rate ($\dot{m}_{\rm w}$), which can be calculated by:
\begin{equation}
    \dot{m}_{\rm w} = \frac{P_{\rm shell}}{v_{\rm w} \tau_{\rm w}},
    \label{eq:moment}
\end{equation} 
where $v_{\rm w}$ is the wind velocity, and $\tau_{\rm w}$ is the wind timescale (i.e., the duration of wind activity). Here, the kinetic timescale ($t_{\rm kinetic}$) is employed to estimate the mass-loss rate of the driving source roughly. Given the typical wind velocity of O- or B-type stars is about 1\,000--2\,000~\kms~\citep{Chen+etal+2013}, the mass-loss rate would be $ 10^{-6}-10^{-5}~{M_{\odot}}~{\rm yr^{-1}}$. The value is much larger than the typical mass-loss rate of B-type stars ($10^{-8}-10^{-11}~M_{\odot}~{\rm yr^{-1}}$) but close to that of O-type stars~\citep[$10^{-6}-10^{-7}~M_{\odot}~{\rm yr^{-1}}$,][]{Chen+etal+2013}. Therefore, it is improbable that the B-type stars are the main driving source of the bubble. Although 15~Mon is not precisely at the center of the bubble, it stands out as the closest high-mass star to the bubble's center, apart from the two B-type stars. Therefore, 15~Mon might be the main driving source of the bubble.

The dynamic age of the \hii~region can be used to judge whether the bubble is triggered by 15~Mon. Assuming that the \hii~region expands in a homogeneous medium, its dynamical age, $t_{\rm dyn}$ is~\citep[see the model described by][]{Dyson+Williams+1980}:

\begin{equation}
    t_{\rm dyn} = \frac{4 R_{\rm s}}{7 c_{s}} \left[ \left( \frac{R_{\rm H_{II}}}{R_{\rm s}}\right)^{7 / 4} - 1\right],
\end{equation}
where $R_{\rm H_{II}}$ is the radius of the \hii~region~\citep{Xu+etal+2017}, for which we adopt an inner radius of 0.7 $\pm$ 0.1~pc. $c_{s}$ is the isothermal sound speed of the ionized gas, and a value of 10~\kms~is adopted. The radius of the associated Str\"omgren sphere ($R_{\rm s}$) reads:

\begin{equation}
    R_{\rm s} = \left(\frac{3 Q_{\rm Ly}}{4 \pi n_{0}^{2} \alpha_{B}}\right)^{1 / 3},
\end{equation}
where $\alpha_{B} = 2.6 \times 10^{-13}~{\rm cm^{3}~s^{-1}}$ is the hydrogen recombination coefficient to all levels above the ground level, $n_{0}$ is the initial number density of the gas, and $Q_{\rm Ly}$ is the ionizing luminosity. We only consider the number of atoms that can be ionized, so $n_{0} = \frac{M_{\rm shell}}{4 / 3 \pi R_{\rm bub}^3} \approx$ (2.6--6.3) $\times 10^{4}~{\rm cm^{-3}}$. Since the actual ionized medium within the bubble is likely less massive as compared to the whole shell mass determined from the broad CO ring, the number of atoms that can be ionized might be overestimated. For an O7V star, the ionizing luminosity ($Q_{\rm Ly}$) is $10^{47.62}~{\rm s^{-1}}$~\citep{Martins+etal+2005}. Based on these calculations, the obtained dynamical age is (3.5--8.9) $\times 10^{5}$~yr. Even if the secondary member of 15 Mon, an O9.5V star, is also taken into account, the dynamical timescale is only reduced by less than 0.1\%. We conclude that the kinetic timescale of the bubble is comparable to the dynamical age of the \hii~region.

\cite{Montillaud+etal+2019} estimated the age of 15~Mon using two methods. The first method, based on SED fitting, led to a median age of $1.5-2$~Myr, with a dispersion between 0.2 and 3~Myr. The second method was based on isochrone fitting in color--magnitude diagrams and led to a median age of 3~Myr, with a dispersion between 1 and 6~Myr. Our estimation on the dynamical timescale of the \hii~region may be less than the age of 15~Mon. Hence, the bubble is likely to be driven by the expanding \hii~region of 15~Mon. As the projected position of 15~Mon is not located at the center of the bubble, we presume that 15~Mon may have moved away from its once central position in the bubble.

\subsection{Energy Cascade in S~Mon}
\label{subsect:energy}
Stellar winds inject energy and momentum into the ambient clouds that may help sustain turbulence and disrupt their surroundings~\citep{Nakamura+Li+2007,Arce+etal+2011}. To assess whether winds can sustain the turbulence in S~Mon, we compare the wind energy injection rate into the cloud, $\dot{E}_{\rm w}$, with the cloud turbulent dissipation rate, $L_{\rm turb, cloud}$. Based on the \cob~emission, we can derive the 3D turbulent velocity dispersion ($\sigma_{\rm 3d}$) with an approximation of $\sigma_{\rm 3d} = \sqrt{3} \sigma_{\rm 1d}$~\citep{Li+etal+2018}, where $\sigma_{\rm 1d}$ is the 1D turbulent velocity dispersion along the line of sight and we consider the mass ($M_{\rm cloud}$) of S~Mon. The obtained $\sigma_{\rm 3d}$ is 1.7 $\pm$ 0.2~\kms~and $M_{\rm cloud}$ is 2\,200 $\pm$ 500~$M_{\odot}$. The cloud turbulent energy, $E_{\rm turb, cloud}$, is about (6.3 $\pm$ 2.5) $\times 10^{46}$~erg (see the calculation in Appendix~\ref{sect:phy_para}). The cloud turbulent dissipation rate can be calculated as:

\begin{equation}
    L_{\rm turb, cloud} = \frac{E_{\rm turb, cloud}}{t_{\rm diss}},
    \label{eq:L_turb}
\end{equation}
where $t_{\rm diss}$ is the cloud turbulent dissipation time. We roughly estimate $t_{\rm diss}$ via~\citep{McKee+Ostriker+2007}:

\begin{equation}
    t_{\rm diss} \sim 0.5 \frac{d_{\rm cloud}}{\sigma_{\rm 1d}},
    \label{eq:t_diss}
\end{equation}
where the cloud diameter is $d_{\rm cloud} \sim$ (4.0 $\pm$ 0.4)~pc and $\sigma_{\rm 1d} \sim$ (1.0 $\pm$ 0.1)~\kms. Based on Equations (\ref{eq:L_turb})--(\ref{eq:t_diss}), we obtain a cloud turbulent dissipation time of $t_{\rm diss} \sim $(2.0 $\pm$ 0.4) $\times 10^{6}$~yr, and a cloud turbulent dissipation rate of $\sim$ (0.5--1.7) $\times 10^{33}~{\rm erg~s^{-1}}$.

Since the bubble might be driven by the expanding \hii~region of 15~Mon, the dynamical age ($t_{\rm dyn}$) of the bubble is more suitable than the kinemtaic timescale ($t_{\rm kinetic}$) to estimate the mass-loss rate ($\dot{m}_{\rm w}$). Therefore, we reestimated this value, which is approximately (1.9--8.6)$\times 10^{-6}~{M_{\odot}~{\rm yr}^{-1}}$, and the wind energy injection rate can be estimated using the following equation from~\citet{McKee+1989}:
\begin{equation}
    \dot{E}_{\rm w} = \frac{1}{2}(\dot{m}_{\rm w} v_{\rm w}) \sigma_{\rm 3d}.
\end{equation}
The derived wind injection rate is about (4.5--8.1)$\times 10^{33}~{\rm erg~s^{-1}}$, which is larger than the cloud turbulent dissipation rate. This indicates that strong winds serve as a significant energy origin for S~Mon, consisting of sustaining the observed turbulence. This result agrees with the findings of other studies~\citep{Arce+etal+2011,Li+etal+2015,Feddersen+etal+2018}.

In the following, we investigate the turbulent energy in this region.
The total wind energy input into the interstellar medium is about $E_{\rm w} = \int 0.5 \dot{m}_{\rm w} (t) v_{\rm w}^{2} dt \simeq \Delta M v_{\rm w}^{2} / 2 \simeq$ (0.7--1.7) $\times 10^{50}$~erg~\citep{Lamers+Cassinelli+1999}. 
Both the wind energy of 15~Mon and the kinetic energy of the bubble ($E_{\rm shell} \sim$ (1.5 $\pm$ 0.4) $\times 10^{47}$~erg) are higher than the turbulent energy of the cloud ($E_{\rm turb, cloud} \sim$ (6.3 $\pm$ 2.5) $\times 10^{46}$~erg), thus both of them are help to sustain the turbulence in the S~Mon region. 
Moreover, the wind energy probably plays a crucial role in driving the bubble, since it is about three orders of magnitude higher than the kinetic energy of the bubble. Despite the substantial energy carried away by ionizing photons from the star throughout its lifetime, only 0.01--0.1 percent of this energy is converted into kinetic energy in the shell, with the majority being lost as radiation~\citep{Geen+etal+2015}. \citet{Walch+etal+2012} conducted hydrodynamic simulations to explore the dynamical effects of a single O7 star. They determined that the total radiative energy injected by the ionizing source reaches $7 \times 10^{51}$~erg within approximately 1~Myr. A portion of this injected energy is then transformed into kinetic energy, amounting to around $3 \times 10^{48}$ erg, which surpasses the kinetic energy we estimated in this work ($E_{\rm shell} \sim$ (1.5 $\pm$ 0.4) $\times 10^{47}$~erg). Thus, radiative energy contributes to driving the bubble.

Next, we studied the turbulent energy of the outflow of clump E. The total kinetic energy of the outflow in clump E ($\sum {E_{\rm lobe}} \sim 4.4 \times 10^{45}$~erg) is higher than its turbulent energy ($E_{\rm turb, clump} \sim$ (5.2 $\pm$ 1.1) $\times 10^{44}$~erg), and it accounts for $\sum {E_{\rm lobe}} / E_{\rm turb, cloud} \sim 4-10 \%$ of the turbulent energy of the cloud. This suggests that the outflow in clump E helps to maintain the turbulence in this clump and introduces additional energy to surrounding gas, thereby affecting the turbulence of the local environment~\citep{Arce+etal+2010,Li+etal+2018}. This result is also consistent with the conclusion made by~\citet{Li+etal+2020} that the outflow is sufficient to maintain turbulence on a scale of 0.1--0.4~pc. 

Moreover, we also estimate the gravitational binding energy of the S~Mon region. The detailed calculation is presented in Appendix~\ref{sect:phy_para}. We find that the total wind energy of 15~Mon is over two orders of magnitude higher than the gravitational binding energy of the cloud, i.e., (2.1 $\pm$ 1.0)$\times 10^{47}$~erg, indicating that the wind can disrupt the cloud. Based on the above calculations, we find that half of the gravitational binding energy of the cloud is approximately equal to the kinetic energy of the bubble, and much higher than the total kinetic energy of the outflow in clump E. According to the virial theorem~\citep{Nakamura+Li+2014}, the bubble has the potential to disrupt the cloud, while the outflow is unable to disrupt the cloud. 

We also investigated the gravitational binding energy of clump E ($E_{\rm grav, clump} \sim $(2.1 $\pm$ 0.2)$\times 10^{44}$~erg), which is lower than the total kinetic energy of the outflow, indicating that the outflow contributes to disrupting the clump. The results for the outflow found here are consistent with the conclusions made by~\citet{Li+etal+2020} in that the outflow activity could potentially disperse material away from the parent cloud on a scale of $0.2 - 1.0$~pc.

In conclusion, we find the bubble produced by 15~Mon helps to maintain the turbulence in S~Mon and is potentially disrupting the cloud, and while the outflow from clump E only adds a small fraction of the overall turbulence and energy of the system, it still may be able to perturb the local gas.

\subsection{Bubble from the YSO Perspective}
\label{subsect:expansion}
Since some YSOs may be too young to have moved away from their birth sites, they often have the same kinematic properties as their parent clouds~\citep{Tobin+etal+2015,Hacar+etal+2016,Groschedl+etal+2021}. Thus, we can utilize the motions of YSOs associated with a cloud to represent the motions of the cloud. According to the correlation between YSOs and gas (see Section~\ref{subsect:gas_ysos} for details), the Class~II YSOs, whose parallax accuracies are better than 10\% and projected locations are approximately aligned with the cloud were initially selected to investigate the kinematic characteristics of the bubble. 

\begin{figure}
    \centering
    \subfigure[]{\includegraphics[width = 0.45\textwidth]{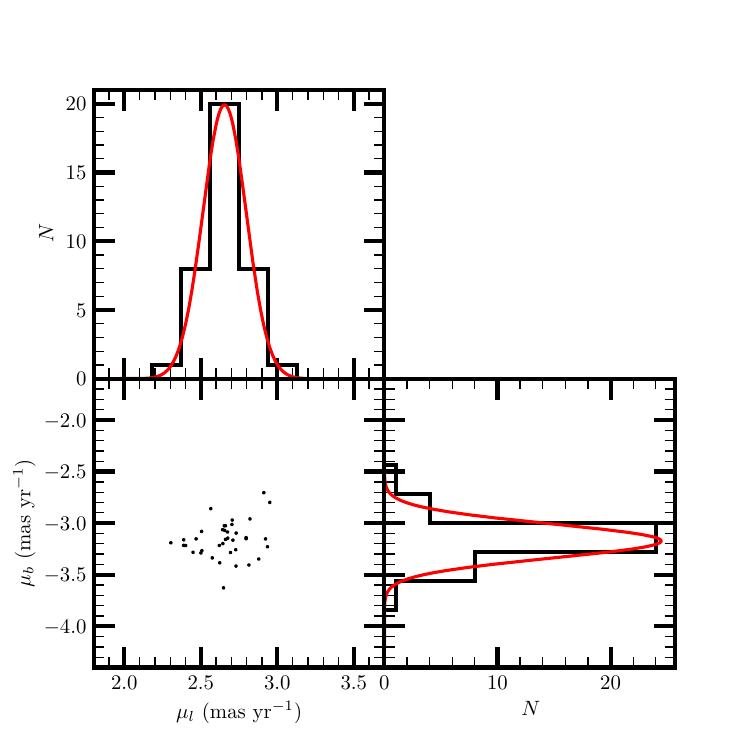}}
    \subfigure[]{\includegraphics[width = 0.51\textwidth]{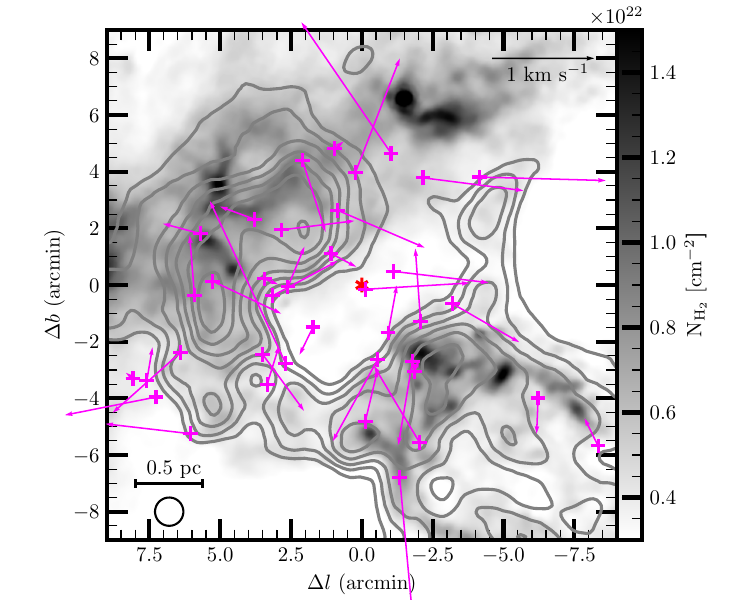}}
    \caption{(a) Distribution of the proper motions of the YSOs, where the average proper motion of these sources is (2.65, -3.17)~${\rm mas~yr^{-1}}$, and the $1 \sigma$ uncertainty range of the Gaussian fitting is (0.14, 0.17)~${\rm mas~yr^{-1}}$. (b) Projected distribution and tangential velocity of the YSOs. The magenta arrows depict the velocity for each YSO with respect to the mean velocity of the YSO sample. The red asterisk (0, 0) is the center of the bubble.}
    \label{fig:pm}
\end{figure}

A Gaussian function was used to estimate the average proper motion of the YSO sample and to remove outliers (beyond a 3$\sigma$ threshold). A total of 38 YSOs were ultimately selected to investigate the kinematic features of the bubble on the plane of the sky. 
Figure~\ref{fig:pm} (a) shows the distribution of proper motion of the selected YSOs. The mean proper motion of these sources is $(\bar{\mu_{l}}, \bar{\mu_b})$ = (2.65, -3.17)~${\rm mas~yr^{-1}}$, and the $1 \sigma$ uncertainty range of the Gaussian fitting is (0.14, 0.17)~${\rm mas~yr^{-1}}$. 

The intrinsic movement of the Class~II YSOs on the plane of the sky was calculated by subtracting the mean proper motions. Then the tangential velocities in $l$ and $b$ directions (in units of \kms) are calculated as follows: 

\begin{equation}
    \begin{aligned}
    \upsilon_{l} = \kappa \frac{1}{\varpi} * (\mu_{l} - \bar{\mu_{l}}), \\
    \upsilon_{b} = \kappa \frac{1}{\varpi} * (\mu_{b} - \bar{\mu_{b}}),
    \end{aligned}
\end{equation}
where $\kappa = 4.74~{\rm km~yr~s^{-1}}$, $\varpi$ is the parallax in units of mas, and $\mu_{l}$ and $\mu_{b}$ are the proper motion along the $l$ and $b$ directions, respectively, in units of mas~yr$^{-1}$.

\begin{figure}
    \centering
    \subfigure[]{\includegraphics[width = 0.45\textwidth]{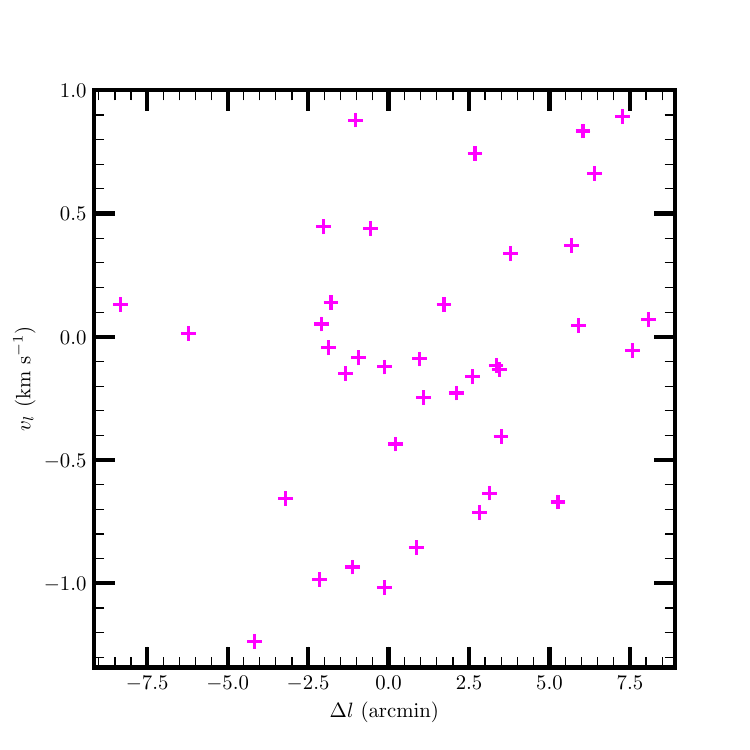}}
    \subfigure[]{\includegraphics[width = 0.45\textwidth]{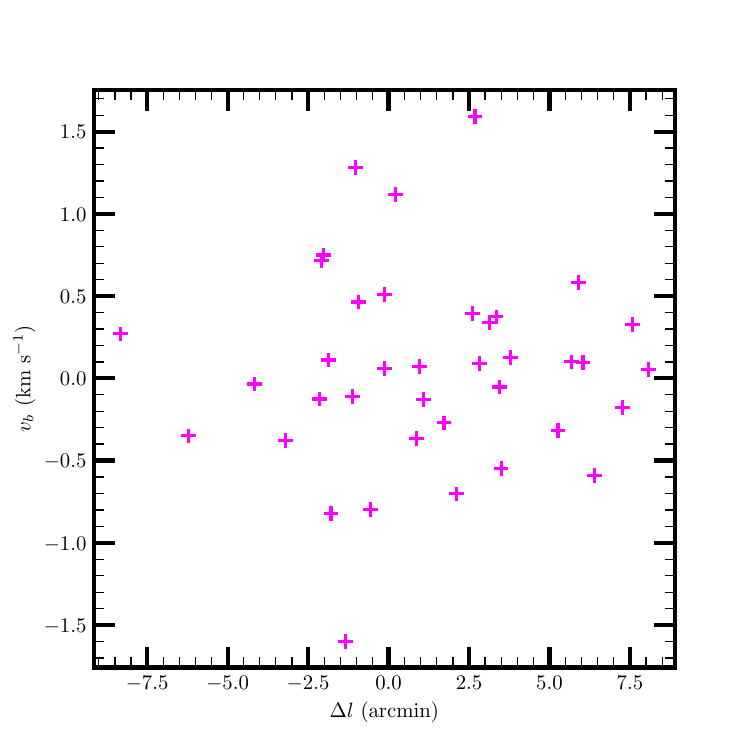}}
    \subfigure[]{\includegraphics[width = 0.45\textwidth]{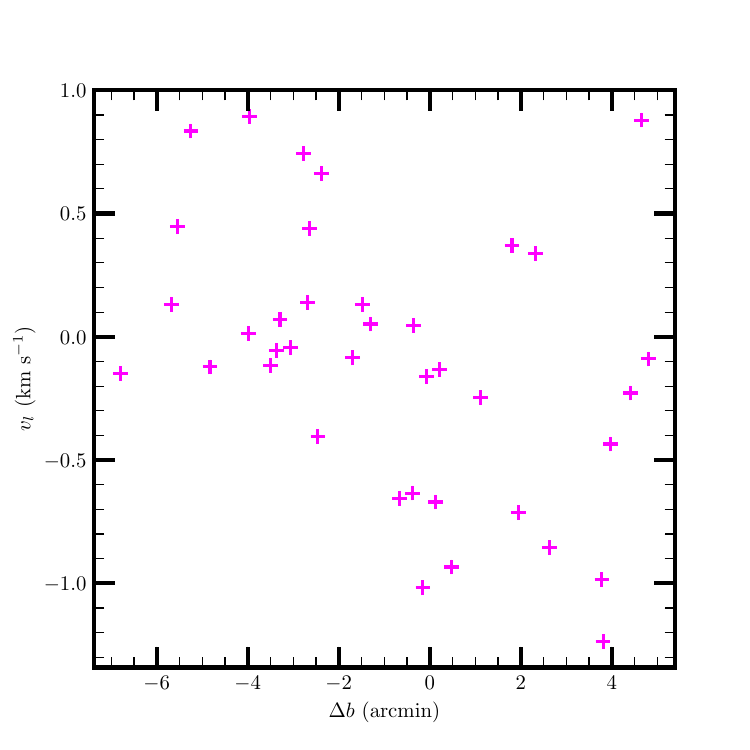}}
    \subfigure[]{\includegraphics[width = 0.45\textwidth]{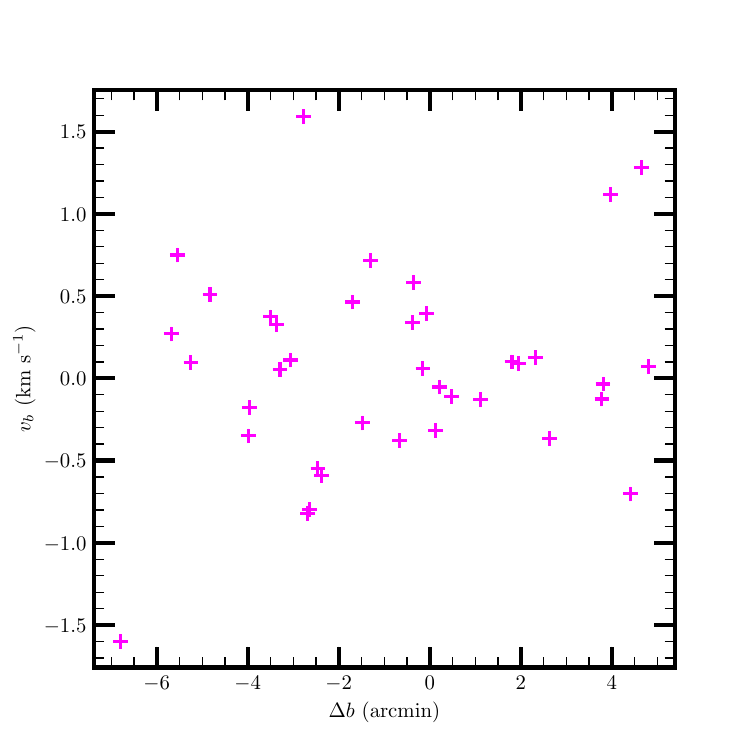}}
    \caption{P-V diagrams of the Class~II YSOs used to investigate the kinematic features.}
    \label{fig:pv_YSO}
\end{figure}

Figure~\ref{fig:pm} (b) shows the projected distributions and tangential velocities of the Class~II YSOs. The Class~II YSOs are mainly located in the molecular gas surrounding the bubble, and their motions are relatively random on the plane of the sky. Figure~\ref{fig:pv_YSO} shows P-V diagrams of the selected YSOs. There is no clear evidence of expansion in the figure. In order to determine whether the YSOs are expanding, we regard the expansion velocity as the radial outward component of the intrinsic velocity from the center of the bubble. The median expansion velocity of the Class~II YSOs is found to be 0.0 $\pm$ 0.1~\kms. Therefore, the YSOs associated with the gas may not exhibit expansion on the plane of the sky.

\clearpage
\section{Summary}
\label{sect:sum}
We observed four molecular lines (\coa, \cob, \coc, and \hco~$J$ = 1--0), and collected YSOs in S~Mon to investigate the bubble, an outflow identified in one of the clumps, as well as the energy cascade in the S Mon region. This includes the dynamical timescale, possible driving sources, and feedback of the outflow and bubble. The main results are summarized as follows.

\begin{itemize}
    \item[1] The Class~II YSOs are closely associated with the molecular gas in terms of spatial distribution and kinematic properties. According to the \emph{Gaia} DR3 parallaxes of the Class~II YSOs, we estimate the distance to S~Mon to be $\sim$ 722 $\pm$ 9~pc.
    \item[2] We discover a molecular bubble in the S~Mon region which is consistent with the {\it Herschel} H$_{2}$ column density maps. The molecular bubble has a radius of $\sim 1.1 \pm 0.1$~pc, a mass of $\sim 1\,400 \pm 400~M_{\odot}$, a momentum of $\sim$ 4\,600 $\pm$ 1\,300~$M_{\odot}$~\kms, and kinetic energy of $\sim (1.5 \pm 0.4) \times 10^{47}$~erg. The dynamical timescale of the \hii~region of 15~Mon is (3.5--8.9) $\times 10^{5}$~yr, which is likely to be the main driving source of the bubble.
    \item[3] We detect a molecular outflow in the shell of the bubble, which indicates that star formation activity is ongoing around the bubble. The outflow has a mass of a few $M_{\odot}$, a momentum of a few tens of $M_{\odot}~{\rm km~s^{-1}}$, kinetic energy of $\sim 10^{45}$~erg, a dynamical timescale of $\sim 10^{5}$~yr, and a mechanical luminosity of a few times $10^{-1}~L_{\odot}$. 
    \item[4] The wind energy helps to sustain the turbulence in the S~Mon region and drive the bubble. The bubble can also provide sufficient energy input to sustain the cloud turbulence. The measured outflow in clump E is enough to maintain the turbulence of the clump E. Outflows (like the one detected in clump E) could generally contribute to the turbulence of a cloud, while in S~Mon further outflows are not yet confirmed.
    \item[5] The Class~II YSOs correlated to the molecular bubble do not show expansion on the plane of the sky.
\end{itemize} 

\begin{acknowledgements}
We appreciate the anonymous referee for the instructive comments that helped us to improve the paper. This work was funded by the NSFC Grands 11933011, National SKA Program of China (Grant No. 2022SKA0120103) and the Key Laboratory for Radio Astronomy. L.Y.J. thanks the support of the NSFC grant No. 12203104, the Natural Science Foundation of Jiangsu Province (grant No. BK20210999), the Entrepreneurship and Innovation Program of Jiangsu Province. This work has made use of data from the European Space Agency (ESA) mission \emph{Gaia} (\url{https://www.cosmos.esa.int/gaia}) processed by the \emph{Gaia} Data Processing and Analysis Consortium (DPAC, \url{https://www.cosmos.esa.int/web/gaia/dpac/consortium}). Funding for the DPAC has been provided by national institutions, in particular the institutions participating in the \emph{Gaia} Multilateral Agreement. 
\end{acknowledgements}
\clearpage
\bibliographystyle{aasjournal}

\begin{thebibliography}{}
    \expandafter\ifx\csname natexlab\endcsname\relax\def\natexlab#1{#1}\fi
    \providecommand{\url}[1]{\href{#1}{#1}}
    \providecommand{\dodoi}[1]{doi:~\href{http://doi.org/#1}{\nolinkurl{#1}}}
    \providecommand{\doeprint}[1]{\href{http://ascl.net/#1}{\nolinkurl{http://ascl.net/#1}}}
    \providecommand{\doarXiv}[1]{\href{https://arxiv.org/abs/#1}{\nolinkurl{https://arxiv.org/abs/#1}}}
    
    \bibitem[{{Abdurro'uf} {et~al.}(2022){Abdurro'uf}, {Accetta}, {Aerts}, {Silva
      Aguirre}, {Ahumada}, {Ajgaonkar}, {Filiz Ak}, {Alam}, {Allende Prieto},
      {Almeida}, {Anders}, {Anderson}, {Andrews}, {Anguiano}, {Aquino-Ort{\'\i}z},
      {Arag{\'o}n-Salamanca}, {Argudo-Fern{\'a}ndez}, {Ata}, {Aubert},
      {Avila-Reese}, {Badenes}, {Barb{\'a}}, {Barger}, {Barrera-Ballesteros},
      {Beaton}, {Beers}, {Belfiore}, {Bender}, {Bernardi}, {Bershady}, {Beutler},
      {Bidin}, {Bird}, {Bizyaev}, {Blanc}, {Blanton}, {Boardman}, {Bolton},
      {Boquien}, {Borissova}, {Bovy}, {Brandt}, {Brown}, {Brownstein}, {Brusa},
      {Buchner}, {Bundy}, {Burchett}, {Bureau}, {Burgasser}, {Cabang}, {Campbell},
      {Cappellari}, {Carlberg}, {Wanderley}, {Carrera}, {Cash}, {Chen}, {Chen},
      {Cherinka}, {Chiappini}, {Choi}, {Chojnowski}, {Chung}, {Clerc}, {Cohen},
      {Comerford}, {Comparat}, {da Costa}, {Covey}, {Crane}, {Cruz-Gonzalez},
      {Culhane}, {Cunha}, {Dai}, {Damke}, {Darling}, {Davidson}, {Davies},
      {Dawson}, {De Lee}, {Diamond-Stanic}, {Cano-D{\'\i}az}, {S{\'a}nchez},
      {Donor}, {Duckworth}, {Dwelly}, {Eisenstein}, {Elsworth}, {Emsellem},
      {Eracleous}, {Escoffier}, {Fan}, {Farr}, {Feng}, {Fern{\'a}ndez-Trincado},
      {Feuillet}, {Filipp}, {Fillingham}, {Frinchaboy}, {Fromenteau}, {Galbany},
      {Garc{\'\i}a}, {Garc{\'\i}a-Hern{\'a}ndez}, {Ge}, {Geisler}, {Gelfand},
      {G{\'e}ron}, {Gibson}, {Goddy}, {Godoy-Rivera}, {Grabowski}, {Green},
      {Greener}, {Grier}, {Griffith}, {Guo}, {Guy}, {Hadjara}, {Harding},
      {Hasselquist}, {Hayes}, {Hearty}, {Hern{\'a}ndez}, {Hill}, {Hogg},
      {Holtzman}, {Horta}, {Hsieh}, {Hsu}, {Hsu}, {Huber}, {Huertas-Company},
      {Hutchinson}, {Hwang}, {Ibarra-Medel}, {Chitham}, {Ilha}, {Imig}, {Jaekle},
      {Jayasinghe}, {Ji}, {Johnson}, {Jones}, {J{\"o}nsson}, {Katkov}, {Khalatyan},
      {Kinemuchi}, {Kisku}, {Knapen}, {Kneib}, {Kollmeier}, {Kong}, {Kounkel},
      {Kreckel}, {Krishnarao}, {Lacerna}, {Lane}, {Langgin}, {Lavender}, {Law},
      {Lazarz}, {Leung}, {Leung}, {Lewis}, {Li}, {Li}, {Lian}, {Liang}, {Lin},
      {Lin}, {Lin}, {Lintott}, {Long}, {Longa-Pe{\~n}a}, {L{\'o}pez-Cob{\'a}},
      {Lu}, {Lundgren}, {Luo}, {Mackereth}, {de la Macorra}, {Mahadevan},
      {Majewski}, {Manchado}, {Mandeville}, {Maraston}, {Margalef-Bentabol},
      {Masseron}, {Masters}, {Mathur}, {McDermid}, {Mckay}, {Merloni},
      {Merrifield}, {Meszaros}, {Miglio}, {Di Mille}, {Minniti}, {Minsley},
      {Monachesi}, {Moon}, {Mosser}, {Mulchaey}, {Muna}, {Mu{\~n}oz}, {Myers},
      {Myers}, {Nadathur}, {Nair}, {Nandra}, {Neumann}, {Newman}, {Nidever},
      {Nikakhtar}, {Nitschelm}, {O'Connell}, {Garma-Oehmichen}, {Luan Souza de
      Oliveira}, {Olney}, {Oravetz}, {Ortigoza-Urdaneta}, {Osorio}, {Otter},
      {Pace}, {Padilla}, {Pan}, {Pan}, {Parikh}, {Parker}, {Peirani}, {Pe{\~n}a
      Ram{\'\i}rez}, {Penny}, {Percival}, {Perez-Fournon}, {Pinsonneault},
      {Poidevin}, {Poovelil}, {Price-Whelan}, {B{\'a}rbara de Andrade Queiroz},
      {Raddick}, {Ray}, {Rembold}, {Riddle}, {Riffel}, {Riffel}, {Rix}, {Robin},
      {Rodr{\'\i}guez-Puebla}, {Roman-Lopes}, {Rom{\'a}n-Z{\'u}{\~n}iga}, {Rose},
      {Ross}, {Rossi}, {Rubin}, {Salvato}, {S{\'a}nchez}, {S{\'a}nchez-Gallego},
      {Sanderson}, {Santana Rojas}, {Sarceno}, {Sarmiento}, {Sayres}, {Sazonova},
      {Schaefer}, {Schiavon}, {Schlegel}, {Schneider}, {Schultheis}, {Schwope},
      {Serenelli}, {Serna}, {Shao}, {Shapiro}, {Sharma}, {Shen}, {Shetrone}, {Shu},
      {Simon}, {Skrutskie}, {Smethurst}, {Smith}, {Sobeck}, {Spoo}, {Sprague},
      {Stark}, {Stassun}, {Steinmetz}, {Stello}, {Stone-Martinez},
      {Storchi-Bergmann}, {Stringfellow}, {Stutz}, {Su}, {Taghizadeh-Popp},
      {Talbot}, {Tayar}, {Telles}, {Teske}, {Thakar}, {Theissen}, {Tkachenko},
      {Thomas}, {Tojeiro}, {Hernandez Toledo}, {Troup}, {Trump}, {Trussler},
      {Turner}, {Tuttle}, {Unda-Sanzana}, {V{\'a}zquez-Mata}, {Valentini},
      {Valenzuela}, {Vargas-Gonz{\'a}lez}, {Vargas-Maga{\~n}a}, {Alfaro},
      {Villanova}, {Vincenzo}, {Wake}, {Warfield}, {Washington}, {Weaver},
      {Weijmans}, {Weinberg}, {Weiss}, {Westfall}, {Wild}, {Wilde}, {Wilson},
      {Wilson}, {Wilson}, {Wolf}, {Wood-Vasey}, {Yan}, {Zamora}, {Zasowski},
      {Zhang}, {Zhao}, {Zheng}, {Zheng}, \& {Zhu}}]{Abdurrouf+etal+2022}
    {Abdurro'uf}, {Accetta}, K., {Aerts}, C., {et~al.} 2022, \apjs, 259, 35,
      \dodoi{10.3847/1538-4365/ac4414}
    
    \bibitem[{{Arce} {et~al.}(2011){Arce}, {Borkin}, {Goodman}, {Pineda}, \&
      {Beaumont}}]{Arce+etal+2011}
    {Arce}, H.~G., {Borkin}, M.~A., {Goodman}, A.~A., {Pineda}, J.~E., \&
      {Beaumont}, C.~N. 2011, \apj, 742, 105, \dodoi{10.1088/0004-637X/742/2/105}
    
    \bibitem[{{Arce} {et~al.}(2010){Arce}, {Borkin}, {Goodman}, {Pineda}, \&
      {Halle}}]{Arce+etal+2010}
    {Arce}, H.~G., {Borkin}, M.~A., {Goodman}, A.~A., {Pineda}, J.~E., \& {Halle},
      M.~W. 2010, \apj, 715, 1170, \dodoi{10.1088/0004-637X/715/2/1170}
    
    \bibitem[{{Bally}(2016)}]{Bally+2016}
    {Bally}, J. 2016, \araa, 54, 491, \dodoi{10.1146/annurev-astro-081915-023341}
    
    \bibitem[{{Bontemps} {et~al.}(1996){Bontemps}, {Andre}, {Terebey}, \&
      {Cabrit}}]{Bontemps+etal+1996}
    {Bontemps}, S., {Andre}, P., {Terebey}, S., \& {Cabrit}, S. 1996, \aap, 311,
      858
    
    \bibitem[{{Broos} {et~al.}(2013){Broos}, {Getman}, {Povich}, {Feigelson},
      {Townsley}, {Naylor}, {Kuhn}, {King}, \& {Busk}}]{Broos+etal+2013}
    {Broos}, P.~S., {Getman}, K.~V., {Povich}, M.~S., {et~al.} 2013, \apjs, 209,
      32, \dodoi{10.1088/0067-0049/209/2/32}
    
    \bibitem[{{Buckle} {et~al.}(2012){Buckle}, {Richer}, \&
      {Davis}}]{Buckle+etal+2012}
    {Buckle}, J.~V., {Richer}, J.~S., \& {Davis}, C.~J. 2012, \mnras, 423, 1127,
      \dodoi{10.1111/j.1365-2966.2012.20941.x}
    
    \bibitem[{{Castor} {et~al.}(1975){Castor}, {McCray}, \&
      {Weaver}}]{Castor+etal+1975}
    {Castor}, J., {McCray}, R., \& {Weaver}, R. 1975, \apjl, 200, L107,
      \dodoi{10.1086/181908}
    
    \bibitem[{{Chen} {et~al.}(2013){Chen}, {Zhou}, \& {Chu}}]{Chen+etal+2013}
    {Chen}, Y., {Zhou}, P., \& {Chu}, Y.-H. 2013, \apjl, 769, L16,
      \dodoi{10.1088/2041-8205/769/1/L16}
    
    \bibitem[{{Churchwell} {et~al.}(2006){Churchwell}, {Povich}, {Allen}, {Taylor},
      {Meade}, {Babler}, {Indebetouw}, {Watson}, {Whitney}, {Wolfire}, {Bania},
      {Benjamin}, {Clemens}, {Cohen}, {Cyganowski}, {Jackson}, {Kobulnicky},
      {Mathis}, {Mercer}, {Stolovy}, {Uzpen}, {Watson}, \&
      {Wolff}}]{Churchwell+etal+2006}
    {Churchwell}, E., {Povich}, M.~S., {Allen}, D., {et~al.} 2006, \apj, 649, 759,
      \dodoi{10.1086/507015}
    
    \bibitem[{{Cody} {et~al.}(2014){Cody}, {Stauffer}, {Baglin}, {Micela},
      {Rebull}, {Flaccomio}, {Morales-Calder{\'o}n}, {Aigrain}, {Bouvier},
      {Hillenbrand}, {Gutermuth}, {Song}, {Turner}, {Alencar}, {Zwintz},
      {Plavchan}, {Carpenter}, {Findeisen}, {Carey}, {Terebey}, {Hartmann},
      {Calvet}, {Teixeira}, {Vrba}, {Wolk}, {Covey}, {Poppenhaeger}, {G{\"u}nther},
      {Forbrich}, {Whitney}, {Affer}, {Herbst}, {Hora}, {Barrado}, {Holtzman},
      {Marchis}, {Wood}, {Medeiros Guimar{\~a}es}, {Lillo Box}, {Gillen},
      {McQuillan}, {Espaillat}, {Allen}, {D'Alessio}, \& {Favata}}]{Cody+etal+2014}
    {Cody}, A.~M., {Stauffer}, J., {Baglin}, A., {et~al.} 2014, \aj, 147, 82,
      \dodoi{10.1088/0004-6256/147/4/82}
    
    \bibitem[{{Dahm}(2008)}]{Dahm+2008}
    {Dahm}, S.~E. 2008, in Handbook of Star Forming Regions, Volume I, ed.
      B.~{Reipurth}, Vol.~4, 966
    
    \bibitem[{{Dale}(2015)}]{Dale+2015}
    {Dale}, J.~E. 2015, \nar, 68, 1, \dodoi{10.1016/j.newar.2015.06.001}
    
    \bibitem[{{Dyson} \& {Williams}(1980)}]{Dyson+Williams+1980}
    {Dyson}, J.~E., \& {Williams}, D.~A. 1980, {Physics of the interstellar medium}
    
    \bibitem[{{Fazio} {et~al.}(2004){Fazio}, {Hora}, {Allen}, {Ashby}, {Barmby},
      {Deutsch}, {Huang}, {Kleiner}, {Marengo}, {Megeath}, {Melnick}, {Pahre},
      {Patten}, {Polizotti}, {Smith}, {Taylor}, {Wang}, {Willner}, {Hoffmann},
      {Pipher}, {Forrest}, {McMurty}, {McCreight}, {McKelvey}, {McMurray}, {Koch},
      {Moseley}, {Arendt}, {Mentzell}, {Marx}, {Losch}, {Mayman}, {Eichhorn},
      {Krebs}, {Jhabvala}, {Gezari}, {Fixsen}, {Flores}, {Shakoorzadeh}, {Jungo},
      {Hakun}, {Workman}, {Karpati}, {Kichak}, {Whitley}, {Mann}, {Tollestrup},
      {Eisenhardt}, {Stern}, {Gorjian}, {Bhattacharya}, {Carey}, {Nelson},
      {Glaccum}, {Lacy}, {Lowrance}, {Laine}, {Reach}, {Stauffer}, {Surace},
      {Wilson}, {Wright}, {Hoffman}, {Domingo}, \& {Cohen}}]{Fazio+etal+2004}
    {Fazio}, G.~G., {Hora}, J.~L., {Allen}, L.~E., {et~al.} 2004, \apjs, 154, 10,
      \dodoi{10.1086/422843}
    
    \bibitem[{{Feddersen} {et~al.}(2018){Feddersen}, {Arce}, {Kong}, {Shimajiri},
      {Nakamura}, {Hara}, {Ishii}, {Sasaki}, \& {Kawabe}}]{Feddersen+etal+2018}
    {Feddersen}, J.~R., {Arce}, H.~G., {Kong}, S., {et~al.} 2018, \apj, 862, 121,
      \dodoi{10.3847/1538-4357/aacaf2}
    
    \bibitem[{{Feigelson} {et~al.}(2013){Feigelson}, {Townsley}, {Broos}, {Busk},
      {Getman}, {King}, {Kuhn}, {Naylor}, {Povich}, {Baddeley}, {Bate},
      {Indebetouw}, {Luhman}, {McCaughrean}, {Pittard}, {Pudritz}, {Sills}, {Song},
      \& {Wadsley}}]{Feigelson+etal+2013}
    {Feigelson}, E.~D., {Townsley}, L.~K., {Broos}, P.~S., {et~al.} 2013, \apjs,
      209, 26, \dodoi{10.1088/0067-0049/209/2/26}
    
    \bibitem[{{F{\H{u}}r{\'e}sz} {et~al.}(2008){F{\H{u}}r{\'e}sz}, {Hartmann},
      {Megeath}, {Szentgyorgyi}, \& {Hamden}}]{Furesz+etal+2008}
    {F{\H{u}}r{\'e}sz}, G., {Hartmann}, L.~W., {Megeath}, S.~T., {Szentgyorgyi},
      A.~H., \& {Hamden}, E.~T. 2008, \apj, 676, 1109, \dodoi{10.1086/525844}
    
    \bibitem[{{Flaccomio} {et~al.}(2023){Flaccomio}, {Micela}, {Peres},
      {Sciortino}, {Salvaggio}, {Prisinzano}, {Guarcello}, {Venuti}, {Bonito}, \&
      {Pillitteri}}]{Flaccomio+etal+2023}
    {Flaccomio}, E., {Micela}, G., {Peres}, G., {et~al.} 2023, \aap, 670, A37,
      \dodoi{10.1051/0004-6361/202244872}
    
    \bibitem[{{Frank} {et~al.}(2014){Frank}, {Ray}, {Cabrit}, {Hartigan}, {Arce},
      {Bacciotti}, {Bally}, {Benisty}, {Eisl{\"o}ffel}, {G{\"u}del}, {Lebedev},
      {Nisini}, \& {Raga}}]{Frank+etal+2014}
    {Frank}, A., {Ray}, T.~P., {Cabrit}, S., {et~al.} 2014, in Protostars and
      Planets VI, ed. H.~{Beuther}, R.~S. {Klessen}, C.~P. {Dullemond}, \&
      T.~{Henning}, 451, \dodoi{10.2458/azu_uapress_9780816531240-ch020}
    
    \bibitem[{{Gaia Collaboration} {et~al.}(2016){Gaia Collaboration}, {Prusti},
      {de Bruijne}, {Brown}, {Vallenari}, {Babusiaux}, {Bailer-Jones}, {Bastian},
      {Biermann}, {Evans}, {Eyer}, {Jansen}, {Jordi}, {Klioner}, {Lammers},
      {Lindegren}, {Luri}, {Mignard}, {Milligan}, {Panem}, {Poinsignon},
      {Pourbaix}, {Randich}, {Sarri}, {Sartoretti}, {Siddiqui}, {Soubiran},
      {Valette}, {van Leeuwen}, {Walton}, {Aerts}, {Arenou}, {Cropper}, {Drimmel},
      {H{\o}g}, {Katz}, {Lattanzi}, {O'Mullane}, {Grebel}, {Holland}, {Huc},
      {Passot}, {Bramante}, {Cacciari}, {Casta{\~n}eda}, {Chaoul}, {Cheek}, {De
      Angeli}, {Fabricius}, {Guerra}, {Hern{\'a}ndez}, {Jean-Antoine-Piccolo},
      {Masana}, {Messineo}, {Mowlavi}, {Nienartowicz}, {Ord{\'o}{\~n}ez-Blanco},
      {Panuzzo}, {Portell}, {Richards}, {Riello}, {Seabroke}, {Tanga},
      {Th{\'e}venin}, {Torra}, {Els}, {Gracia-Abril}, {Comoretto},
      {Garcia-Reinaldos}, {Lock}, {Mercier}, {Altmann}, {Andrae}, {Astraatmadja},
      {Bellas-Velidis}, {Benson}, {Berthier}, {Blomme}, {Busso}, {Carry},
      {Cellino}, {Clementini}, {Cowell}, {Creevey}, {Cuypers}, {Davidson}, {De
      Ridder}, {de Torres}, {Delchambre}, {Dell'Oro}, {Ducourant}, {Fr{\'e}mat},
      {Garc{\'\i}a-Torres}, {Gosset}, {Halbwachs}, {Hambly}, {Harrison}, {Hauser},
      {Hestroffer}, {Hodgkin}, {Huckle}, {Hutton}, {Jasniewicz}, {Jordan},
      {Kontizas}, {Korn}, {Lanzafame}, {Manteiga}, {Moitinho}, {Muinonen},
      {Osinde}, {Pancino}, {Pauwels}, {Petit}, {Recio-Blanco}, {Robin}, {Sarro},
      {Siopis}, {Smith}, {Smith}, {Sozzetti}, {Thuillot}, {van Reeven}, {Viala},
      {Abbas}, {Abreu Aramburu}, {Accart}, {Aguado}, {Allan}, {Allasia},
      {Altavilla}, {{\'A}lvarez}, {Alves}, {Anderson}, {Andrei}, {Anglada Varela},
      {Antiche}, {Antoja}, {Ant{\'o}n}, {Arcay}, {Atzei}, {Ayache}, {Bach},
      {Baker}, {Balaguer-N{\'u}{\~n}ez}, {Barache}, {Barata}, {Barbier}, {Barblan},
      {Baroni}, {Barrado y Navascu{\'e}s}, {Barros}, {Barstow}, {Becciani},
      {Bellazzini}, {Bellei}, {Bello Garc{\'\i}a}, {Belokurov}, {Bendjoya},
      {Berihuete}, {Bianchi}, {Bienaym{\'e}}, {Billebaud}, {Blagorodnova},
      {Blanco-Cuaresma}, {Boch}, {Bombrun}, {Borrachero}, {Bouquillon}, {Bourda},
      {Bouy}, {Bragaglia}, {Breddels}, {Brouillet}, {Br{\"u}semeister},
      {Bucciarelli}, {Budnik}, {Burgess}, {Burgon}, {Burlacu}, {Busonero}, {Buzzi},
      {Caffau}, {Cambras}, {Campbell}, {Cancelliere}, {Cantat-Gaudin}, {Carlucci},
      {Carrasco}, {Castellani}, {Charlot}, {Charnas}, {Charvet}, {Chassat},
      {Chiavassa}, {Clotet}, {Cocozza}, {Collins}, {Collins}, {Costigan}, {Crifo},
      {Cross}, {Crosta}, {Crowley}, {Dafonte}, {Damerdji}, {Dapergolas}, {David},
      {David}, {De Cat}, {de Felice}, {de Laverny}, {De Luise}, {De March}, {de
      Martino}, {de Souza}, {Debosscher}, {del Pozo}, {Delbo}, {Delgado},
      {Delgado}, {di Marco}, {Di Matteo}, {Diakite}, {Distefano}, {Dolding}, {Dos
      Anjos}, {Drazinos}, {Dur{\'a}n}, {Dzigan}, {Ecale}, {Edvardsson}, {Enke},
      {Erdmann}, {Escolar}, {Espina}, {Evans}, {Eynard Bontemps}, {Fabre},
      {Fabrizio}, {Faigler}, {Falc{\~a}o}, {Farr{\`a}s Casas}, {Faye}, {Federici},
      {Fedorets}, {Fern{\'a}ndez-Hern{\'a}ndez}, {Fernique}, {Fienga}, {Figueras},
      {Filippi}, {Findeisen}, {Fonti}, {Fouesneau}, {Fraile}, {Fraser}, {Fuchs},
      {Furnell}, {Gai}, {Galleti}, {Galluccio}, {Garabato}, {Garc{\'\i}a-Sedano},
      {Gar{\'e}}, {Garofalo}, {Garralda}, {Gavras}, {Gerssen}, {Geyer}, {Gilmore},
      {Girona}, {Giuffrida}, {Gomes}, {Gonz{\'a}lez-Marcos},
      {Gonz{\'a}lez-N{\'u}{\~n}ez}, {Gonz{\'a}lez-Vidal}, {Granvik}, {Guerrier},
      {Guillout}, {Guiraud}, {G{\'u}rpide}, {Guti{\'e}rrez-S{\'a}nchez}, {Guy},
      {Haigron}, {Hatzidimitriou}, {Haywood}, {Heiter}, {Helmi}, {Hobbs},
      {Hofmann}, {Holl}, {Holland}, {Hunt}, {Hypki}, {Icardi}, {Irwin}, {Jevardat
      de Fombelle}, {Jofr{\'e}}, {Jonker}, {Jorissen}, {Julbe}, {Karampelas},
      {Kochoska}, {Kohley}, {Kolenberg}, {Kontizas}, {Koposov}, {Kordopatis},
      {Koubsky}, {Kowalczyk}, {Krone-Martins}, {Kudryashova}, {Kull}, {Bachchan},
      {Lacoste-Seris}, {Lanza}, {Lavigne}, {Le Poncin-Lafitte}, {Lebreton},
      {Lebzelter}, {Leccia}, {Leclerc}, {Lecoeur-Taibi}, {Lemaitre}, {Lenhardt},
      {Leroux}, {Liao}, {Licata}, {Lindstr{\o}m}, {Lister}, {Livanou}, {Lobel},
      {L{\"o}ffler}, {L{\'o}pez}, {Lopez-Lozano}, {Lorenz}, {Loureiro},
      {MacDonald}, {Magalh{\~a}es Fernandes}, {Managau}, {Mann}, {Mantelet},
      {Marchal}, {Marchant}, {Marconi}, {Marie}, {Marinoni}, {Marrese},
      {Marschalk{\'o}}, {Marshall}, {Mart{\'\i}n-Fleitas}, {Martino}, {Mary},
      {Matijevi{\v{c}}}, {Mazeh}, {McMillan}, {Messina}, {Mestre}, {Michalik},
      {Millar}, {Miranda}, {Molina}, {Molinaro}, {Molinaro}, {Moln{\'a}r},
      {Moniez}, {Montegriffo}, {Monteiro}, {Mor}, {Mora}, {Morbidelli}, {Morel},
      {Morgenthaler}, {Morley}, {Morris}, {Mulone}, {Muraveva}, {Musella},
      {Narbonne}, {Nelemans}, {Nicastro}, {Noval}, {Ord{\'e}novic},
      {Ordieres-Mer{\'e}}, {Osborne}, {Pagani}, {Pagano}, {Pailler}, {Palacin},
      {Palaversa}, {Parsons}, {Paulsen}, {Pecoraro}, {Pedrosa}, {Pentik{\"a}inen},
      {Pereira}, {Pichon}, {Piersimoni}, {Pineau}, {Plachy}, {Plum}, {Poujoulet},
      {Pr{\v{s}}a}, {Pulone}, {Ragaini}, {Rago}, {Rambaux}, {Ramos-Lerate},
      {Ranalli}, {Rauw}, {Read}, {Regibo}, {Renk}, {Reyl{\'e}}, {Ribeiro},
      {Rimoldini}, {Ripepi}, {Riva}, {Rixon}, {Roelens}, {Romero-G{\'o}mez},
      {Rowell}, {Royer}, {Rudolph}, {Ruiz-Dern}, {Sadowski}, {Sagrist{\`a}
      Sell{\'e}s}, {Sahlmann}, {Salgado}, {Salguero}, {Sarasso}, {Savietto},
      {Schnorhk}, {Schultheis}, {Sciacca}, {Segol}, {Segovia}, {Segransan},
      {Serpell}, {Shih}, {Smareglia}, {Smart}, {Smith}, {Solano}, {Solitro},
      {Sordo}, {Soria Nieto}, {Souchay}, {Spagna}, {Spoto}, {Stampa}, {Steele},
      {Steidelm{\"u}ller}, {Stephenson}, {Stoev}, {Suess}, {S{\"u}veges}, {Surdej},
      {Szabados}, {Szegedi-Elek}, {Tapiador}, {Taris}, {Tauran}, {Taylor},
      {Teixeira}, {Terrett}, {Tingley}, {Trager}, {Turon}, {Ulla}, {Utrilla},
      {Valentini}, {van Elteren}, {Van Hemelryck}, {van Leeuwen}, {Varadi},
      {Vecchiato}, {Veljanoski}, {Via}, {Vicente}, {Vogt}, {Voss}, {Votruba},
      {Voutsinas}, {Walmsley}, {Weiler}, {Weingrill}, {Werner}, {Wevers},
      {Whitehead}, {Wyrzykowski}, {Yoldas}, {{\v{Z}}erjal}, {Zucker}, {Zurbach},
      {Zwitter}, {Alecu}, {Allen}, {Allende Prieto}, {Amorim},
      {Anglada-Escud{\'e}}, {Arsenijevic}, {Azaz}, {Balm}, {Beck}, {Bernstein},
      {Bigot}, {Bijaoui}, {Blasco}, {Bonfigli}, {Bono}, {Boudreault}, {Bressan},
      {Brown}, {Brunet}, {Bunclark}, {Buonanno}, {Butkevich}, {Carret}, {Carrion},
      {Chemin}, {Ch{\'e}reau}, {Corcione}, {Darmigny}, {de Boer}, {de Teodoro}, {de
      Zeeuw}, {Delle Luche}, {Domingues}, {Dubath}, {Fodor}, {Fr{\'e}zouls},
      {Fries}, {Fustes}, {Fyfe}, {Gallardo}, {Gallegos}, {Gardiol}, {Gebran},
      {Gomboc}, {G{\'o}mez}, {Grux}, {Gueguen}, {Heyrovsky}, {Hoar}, {Iannicola},
      {Isasi Parache}, {Janotto}, {Joliet}, {Jonckheere}, {Keil}, {Kim},
      {Klagyivik}, {Klar}, {Knude}, {Kochukhov}, {Kolka}, {Kos}, {Kutka}, {Lainey},
      {LeBouquin}, {Liu}, {Loreggia}, {Makarov}, {Marseille}, {Martayan},
      {Martinez-Rubi}, {Massart}, {Meynadier}, {Mignot}, {Munari}, {Nguyen},
      {Nordlander}, {Ocvirk}, {O'Flaherty}, {Olias Sanz}, {Ortiz}, {Osorio},
      {Oszkiewicz}, {Ouzounis}, {Palmer}, {Park}, {Pasquato}, {Peltzer}, {Peralta},
      {P{\'e}turaud}, {Pieniluoma}, {Pigozzi}, {Poels}, {Prat}, {Prod'homme},
      {Raison}, {Rebordao}, {Risquez}, {Rocca-Volmerange}, {Rosen}, {Ruiz-Fuertes},
      {Russo}, {Sembay}, {Serraller Vizcaino}, {Short}, {Siebert}, {Silva},
      {Sinachopoulos}, {Slezak}, {Soffel}, {Sosnowska}, {Strai{\v{z}}ys}, {ter
      Linden}, {Terrell}, {Theil}, {Tiede}, {Troisi}, {Tsalmantza}, {Tur},
      {Vaccari}, {Vachier}, {Valles}, {Van Hamme}, {Veltz}, {Virtanen}, {Wallut},
      {Wichmann}, {Wilkinson}, {Ziaeepour}, \& {Zschocke}}]{Gaia+2016}
    {Gaia Collaboration}, {Prusti}, T., {de Bruijne}, J.~H.~J., {et~al.} 2016,
      \aap, 595, A1, \dodoi{10.1051/0004-6361/201629272}
    
    \bibitem[{{Gaia Collaboration} {et~al.}(2023){Gaia Collaboration}, {Vallenari},
      {Brown}, {Prusti}, {de Bruijne}, {Arenou}, {Babusiaux}, {Biermann},
      {Creevey}, {Ducourant}, {Evans}, {Eyer}, {Guerra}, {Hutton}, {Jordi},
      {Klioner}, {Lammers}, {Lindegren}, {Luri}, {Mignard}, {Panem}, {Pourbaix},
      {Randich}, {Sartoretti}, {Soubiran}, {Tanga}, {Walton}, {Bailer-Jones},
      {Bastian}, {Drimmel}, {Jansen}, {Katz}, {Lattanzi}, {van Leeuwen}, {Bakker},
      {Cacciari}, {Casta{\~n}eda}, {De Angeli}, {Fabricius}, {Fouesneau},
      {Fr{\'e}mat}, {Galluccio}, {Guerrier}, {Heiter}, {Masana}, {Messineo},
      {Mowlavi}, {Nicolas}, {Nienartowicz}, {Pailler}, {Panuzzo}, {Riclet}, {Roux},
      {Seabroke}, {Sordo}, {Th{\'e}venin}, {Gracia-Abril}, {Portell}, {Teyssier},
      {Altmann}, {Andrae}, {Audard}, {Bellas-Velidis}, {Benson}, {Berthier},
      {Blomme}, {Burgess}, {Busonero}, {Busso}, {C{\'a}novas}, {Carry}, {Cellino},
      {Cheek}, {Clementini}, {Damerdji}, {Davidson}, {de Teodoro}, {Nu{\~n}ez
      Campos}, {Delchambre}, {Dell'Oro}, {Esquej}, {Fern{\'a}ndez-Hern{\'a}ndez},
      {Fraile}, {Garabato}, {Garc{\'\i}a-Lario}, {Gosset}, {Haigron}, {Halbwachs},
      {Hambly}, {Harrison}, {Hern{\'a}ndez}, {Hestroffer}, {Hodgkin}, {Holl},
      {Jan{\ss}en}, {Jevardat de Fombelle}, {Jordan}, {Krone-Martins}, {Lanzafame},
      {L{\"o}ffler}, {Marchal}, {Marrese}, {Moitinho}, {Muinonen}, {Osborne},
      {Pancino}, {Pauwels}, {Recio-Blanco}, {Reyl{\'e}}, {Riello}, {Rimoldini},
      {Roegiers}, {Rybizki}, {Sarro}, {Siopis}, {Smith}, {Sozzetti}, {Utrilla},
      {van Leeuwen}, {Abbas}, {{\'A}brah{\'a}m}, {Abreu Aramburu}, {Aerts},
      {Aguado}, {Ajaj}, {Aldea-Montero}, {Altavilla}, {{\'A}lvarez}, {Alves},
      {Anders}, {Anderson}, {Anglada Varela}, {Antoja}, {Baines}, {Baker},
      {Balaguer-N{\'u}{\~n}ez}, {Balbinot}, {Balog}, {Barache}, {Barbato},
      {Barros}, {Barstow}, {Bartolom{\'e}}, {Bassilana}, {Bauchet}, {Becciani},
      {Bellazzini}, {Berihuete}, {Bernet}, {Bertone}, {Bianchi}, {Binnenfeld},
      {Blanco-Cuaresma}, {Blazere}, {Boch}, {Bombrun}, {Bossini}, {Bouquillon},
      {Bragaglia}, {Bramante}, {Breedt}, {Bressan}, {Brouillet}, {Brugaletta},
      {Bucciarelli}, {Burlacu}, {Butkevich}, {Buzzi}, {Caffau}, {Cancelliere},
      {Cantat-Gaudin}, {Carballo}, {Carlucci}, {Carnerero}, {Carrasco},
      {Casamiquela}, {Castellani}, {Castro-Ginard}, {Chaoul}, {Charlot}, {Chemin},
      {Chiaramida}, {Chiavassa}, {Chornay}, {Comoretto}, {Contursi}, {Cooper},
      {Cornez}, {Cowell}, {Crifo}, {Cropper}, {Crosta}, {Crowley}, {Dafonte},
      {Dapergolas}, {David}, {David}, {de Laverny}, {De Luise}, {De March}, {De
      Ridder}, {de Souza}, {de Torres}, {del Peloso}, {del Pozo}, {Delbo},
      {Delgado}, {Delisle}, {Demouchy}, {Dharmawardena}, {Di Matteo}, {Diakite},
      {Diener}, {Distefano}, {Dolding}, {Edvardsson}, {Enke}, {Fabre}, {Fabrizio},
      {Faigler}, {Fedorets}, {Fernique}, {Fienga}, {Figueras}, {Fournier},
      {Fouron}, {Fragkoudi}, {Gai}, {Garcia-Gutierrez}, {Garcia-Reinaldos},
      {Garc{\'\i}a-Torres}, {Garofalo}, {Gavel}, {Gavras}, {Gerlach}, {Geyer},
      {Giacobbe}, {Gilmore}, {Girona}, {Giuffrida}, {Gomel}, {Gomez},
      {Gonz{\'a}lez-N{\'u}{\~n}ez}, {Gonz{\'a}lez-Santamar{\'\i}a},
      {Gonz{\'a}lez-Vidal}, {Granvik}, {Guillout}, {Guiraud},
      {Guti{\'e}rrez-S{\'a}nchez}, {Guy}, {Hatzidimitriou}, {Hauser}, {Haywood},
      {Helmer}, {Helmi}, {Sarmiento}, {Hidalgo}, {Hilger}, {H{\l}adczuk}, {Hobbs},
      {Holland}, {Huckle}, {Jardine}, {Jasniewicz}, {Jean-Antoine Piccolo},
      {Jim{\'e}nez-Arranz}, {Jorissen}, {Juaristi Campillo}, {Julbe}, {Karbevska},
      {Kervella}, {Khanna}, {Kontizas}, {Kordopatis}, {Korn}, {K{\'o}sp{\'a}l},
      {Kostrzewa-Rutkowska}, {Kruszy{\'n}ska}, {Kun}, {Laizeau}, {Lambert},
      {Lanza}, {Lasne}, {Le Campion}, {Lebreton}, {Lebzelter}, {Leccia}, {Leclerc},
      {Lecoeur-Taibi}, {Liao}, {Licata}, {Lindstr{\o}m}, {Lister}, {Livanou},
      {Lobel}, {Lorca}, {Loup}, {Madrero Pardo}, {Magdaleno Romeo}, {Managau},
      {Mann}, {Manteiga}, {Marchant}, {Marconi}, {Marcos}, {Marcos Santos},
      {Mar{\'\i}n Pina}, {Marinoni}, {Marocco}, {Marshall}, {Martin Polo},
      {Mart{\'\i}n-Fleitas}, {Marton}, {Mary}, {Masip}, {Massari},
      {Mastrobuono-Battisti}, {Mazeh}, {McMillan}, {Messina}, {Michalik}, {Millar},
      {Mints}, {Molina}, {Molinaro}, {Moln{\'a}r}, {Monari}, {Mongui{\'o}},
      {Montegriffo}, {Montero}, {Mor}, {Mora}, {Morbidelli}, {Morel}, {Morris},
      {Muraveva}, {Murphy}, {Musella}, {Nagy}, {Noval}, {Oca{\~n}a}, {Ogden},
      {Ordenovic}, {Osinde}, {Pagani}, {Pagano}, {Palaversa}, {Palicio},
      {Pallas-Quintela}, {Panahi}, {Payne-Wardenaar}, {Pe{\~n}alosa Esteller},
      {Penttil{\"a}}, {Pichon}, {Piersimoni}, {Pineau}, {Plachy}, {Plum}, {Poggio},
      {Pr{\v{s}}a}, {Pulone}, {Racero}, {Ragaini}, {Rainer}, {Raiteri}, {Rambaux},
      {Ramos}, {Ramos-Lerate}, {Re Fiorentin}, {Regibo}, {Richards}, {Rios Diaz},
      {Ripepi}, {Riva}, {Rix}, {Rixon}, {Robichon}, {Robin}, {Robin}, {Roelens},
      {Rogues}, {Rohrbasser}, {Romero-G{\'o}mez}, {Rowell}, {Royer}, {Ruz Mieres},
      {Rybicki}, {Sadowski}, {S{\'a}ez N{\'u}{\~n}ez}, {Sagrist{\`a} Sell{\'e}s},
      {Sahlmann}, {Salguero}, {Samaras}, {Sanchez Gimenez}, {Sanna},
      {Santove{\~n}a}, {Sarasso}, {Schultheis}, {Sciacca}, {Segol}, {Segovia},
      {S{\'e}gransan}, {Semeux}, {Shahaf}, {Siddiqui}, {Siebert}, {Siltala},
      {Silvelo}, {Slezak}, {Slezak}, {Smart}, {Snaith}, {Solano}, {Solitro},
      {Souami}, {Souchay}, {Spagna}, {Spina}, {Spoto}, {Steele},
      {Steidelm{\"u}ller}, {Stephenson}, {S{\"u}veges}, {Surdej}, {Szabados},
      {Szegedi-Elek}, {Taris}, {Taylor}, {Teixeira}, {Tolomei}, {Tonello}, {Torra},
      {Torra}, {Torralba Elipe}, {Trabucchi}, {Tsounis}, {Turon}, {Ulla}, {Unger},
      {Vaillant}, {van Dillen}, {van Reeven}, {Vanel}, {Vecchiato}, {Viala},
      {Vicente}, {Voutsinas}, {Weiler}, {Wevers}, {Wyrzykowski}, {Yoldas}, {Yvard},
      {Zhao}, {Zorec}, {Zucker}, \& {Zwitter}}]{gaia2022}
    {Gaia Collaboration}, {Vallenari}, A., {Brown}, A.~G.~A., {et~al.} 2023, \aap,
      674, A1, \dodoi{10.1051/0004-6361/202243940}
    
    \bibitem[{{Garden} {et~al.}(1991){Garden}, {Hayashi}, {Gatley}, {Hasegawa}, \&
      {Kaifu}}]{Garden+etal+1991}
    {Garden}, R.~P., {Hayashi}, M., {Gatley}, I., {Hasegawa}, T., \& {Kaifu}, N.
      1991, \apj, 374, 540, \dodoi{10.1086/170143}
    
    \bibitem[{{Geen} {et~al.}(2015){Geen}, {Rosdahl}, {Blaizot}, {Devriendt}, \&
      {Slyz}}]{Geen+etal+2015}
    {Geen}, S., {Rosdahl}, J., {Blaizot}, J., {Devriendt}, J., \& {Slyz}, A. 2015,
      \mnras, 448, 3248, \dodoi{10.1093/mnras/stv251}
    
    \bibitem[{{Gilmore} {et~al.}(2012){Gilmore}, {Randich}, {Asplund}, {Binney},
      {Bonifacio}, {Drew}, {Feltzing}, {Ferguson}, {Jeffries}, {Micela},
      {Negueruela}, {Prusti}, {Rix}, {Vallenari}, {Alfaro}, {Allende-Prieto},
      {Babusiaux}, {Bensby}, {Blomme}, {Bragaglia}, {Flaccomio}, {Fran{\c{c}}ois},
      {Irwin}, {Koposov}, {Korn}, {Lanzafame}, {Pancino}, {Paunzen},
      {Recio-Blanco}, {Sacco}, {Smiljanic}, {Van Eck}, {Walton}, {Aden}, {Aerts},
      {Affer}, {Alcala}, {Altavilla}, {Alves}, {Antoja}, {Arenou}, {Argiroffi},
      {Asensio Ramos}, {Bailer-Jones}, {Balaguer-Nunez}, {Bayo}, {Barbuy},
      {Barisevicius}, {Barrado y Navascues}, {Battistini}, {Bellas Velidis},
      {Bellazzini}, {Belokurov}, {Bergemann}, {Bertelli}, {Biazzo}, {Bienayme},
      {Bland-Hawthorn}, {Boeche}, {Bonito}, {Boudreault}, {Bouvier}, {Brandao},
      {Brown}, {de Bruijne}, {Burleigh}, {Caballero}, {Caffau}, {Calura},
      {Capuzzo-Dolcetta}, {Caramazza}, {Carraro}, {Casagrande}, {Casewell},
      {Chapman}, {Chiappini}, {Chorniy}, {Christlieb}, {Cignoni}, {Cocozza},
      {Colless}, {Collet}, {Collins}, {Correnti}, {Covino}, {Crnojevic}, {Cropper},
      {Cunha}, {Damiani}, {David}, {Delgado}, {Duffau}, {Edvardsson}, {Eldridge},
      {Enke}, {Eriksson}, {Evans}, {Eyer}, {Famaey}, {Fellhauer}, {Ferreras},
      {Figueras}, {Fiorentino}, {Flynn}, {Folha}, {Franciosini}, {Frasca},
      {Freeman}, {Fremat}, {Friel}, {Gaensicke}, {Gameiro}, {Garzon}, {Geier},
      {Geisler}, {Gerhard}, {Gibson}, {Gomboc}, {Gomez}, {Gonzalez-Fernandez},
      {Gonzalez Hernandez}, {Gosset}, {Grebel}, {Greimel}, {Groenewegen},
      {Grundahl}, {Guarcello}, {Gustafsson}, {Hadrava}, {Hatzidimitriou}, {Hambly},
      {Hammersley}, {Hansen}, {Haywood}, {Heber}, {Heiter}, {Held}, {Helmi},
      {Hensler}, {Herrero}, {Hill}, {Hodgkin}, {Huelamo}, {Huxor}, {Ibata},
      {Jackson}, {de Jong}, {Jonker}, {Jordan}, {Jordi}, {Jorissen}, {Katz},
      {Kawata}, {Keller}, {Kharchenko}, {Klement}, {Klutsch}, {Knude}, {Koch},
      {Kochukhov}, {Kontizas}, {Koubsky}, {Lallement}, {de Laverny}, {van Leeuwen},
      {Lemasle}, {Lewis}, {Lind}, {Lindstrom}, {Lobel}, {Lopez Santiago}, {Lucas},
      {Ludwig}, {Lueftinger}, {Magrini}, {Maiz Apellaniz}, {Maldonado}, {Marconi},
      {Marino}, {Martayan}, {Martinez-Valpuesta}, {Matijevic}, {McMahon},
      {Messina}, {Meyer}, {Miglio}, {Mikolaitis}, {Minchev}, {Minniti}, {Moitinho},
      {Momany}, {Monaco}, {Montalto}, {Monteiro}, {Monier}, {Montes}, {Mora},
      {Moraux}, {Morel}, {Mowlavi}, {Mucciarelli}, {Munari}, {Napiwotzki},
      {Nardetto}, {Naylor}, {Naze}, {Nelemans}, {Okamoto}, {Ortolani}, {Pace},
      {Palla}, {Palous}, {Parker}, {Penarrubia}, {Pillitteri}, {Piotto}, {Posbic},
      {Prisinzano}, {Puzeras}, {Quirrenbach}, {Ragaini}, {Read}, {Read}, {Reyle},
      {De Ridder}, {Robichon}, {Robin}, {Roeser}, {Romano}, {Royer}, {Ruchti},
      {Ruzicka}, {Ryan}, {Ryde}, {Santos}, {Sanz Forcada}, {Sarro Baro},
      {Sbordone}, {Schilbach}, {Schmeja}, {Schnurr}, {Schoenrich}, {Scholz},
      {Seabroke}, {Sharma}, {De Silva}, {Smith}, {Solano}, {Sordo}, {Soubiran},
      {Sousa}, {Spagna}, {Steffen}, {Steinmetz}, {Stelzer}, {Stempels},
      {Tabernero}, {Tautvaisiene}, {Thevenin}, {Torra}, {Tosi}, {Tolstoy}, {Turon},
      {Walker}, {Wambsganss}, {Worley}, {Venn}, {Vink}, {Wyse}, {Zaggia},
      {Zeilinger}, {Zoccali}, {Zorec}, {Zucker}, {Zwitter}, \& {Gaia-ESO Survey
      Team}}]{Gilmore+etal+2012}
    {Gilmore}, G., {Randich}, S., {Asplund}, M., {et~al.} 2012, The Messenger, 147,
      25
    
    \bibitem[{{Grady} {et~al.}(1984){Grady}, {Snow}, \& {Cash}}]{Grady+etal+1984}
    {Grady}, C.~A., {Snow}, T.~P., \& {Cash}, W.~C. 1984, \apj, 283, 218,
      \dodoi{10.1086/162296}
    
    \bibitem[{{Green} {et~al.}(2019){Green}, {Schlafly}, {Zucker}, {Speagle}, \&
      {Finkbeiner}}]{Green+etal+2019}
    {Green}, G.~M., {Schlafly}, E., {Zucker}, C., {Speagle}, J.~S., \&
      {Finkbeiner}, D. 2019, \apj, 887, 93, \dodoi{10.3847/1538-4357/ab5362}
    
    \bibitem[{{Greene} {et~al.}(1994){Greene}, {Wilking}, {Andre}, {Young}, \&
      {Lada}}]{Greene+etal+1994}
    {Greene}, T.~P., {Wilking}, B.~A., {Andre}, P., {Young}, E.~T., \& {Lada},
      C.~J. 1994, \apj, 434, 614, \dodoi{10.1086/174763}
    
    \bibitem[{{Gro{\ss}schedl} {et~al.}(2021){Gro{\ss}schedl}, {Alves}, {Meingast},
      \& {Herbst-Kiss}}]{Groschedl+etal+2021}
    {Gro{\ss}schedl}, J.~E., {Alves}, J., {Meingast}, S., \& {Herbst-Kiss}, G.
      2021, \aap, 647, A91, \dodoi{10.1051/0004-6361/202038913}
    
    \bibitem[{{Gro{\ss}schedl} {et~al.}(2018){Gro{\ss}schedl}, {Alves}, {Meingast},
      {Ackerl}, {Ascenso}, {Bouy}, {Burkert}, {Forbrich}, {F{\"u}rnkranz},
      {Goodman}, {Hacar}, {Herbst-Kiss}, {Lada}, {Larreina}, {Leschinski},
      {Lombardi}, {Moitinho}, {Mortimer}, \& {Zari}}]{Groschedl+etal+2018}
    {Gro{\ss}schedl}, J.~E., {Alves}, J., {Meingast}, S., {et~al.} 2018, \aap, 619,
      A106, \dodoi{10.1051/0004-6361/201833901}
    
    \bibitem[{{Guo} {et~al.}(2021){Guo}, {Chen}, {Yuan}, {Huang}, {Liu}, {Yang},
      {Li}, {Sun}, \& {Liu}}]{Guo+etal+2021}
    {Guo}, H.~L., {Chen}, B.~Q., {Yuan}, H.~B., {et~al.} 2021, \apj, 906, 47,
      \dodoi{10.3847/1538-4357/abc68a}
    
    \bibitem[{{Hacar} {et~al.}(2016){Hacar}, {Alves}, {Forbrich}, {Meingast},
      {Kubiak}, \& {Gro{\ss}schedl}}]{Hacar+etal+2016}
    {Hacar}, A., {Alves}, J., {Forbrich}, J., {et~al.} 2016, \aap, 589, A80,
      \dodoi{10.1051/0004-6361/201527805}
    
    \bibitem[{{Kerr} \& {Lynden-Bell}(1986)}]{Kerr+Bell+1986}
    {Kerr}, F.~J., \& {Lynden-Bell}, D. 1986, \mnras, 221, 1023,
      \dodoi{10.1093/mnras/221.4.1023}
    
    \bibitem[{{Lada}(1985)}]{Lada+1985}
    {Lada}, C.~J. 1985, \araa, 23, 267, \dodoi{10.1146/annurev.aa.23.090185.001411}
    
    \bibitem[{{Lada}(1987)}]{Lada+1987}
    {Lada}, C.~J. 1987, in Star Forming Regions, ed. M.~{Peimbert} \& J.~{Jugaku},
      Vol. 115, 1
    
    \bibitem[{{Lada} \& {Wilking}(1984)}]{Lada+Wilking+1984}
    {Lada}, C.~J., \& {Wilking}, B.~A. 1984, \apj, 287, 610, \dodoi{10.1086/162719}
    
    \bibitem[{{Lada} {et~al.}(2006){Lada}, {Muench}, {Luhman}, {Allen}, {Hartmann},
      {Megeath}, {Myers}, {Fazio}, {Wood}, {Muzerolle}, {Rieke}, {Siegler}, \&
      {Young}}]{Lada+etal+2006}
    {Lada}, C.~J., {Muench}, A.~A., {Luhman}, K.~L., {et~al.} 2006, \aj, 131, 1574,
      \dodoi{10.1086/499808}
    
    \bibitem[{{Lallement} {et~al.}(2019){Lallement}, {Babusiaux}, {Vergely},
      {Katz}, {Arenou}, {Valette}, {Hottier}, \& {Capitanio}}]{Lallement+etal+2019}
    {Lallement}, R., {Babusiaux}, C., {Vergely}, J.~L., {et~al.} 2019, \aap, 625,
      A135, \dodoi{10.1051/0004-6361/201834695}
    
    \bibitem[{{Lamers} \& {Cassinelli}(1999)}]{Lamers+Cassinelli+1999}
    {Lamers}, H. J.~G.~L.~M., \& {Cassinelli}, J.~P. 1999, {Introduction to Stellar
      Winds}
    
    \bibitem[{{Li} {et~al.}(2015){Li}, {Li}, {Qian}, {Xu}, {Goldsmith},
      {Noriega-Crespo}, {Wu}, {Song}, \& {Nan}}]{Li+etal+2015}
    {Li}, H., {Li}, D., {Qian}, L., {et~al.} 2015, \apjs, 219, 20,
      \dodoi{10.1088/0067-0049/219/2/20}
    
    \bibitem[{{Li} {et~al.}(2018){Li}, {Li}, {Xu}, {Wang}, {Du}, {Yang}, \&
      {Yang}}]{Li+etal+2018}
    {Li}, Y., {Li}, F.-C., {Xu}, Y., {et~al.} 2018, \apjs, 235, 15,
      \dodoi{10.3847/1538-4365/aaab67}
    
    \bibitem[{{Li} {et~al.}(2020){Li}, {Xu}, {Sun}, \& {Yang}}]{Li+etal+2020}
    {Li}, Y., {Xu}, Y., {Sun}, Y., \& {Yang}, J. 2020, \apjs, 251, 26,
      \dodoi{10.3847/1538-4365/abc34b}
    
    \bibitem[{{Lindegren} {et~al.}(2021){Lindegren}, {Bastian}, {Biermann},
      {Bombrun}, {de Torres}, {Gerlach}, {Geyer}, {Hern{\'a}ndez}, {Hilger},
      {Hobbs}, {Klioner}, {Lammers}, {McMillan}, {Ramos-Lerate},
      {Steidelm{\"u}ller}, {Stephenson}, \& {van Leeuwen}}]{Lindegren+etal+2021}
    {Lindegren}, L., {Bastian}, U., {Biermann}, M., {et~al.} 2021, \aap, 649, A4,
      \dodoi{10.1051/0004-6361/202039653}
    
    \bibitem[{{Liu} {et~al.}(2021){Liu}, {Xu}, {Li}, {Zheng}, {Lu}, {Hao}, {Lin},
      {Bian}, \& {Liu}}]{Liu+etal+2021}
    {Liu}, D.-J., {Xu}, Y., {Li}, Y.-J., {et~al.} 2021, \apjs, 253, 15,
      \dodoi{10.3847/1538-4365/abcece}
    
    \bibitem[{{Martins} {et~al.}(2005){Martins}, {Schaerer}, \&
      {Hillier}}]{Martins+etal+2005}
    {Martins}, F., {Schaerer}, D., \& {Hillier}, D.~J. 2005, \aap, 436, 1049,
      \dodoi{10.1051/0004-6361:20042386}
    
    \bibitem[{{McKee}(1989)}]{McKee+1989}
    {McKee}, C.~F. 1989, \apj, 345, 782, \dodoi{10.1086/167950}
    
    \bibitem[{{McKee} \& {Ostriker}(2007)}]{McKee+Ostriker+2007}
    {McKee}, C.~F., \& {Ostriker}, E.~C. 2007, \araa, 45, 565,
      \dodoi{10.1146/annurev.astro.45.051806.110602}
    
    \bibitem[{{Montillaud} {et~al.}(2019){Montillaud}, {Juvela}, {Vastel}, {He},
      {Liu}, {Ristorcelli}, {Eden}, {Kang}, {Kim}, {Koch}, {Lee}, {Rawlings},
      {Saajasto}, {Sanhueza}, {Soam}, {Zahorecz}, {Alina}, {B{\"o}gner}, {Cornu},
      {Doi}, {Malinen}, {Marshall}, {Micelotta}, {Pelkonen}, {Viktor T{\'o}th},
      {Traficante}, \& {Wang}}]{Montillaud+etal+2019}
    {Montillaud}, J., {Juvela}, M., {Vastel}, C., {et~al.} 2019, \aap, 631, A3,
      \dodoi{10.1051/0004-6361/201834903}
    
    \bibitem[{Nakamura \& Li(2007)}]{Nakamura+Li+2007}
    Nakamura, F., \& Li, Z.-Y. 2007, The Astrophysical Journal, 662, 395,
      \dodoi{10.1086/517515}
    
    \bibitem[{{Nakamura} \& {Li}(2014)}]{Nakamura+Li+2014}
    {Nakamura}, F., \& {Li}, Z.-Y. 2014, \apj, 783, 115,
      \dodoi{10.1088/0004-637X/783/2/115}
    
    \bibitem[{{Nony} {et~al.}(2021){Nony}, {Robitaille}, {Motte}, {Gonzalez},
      {Joncour}, {Moraux}, {Men'shchikov}, {Didelon}, {Louvet}, {Buckner},
      {Schneider}, {Lumsden}, {Bontemps}, {Pouteau}, {Cunningham}, {Fiorellino},
      {Oudmaijer}, {Andr{\'e}}, \& {Thomasson}}]{Nony+etal+2021}
    {Nony}, T., {Robitaille}, J.~F., {Motte}, F., {et~al.} 2021, \aap, 645, A94,
      \dodoi{10.1051/0004-6361/202039353}
    
    \bibitem[{{Perryman} {et~al.}(1997){Perryman}, {Lindegren}, {Kovalevsky},
      {Hoeg}, {Bastian}, {Bernacca}, {Cr{\'e}z{\'e}}, {Donati}, {Grenon},
      {Grewing}, {van Leeuwen}, {van der Marel}, {Mignard}, {Murray}, {Le Poole},
      {Schrijver}, {Turon}, {Arenou}, {Froeschl{\'e}}, \&
      {Petersen}}]{Perryman+etal+1997}
    {Perryman}, M.~A.~C., {Lindegren}, L., {Kovalevsky}, J., {et~al.} 1997, \aap,
      323, L49
    
    \bibitem[{{Pety}(2005)}]{Pety+2005}
    {Pety}, J. 2005, in SF2A-2005: Semaine de l'Astrophysique Francaise, ed.
      F.~{Casoli}, T.~{Contini}, J.~M. {Hameury}, \& L.~{Pagani}, 721
    
    \bibitem[{{Pilbratt} {et~al.}(2010){Pilbratt}, {Riedinger}, {Passvogel},
      {Crone}, {Doyle}, {Gageur}, {Heras}, {Jewell}, {Metcalfe}, {Ott}, \&
      {Schmidt}}]{Pilbratt+etal+2010}
    {Pilbratt}, G.~L., {Riedinger}, J.~R., {Passvogel}, T., {et~al.} 2010, \aap,
      518, L1, \dodoi{10.1051/0004-6361/201014759}
    
    \bibitem[{{Randich} {et~al.}(2013){Randich}, {Gilmore}, \& {Gaia-ESO
      Consortium}}]{Randich+etal+2013}
    {Randich}, S., {Gilmore}, G., \& {Gaia-ESO Consortium}. 2013, The Messenger,
      154, 47
    
    \bibitem[{{Rapson} {et~al.}(2014){Rapson}, {Pipher}, {Gutermuth}, {Megeath},
      {Allen}, {Myers}, \& {Allen}}]{Rapson+etal+2014}
    {Rapson}, V.~A., {Pipher}, J.~L., {Gutermuth}, R.~A., {et~al.} 2014, \apj, 794,
      124, \dodoi{10.1088/0004-637X/794/2/124}
    
    \bibitem[{{Reid} {et~al.}(2009){Reid}, {Menten}, {Zheng}, {Brunthaler},
      {Moscadelli}, {Xu}, {Zhang}, {Sato}, {Honma}, {Hirota}, {Hachisuka}, {Choi},
      {Moellenbrock}, \& {Bartkiewicz}}]{Reid+etal+2009}
    {Reid}, M.~J., {Menten}, K.~M., {Zheng}, X.~W., {et~al.} 2009, \apj, 700, 137,
      \dodoi{10.1088/0004-637X/700/1/137}
    
    \bibitem[{{Rieke} {et~al.}(2004){Rieke}, {Young}, {Engelbracht}, {Kelly},
      {Low}, {Haller}, {Beeman}, {Gordon}, {Stansberry}, {Misselt}, {Cadien},
      {Morrison}, {Rivlis}, {Latter}, {Noriega-Crespo}, {Padgett}, {Stapelfeldt},
      {Hines}, {Egami}, {Muzerolle}, {Alonso-Herrero}, {Blaylock}, {Dole}, {Hinz},
      {Le Floc'h}, {Papovich}, {P{\'e}rez-Gonz{\'a}lez}, {Smith}, {Su}, {Bennett},
      {Frayer}, {Henderson}, {Lu}, {Masci}, {Pesenson}, {Rebull}, {Rho}, {Keene},
      {Stolovy}, {Wachter}, {Wheaton}, {Werner}, \& {Richards}}]{Rieke+etal+2004}
    {Rieke}, G.~H., {Young}, E.~T., {Engelbracht}, C.~W., {et~al.} 2004, \apjs,
      154, 25, \dodoi{10.1086/422717}
    
    \bibitem[{{Shan} {et~al.}(2012){Shan}, {Yang}, {Shi}, {Yao}, {Zuo}, {Lin},
      {Chen}, {Zhang}, {Duan}, {Cao}, {Li}, {Li}, {Liu}, \&
      {Zhong}}]{Shan+etal+2012}
    {Shan}, W., {Yang}, J., {Shi}, S., {et~al.} 2012, IEEE Transactions on
      Terahertz Science and Technology, 2, 593, \dodoi{10.1109/TTHZ.2012.2213818}
    
    \bibitem[{{Simon} {et~al.}(2001){Simon}, {Jackson}, {Clemens}, {Bania}, \&
      {Heyer}}]{Simon+etal+2001}
    {Simon}, R., {Jackson}, J.~M., {Clemens}, D.~P., {Bania}, T.~M., \& {Heyer},
      M.~H. 2001, \apj, 551, 747, \dodoi{10.1086/320230}
    
    \bibitem[{{Skrutskie} {et~al.}(2006){Skrutskie}, {Cutri}, {Stiening},
      {Weinberg}, {Schneider}, {Carpenter}, {Beichman}, {Capps}, {Chester},
      {Elias}, {Huchra}, {Liebert}, {Lonsdale}, {Monet}, {Price}, {Seitzer},
      {Jarrett}, {Kirkpatrick}, {Gizis}, {Howard}, {Evans}, {Fowler}, {Fullmer},
      {Hurt}, {Light}, {Kopan}, {Marsh}, {McCallon}, {Tam}, {Van Dyk}, \&
      {Wheelock}}]{Skrutskie+etal+2006}
    {Skrutskie}, M.~F., {Cutri}, R.~M., {Stiening}, R., {et~al.} 2006, \aj, 131,
      1163, \dodoi{10.1086/498708}
    
    \bibitem[{{Snow} {et~al.}(1981){Snow}, {Cash}, \& {Grady}}]{Snow+etal+1981}
    {Snow}, T.~P., J., {Cash}, W., \& {Grady}, C.~A. 1981, \apjl, 244, L19,
      \dodoi{10.1086/183470}
    
    \bibitem[{{Sun} {et~al.}(2018){Sun}, {Lu}, {Yang}, {Su}, {Zhang}, {Zhou}, \&
      {Lin}}]{Sun+etal+2018}
    {Sun}, J.~X., {Lu}, D.~R., {Yang}, J., {et~al.} 2018, Acta Astronomica Sinica,
      59, 3
    
    \bibitem[{{Sung} \& {Bessell}(2010)}]{Sung+Michael+2010}
    {Sung}, H., \& {Bessell}, M.~S. 2010, \aj, 140, 2070,
      \dodoi{10.1088/0004-6256/140/6/2070}
    
    \bibitem[{{Sung} {et~al.}(2008){Sung}, {Bessell}, {Chun}, {Karimov}, \&
      {Ibrahimov}}]{Sung+etal+2008}
    {Sung}, H., {Bessell}, M.~S., {Chun}, M.-Y., {Karimov}, R., \& {Ibrahimov}, M.
      2008, \aj, 135, 441, \dodoi{10.1088/0004-6256/135/2/441}
    
    \bibitem[{{Sung} {et~al.}(2009){Sung}, {Stauffer}, \&
      {Bessell}}]{Sung+etal+2009}
    {Sung}, H., {Stauffer}, J.~R., \& {Bessell}, M.~S. 2009, \aj, 138, 1116,
      \dodoi{10.1088/0004-6256/138/4/1116}
    
    \bibitem[{{Tauber} {et~al.}(1993){Tauber}, {Lis}, \&
      {Goldsmith}}]{Tauber+etal+1993}
    {Tauber}, J.~A., {Lis}, D.~C., \& {Goldsmith}, P.~F. 1993, \apj, 403, 202,
      \dodoi{10.1086/172194}
    
    \bibitem[{{Teixeira} {et~al.}(2012){Teixeira}, {Lada}, {Marengo}, \&
      {Lada}}]{Teixeira+etal+2012}
    {Teixeira}, P.~S., {Lada}, C.~J., {Marengo}, M., \& {Lada}, E.~A. 2012, \aap,
      540, A83, \dodoi{10.1051/0004-6361/201015326}
    
    \bibitem[{{Tobin} {et~al.}(2015){Tobin}, {Hartmann}, {F{\H{u}}r{\'e}sz}, {Hsu},
      \& {Mateo}}]{Tobin+etal+2015}
    {Tobin}, J.~J., {Hartmann}, L., {F{\H{u}}r{\'e}sz}, G., {Hsu}, W.-H., \&
      {Mateo}, M. 2015, \aj, 149, 119, \dodoi{10.1088/0004-6256/149/4/119}
    
    \bibitem[{{Tobin} {et~al.}(2009){Tobin}, {Hartmann}, {Furesz}, {Mateo}, \&
      {Megeath}}]{Tobin+etal+2009}
    {Tobin}, J.~J., {Hartmann}, L., {Furesz}, G., {Mateo}, M., \& {Megeath}, S.~T.
      2009, \apj, 697, 1103, \dodoi{10.1088/0004-637X/697/2/1103}
    
    \bibitem[{{Turner} {et~al.}(1997){Turner}, {Pirogov}, \&
      {Minh}}]{Turner+etal+1997}
    {Turner}, B.~E., {Pirogov}, L., \& {Minh}, Y.~C. 1997, \apj, 483, 235,
      \dodoi{10.1086/304228}
    
    \bibitem[{{Venuti} {et~al.}(2018){Venuti}, {Prisinzano}, {Sacco}, {Flaccomio},
      {Bonito}, {Damiani}, {Micela}, {Guarcello}, {Randich}, {Stauffer}, {Cody},
      {Jeffries}, {Alencar}, {Alfaro}, {Lanzafame}, {Pancino}, {Bayo}, {Carraro},
      {Costado}, {Frasca}, {Jofr{\'e}}, {Morbidelli}, {Sousa}, \&
      {Zaggia}}]{Venuti+etal+2018}
    {Venuti}, L., {Prisinzano}, L., {Sacco}, G.~G., {et~al.} 2018, \aap, 609, A10,
      \dodoi{10.1051/0004-6361/201731103}
    
    \bibitem[{{Walch} {et~al.}(2012){Walch}, {Whitworth}, {Bisbas}, {W{\"u}nsch},
      \& {Hubber}}]{Walch+etal+2012}
    {Walch}, S.~K., {Whitworth}, A.~P., {Bisbas}, T., {W{\"u}nsch}, R., \&
      {Hubber}, D. 2012, \mnras, 427, 625, \dodoi{10.1111/j.1365-2966.2012.21767.x}
    
    \bibitem[{{Weisskopf} {et~al.}(2002){Weisskopf}, {Brinkman}, {Canizares},
      {Garmire}, {Murray}, \& {Van Speybroeck}}]{Weisskopf+etal+2002}
    {Weisskopf}, M.~C., {Brinkman}, B., {Canizares}, C., {et~al.} 2002, \pasp, 114,
      1, \dodoi{10.1086/338108}
    
    \bibitem[{{Xu} {et~al.}(2017){Xu}, {Xu}, {Yu}, {Zhang}, {Liu}, {Wang}, {Ning},
      {Ju}, \& {Zhang}}]{Xu+etal+2017}
    {Xu}, J.-L., {Xu}, Y., {Yu}, N., {et~al.} 2017, \apj, 849, 140,
      \dodoi{10.3847/1538-4357/aa8ee0}
    
    \bibitem[{{Yang} {et~al.}(2018){Yang}, {Thompson}, {Urquhart}, \&
      {Tian}}]{Yang+etal+2018}
    {Yang}, A.~Y., {Thompson}, M.~A., {Urquhart}, J.~S., \& {Tian}, W.~W. 2018,
      \apjs, 235, 3, \dodoi{10.3847/1538-4365/aaa297}
    
    \bibitem[{{Zhang}(2023)}]{Zhang+2023}
    {Zhang}, M. 2023, \apjs, 265, 115, \dodoi{10.1088/0004-637X/783/2/115}
    
    \bibitem[{{Zhang} {et~al.}(2015){Zhang}, {Wang}, {Lu}, \&
      {Jim{\'e}nez-Serra}}]{Zhang+etal+2015}
    {Zhang}, Q., {Wang}, K., {Lu}, X., \& {Jim{\'e}nez-Serra}, I. 2015, \apj, 804,
      141, \dodoi{10.1088/0004-637X/804/2/141}
    
    \bibitem[{{Zucker} {et~al.}(2020){Zucker}, {Speagle}, {Schlafly}, {Green},
      {Finkbeiner}, {Goodman}, \& {Alves}}]{Zucker+etal+2020}
    {Zucker}, C., {Speagle}, J.~S., {Schlafly}, E.~F., {et~al.} 2020, \aap, 633,
      A51, \dodoi{10.1051/0004-6361/201936145}
    
    \end{thebibliography}

\clearpage
\appendix
\section{YSO Classification Scheme}
\label{app:data}
We describe the classification scheme for YSOs in this section. The YSO candidates in S~Mon are collected from the catalogs of \citet{Sung+etal+2009}, \citet{Broos+etal+2013}, \citet{Cody+etal+2014}, and \citet{Rapson+etal+2014}, within a field of $202.8^{\circ} < l < 203.1^{\circ}, 2.0^{\circ} < b < 2.3^{\circ}$. We consider two stars separated by less than 1$\arcsec$ to be the same star, and remove duplicated YSO candidates. A total of 1\,023 YSO candidates were initially selected. However, this initial catalog not only included numerous samples that were not YSOs, but also the YSOs in the catalog were obtained through different classification schemes. To maintain the homogeneity of the YSO catalog, we classified the YSO candidates based on a homogeneous method.

\begin{figure}
    \centering
    \includegraphics[width = 8cm]{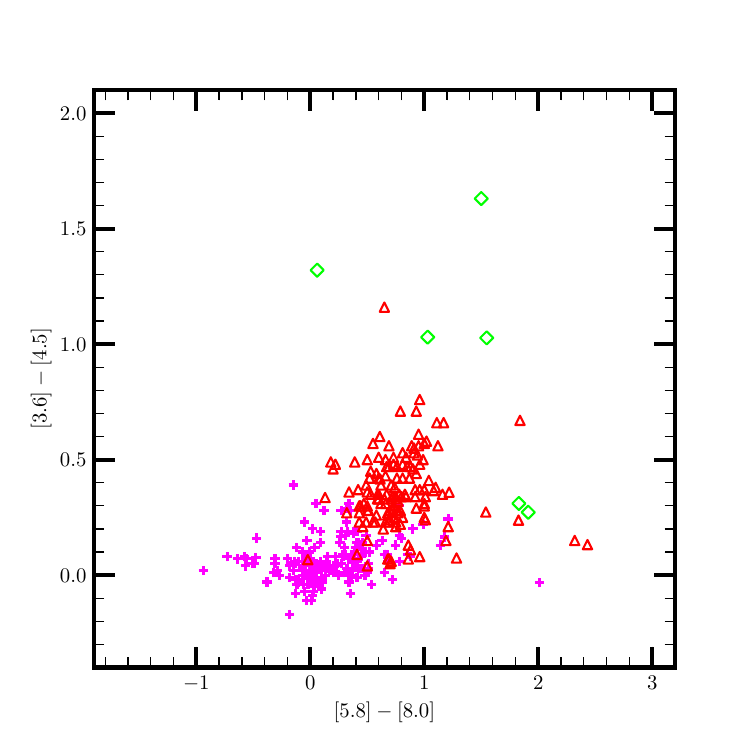}
    \includegraphics[width = 8cm]{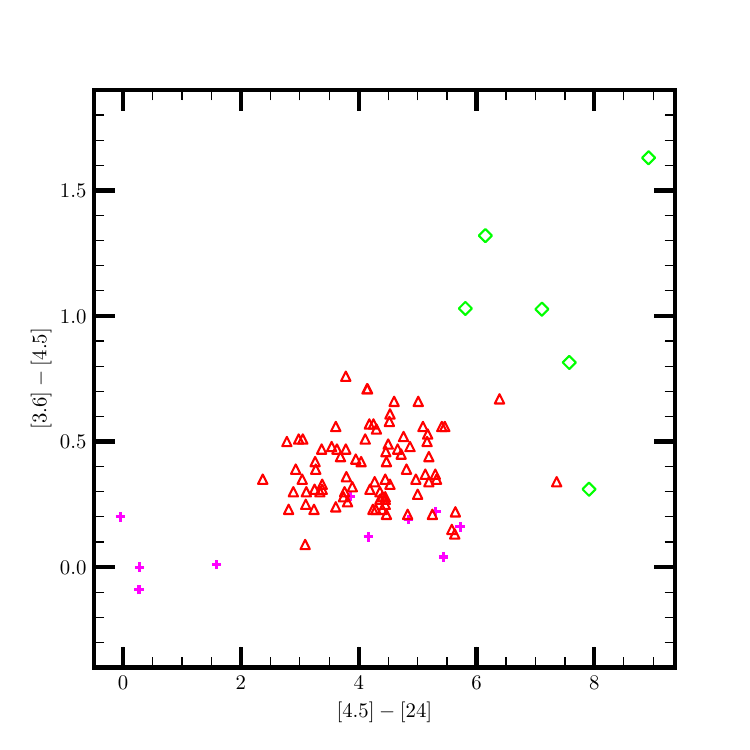}
    \caption{Color--color diagrams showing the classified YSOs. The green diamonds, red triangles, and magenta crosses present Class~I, Class~II, Class~III YSOs, respectively.}
    \label{fig:color_color}
\end{figure}

We applied the spectral index of the SED to classify the evolutionary stage of the YSOs, where $\alpha = {\rm d} \log \lambda F_{\lambda} / {\rm d} \log \lambda$~\citep{Greene+etal+1994}. To build this scheme, we use photometric data from the 2MASS $K_{s}$ band~\citep{Skrutskie+etal+2006}, four IRAC bands (3.6, 4.2, 5.8, and 8.0~$\mu$m) of \textit{Spitzer}~\citep{Fazio+etal+2004}, and the MIPS 24~$\mu$m band~\citep{Rieke+etal+2004}. Note that not all six bands are required simultaneously for the classification; however, at least three IRAC bands are necessary. Only sources with photometric uncertainties $\sigma < 0.2$~mag were selected for classification. Finally, we compiled a sample of seven Class~I, 160 Class~II, and 245 Class~III YSOs. Figure~\ref{fig:color_color} shows color--color diagrams of the classified YSOs in the S~Mon region. 

To obtain the parallaxes and proper motions of the classified YSOs, we cross matched them with stars in \textit{Gaia}~DR3 within 1$\arcsec$. 
One Class~I, 145 Class~II and 231 Class~III YSOs were assigned astrometric parameters from \textit{Gaia}~DR3.

\begin{figure}[htbp]
    \centering
    \includegraphics[width = 8cm]{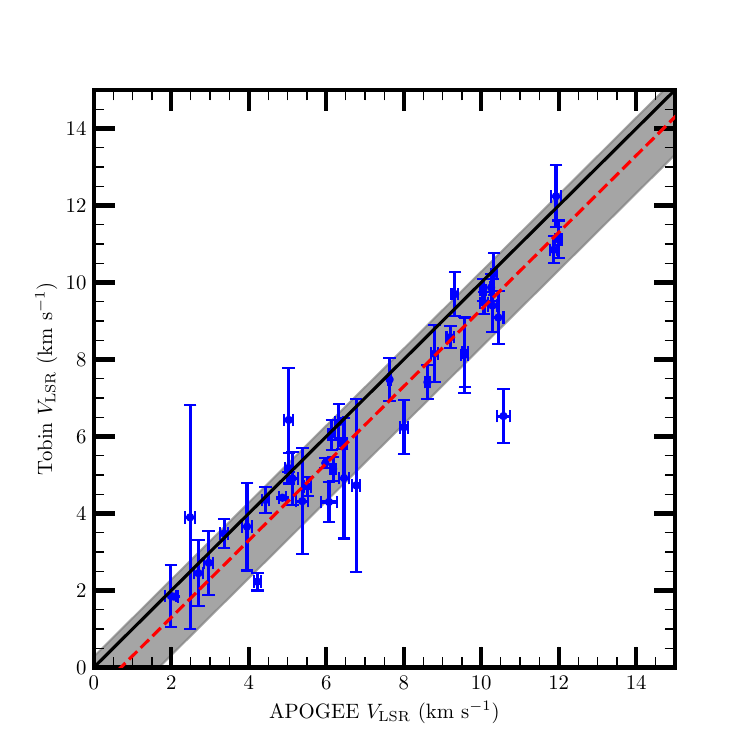}
    \caption{Comparison of the $V_{\rm LSR}$ values in the APOGEE and~\citet{Tobin+etal+2015} samples. The black line represents where $V_{\rm LSR}$ is the same in both catalogs. The red dashed line indicates the mean value of the difference between the APOGEE and~\citet{Tobin+etal+2015} $V_{\rm LSR}$ values, and the gray filled region depicts the corresponding $1 \sigma$ region.}
    \label{fig:Comparison_RVs}
\end{figure}

In the next step we added RV data to our YSO sample. First, we checked the availability of {\emph Gaia} DR3 RVs. There were only 16 YSOs with measured Gaia DR3 RVs, of which most fell outside the velocity range of interest (about 63\% are beyond the range of 0--15~\kms), and/or they had large RV uncertainties. Therefore, we did not use Gaia RVs in our analysis.
The RVs of the classified YSOs were collected from APOGEE-2 DR17~\citep{Abdurrouf+etal+2022} and the catalog of \citet{Tobin+etal+2015}, who measured the RVs of 695 stars in NGC~2264 (using multi-fiber echelle spectroscopy at the 6.5~m MMT and \textit{Magellan} telescopes). We cross matched our YSO candidates (1\,023) with the APOGEE catalog using the {\texttt{source\_id}} in the {\it Gaia}~DR3 catalog, and with the catalog of \citet{Tobin+etal+2015} within 1$\arcsec$, yielding 73 YSOs with RVs from APOGEE and 148 YSOs with RVs from~\citet{Tobin+etal+2015}. A total of 44 YSO candidates were matched in both catalogs. We made a comparison of the $V_{\rm LSR}$ (converted from heliocentric RVs to RVs relative to the local standard of rest, LSR) collected from the two catalogs for the YSO candidates, as shown in Figure~\ref{fig:Comparison_RVs}. The average difference between them is 0.7~\kms, and the $1 \sigma$ dispersion is 1.0~\kms. The Pearson's correlation coefficient between them is 0.99. We therefore added the RVs from~\citet{Tobin+etal+2015} to our sample and applied a systematic shift of 0.7~\kms, applied because the RV errors in the APOGEE catalog are consistently smaller than those in the catalog of~\citet{Tobin+etal+2015}. Finally, we obtained 47 Class II, and 91 Class III YSOs with RVs.

To sum up, we obtained seven Class~I, 160 Class~II, and 245 Class~III YSOs in the S~Mon region. Among them, one Class~I, 145 Class~II, and 231 Class~III YSOs have counterparts in \emph{Gaia}~DR3, where 47 Class~II and 91 Class~III YSOs have RVs. Table~\ref{tab:data} describes the YSO samples.

\begin{deluxetable}{rccl}[htbp]
\label{tab:data}
\tablecaption{Description of the YSO Samples}
\tablehead{
    Column & Format & Unit & Description
}
\startdata
    ID      & A4   &   ---    &  ID \\
    Gaia    & I19  &   ---    &  Gaia DR3 identifier \\
    RAdeg   & F8.4 &   deg    &  Right ascension (J2000) \\
    DEdeg   & F8.4 &   deg    &  Declination (J2000) \\
    plx     & F7.3 &   mas    &  Gaia DR3 parallax \\ 
  e\_plx     & F7.3 &   mas    &  Uncertainty in the parallax \\
    pmRA    & F7.3 &   mas/yr &  Proper motion in the RA* direction \\
  e\_pmRA    & F7.3 &   mas/yr &  Uncertainty in pmRA \\
    pmDE    & F7.3 &   mas/yr &  Proper motion in the DEC direction \\
  e\_pmDE    & F7.3 &   mas/yr &  Uncertainty in pmDE \\
    Vlsr    & F7.1 &   km/s   &  LSR velocity \\
  e\_Vlsr    & F7.1 &   km/s   &  Uncertainty in the LSR velocity \\
    Jmag    & F5.2 &   mag    &  2MASS J band magnitude \\
  e\_Jmag    & F5.2 &   mag    &  Uncertainty in Jmag \\
    Hmag    & F5.2 &   mag    &  2MASS H band magnitude \\
  e\_Hmag    & F5.2 &   mag    &  Uncertainty in Hmag \\
    Ksmag   & F5.2 &   mag    &  2MASS Ks band magnitude \\
  e\_Ksmag   & F5.2 &   mag    &  Uncertainty in Ksmag \\
   3.6mag   & F5.2 &   mag    &  Spitzer/IRAC 3.6 micron band magnitude \\
 e\_3.6mag   & F5.2 &   mag    &  Uncertainty in 3.6mag \\ 
   4.5mag   & F5.2 &   mag    &  Spitzer/IRAC 4.5 micron band magnitude \\
 e\_4.5mag   & F5.2 &   mag    &  Uncertainty in 4.5mag \\
   5.8mag   & F5.2 &   mag    &  Spitzer/IRAC 5.8 micron band magnitude \\
 e\_5.8mag   & F5.2 &   mag    &  Uncertainty in 5.8mag \\
   8.0mag   & F5.2 &   mag    &  Spitzer/IRAC 8.0 micron band magnitude \\
 e\_8.0mag   & F5.2 &   mag    &  Uncertainty in 8.0mag \\
   24mag    & F5.2 &   mag    &  Spitzer/MIPS 24 micron band magnitude \\
 e\_24mag    & F5.2 &   mag    &  Uncertainty in 24mag \\
  Class     & A10  &   ---    &  Source classification \\
  OClass    & A10  &   ---    &  Other source classification \\
  Ref       & A15  &   ---    &  Reference  \\
  Ref\_RV    & A15  &  ---    & Reference for the LSR velocity
\enddata
\tablecomments{This table is available in its entirety in machine-readable form.}
\end{deluxetable}

\section{Calculation of the Physical Parameters}
\label{sect:phy_para}
In this study, the properties of the molecular gas are estimated under the assumption that all of the molecular gas is in LTE. Optically thin \cob~lines are employed to estimate the column density and mass of the molecular gas. According to~\citet{Garden+etal+1991}, the \cob~column density can be estimated via:

\begin{equation}
    N({\rm ^{13}CO}) = 4.71 \times 10^{13} \frac{T_{\rm ex} + 0.88}{\exp{(-5.29 / T_{\rm ex}})} \times \frac{\tau}{1 - \exp{(-\tau)}} \int T_{\rm mb, 13} dv,
    \label{eq:N_13}
\end{equation}
where $\tau$ is the optical depth, $T_{\rm ex}$ is the mean excitation temperature of the molecular gas, and $T_{\rm mb, 13}$ is the main-beam temperature of \cob.

We assume that the \coa~emission is optically thick. The excitation temperature, $T_{\rm ex}$, can be estimated by~\citep{Garden+etal+1991}:
\begin{equation}
    T_{\rm ex} = \frac{5.53}{\ln{[1 + 5.53 / (T_{\rm mb, 12} + 0.82)]}},
    \label{eq:T_ex}
\end{equation}
where $T_{\rm mb, 12}$ is the peak main-beam temperature of \coa. 

The optical depth can be derived using the following equation~\citep{Garden+etal+1991}:
\begin{equation}
    \tau ({\rm ^{13}CO}) = -\ln{\left[ 1 - \frac{T_{\rm mb, 13}}{5.29 / [\exp{(5.29 / T_{\rm ex})} - 1] - 0.89} \right]}.
\end{equation}

The relation $N_{\rm H_{2}} / N_{\rm ^{13}CO} \approx 5 \times 10^{5}$~\citep{Simon+etal+2001} is used to estimate the H$_{2}$ column density. The mass, $M$, can be determined by: 
\begin{equation}
    M =\mu m_{\rm H} N({\rm H_{2}}) S,
\end{equation}
where $\mu = 2.72$ is the mean molecular weight, $m_{\rm H}$ is the mass of the hydrogen atom~\citep{Garden+etal+1991}, and $S$ is the projected 2D area. 

The turbulent energy, $E_{\rm turb}$, can be given approximately by:
\begin{equation}
    E_{\rm turb} = \frac{1}{2} M \sigma_{\rm 3d}^{2},
    \label{eq:E_turb}
\end{equation}
where $\sigma_{\rm 3d}$ is the 3D turbulent velocity dispersion, which can be calculated by: 
\begin{equation}
    \sigma_{\rm 3d} = \sqrt{3} \sigma_{\rm 1d} = \frac{\sqrt{3}}{2 \sqrt{2 \ln{2}}} \Delta V_{\rm FWHM},
\end{equation}
where $\Delta V_{\rm FWHM}$ is the 1D FWHM velocity dispersion based on typical \cob.

The gravitational binding energy, $E_{\rm grav}$, can be calculated by:
\begin{equation}
    E_{\rm grav} = \frac{G M^{2}}{R},
    \label{eq:E_grav}
\end{equation}
where $R$ is the radius. 

The excitation temperatures of the eight clumps are between 21.3 and 33.2~K. Since these clumps are regularly located within the bubble, a mean excitation temperature of about 26.9~K is adopted as that of the bubble. The mean column density of the bubble is $\sim 2.5 \times 10^{20}~{\rm cm^{-2}}$ and its mass is $\sim 1\,400~M_{\odot}$. In this work, we consider the uncertainties in gas temperature, shell region, and distance for the shell mass estimation. The total uncertainty caused by these factors is about $\pm 400~M_{\odot}$.

\begin{figure}[htbp]
    \centering
    \includegraphics{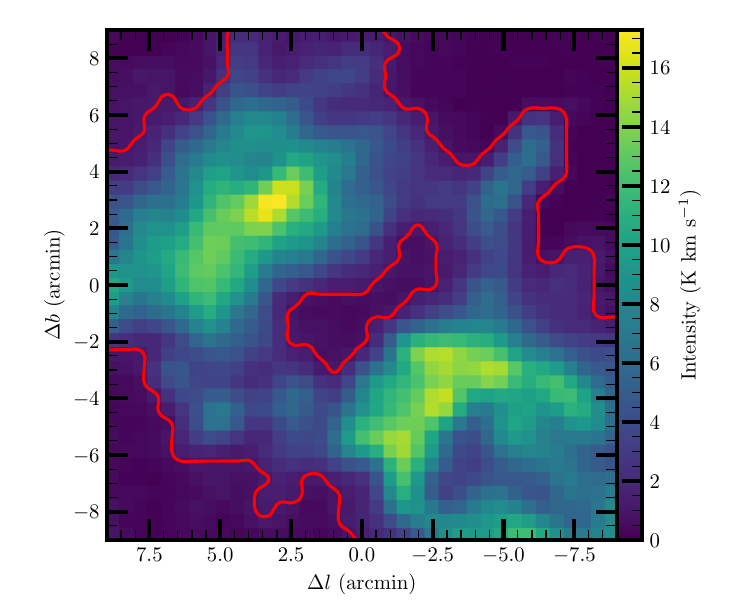}
    \caption{Intensity map of \cob~for the velocity range 7--15~\kms. The red lines depict the boundary of the cloud.}
    \label{fig:cloud}
\end{figure}

Figure~\ref{fig:cloud} shows the projected area of the cloud. The red lines denoted its boundary, determined by pixels whose main-beam brightness temperature is larger than $3 \times {\rm rms}$ in at least three successive channels. The velocity range adopted here is between 7 and 15~\kms~and the mass of the cloud is derived as 2\,200 $\pm$ 500~$M_{\odot}$. The uncertainties in the excitation temperature and distance are considered in the estimation of the cloud mass. $\sigma_{\rm 1d}$ is obtained from Gaussian fitting of the average \cob~line, which is about 1.0 $\pm$ 0.1~\kms. The turbulent energy and gravitational binding energy of the cloud are estimated by Equation~(\ref{eq:E_turb}) and Equation~(\ref{eq:E_grav}), respectively, which correspond to (6.3 $\pm$ 2.5)$\times 10^{46}$~erg and (2.1 $\pm$ 1.0)$\times 10^{47}$~erg.

The clump's column density ($N_{\rm H_{2}}$), mass ($M_{\rm clump}$), turbulent energy ($E_{\rm turb, clump}$), and gravitational binding energy ($E_{\rm grav, clump}$) were calculated using the same equations as applied to the cloud. The uncertainties in the distance and clump region were considered in the calculation.

\section{Distribution of OB stars}
\label{sect:OBstars}
\begin{figure}[htbp]
    \centering
    \includegraphics{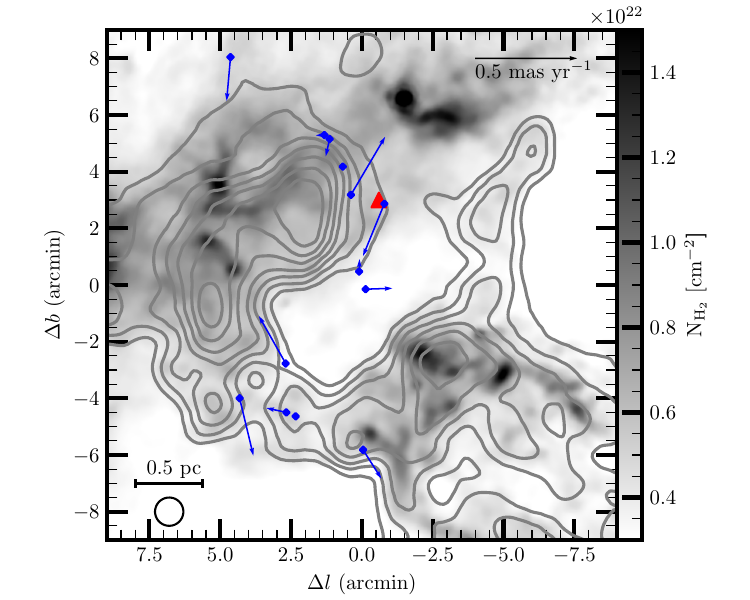}
    \caption{Projected distributions and relative proper motions of the O-(red triangle) and B-type (blue diamonds) stars. The average proper motion utilized being (2.65, -3.17)~mas~yr$^{-1}$.}
    \label{fig:OBstars}
\end{figure}

Figure~\ref{fig:OBstars} shows the projected distributions and relative proper motions of the O- and B-type stars in the S~Mon region. Table~\ref{tab:para_OBstars} lists the parameters of these stars, which are collected from SIMBAD. The astrometry parameters of 15~Mon are from \textit{Hipparcos} catalog~\citep{Perryman+etal+1997}, and those of other sources are from {\emph Gaia}~DR3~\citep{gaia2022}.

\begin{deluxetable}{ccccccccccc}[htbp]
\rotate
\label{tab:para_OBstars}
\tablecaption{Summary of the Parameters of the O- and B-type Stars}
\tablehead{
    \multirow{2}[0]{*}{Identifier} & \multirow{2}[0]{*}{SpType} & $\alpha$ & $\delta$ & $\varpi$ & $\sigma_{\varpi}$ & $\mu_{\alpha} \cos \delta$ & $\sigma_{\mu_{\alpha} \cos \delta}$ & $\mu_{\delta}$ & $\sigma_{\mu_{\delta}}$ & \multirow{2}[0]{*}{Gaia DR3 ID} \\
    & & (deg) & (deg) & (mas) & (mas) & (mas yr$^{-1}$) & (mas yr$^{-1}$) & (mas yr$^{-1}$) & (mas yr$^{-1}$) &  \\
    (1) & (2) & (3) & (4) & (5) & (6) & (7) & (8) & (9) & (10) & (11) 
}
\startdata
    15 Mon$^{\dagger}$  & O7V+B1.5/2V  & 100.2444 & 9.8958 & 3.550 & 0.500 & -2.610 & 0.560 & -1.610 & 0.390 & 3326740865570982400 \\
    15 Mon B  & B1.5/2V  & 100.2440 & 9.8951 & 1.402 & 0.098 & -1.971 & 0.113 & -4.225 & 0.082 & 3326740865571935232 \\
    V* V641 Mon  & B1.5IV+(B2V)  & 100.1191 & 9.8179 & 1.389 & 0.045 & -1.943 & 0.051 & -3.775 & 0.040 & 3326716813754146176 \\
    LS VI +09 13  & B1V   & 100.1552 & 9.7916 & ... & ... & ... & ... & ... & ... & 3326716470156761472 \\
    V* V684 Mon  & B2.5V  & 100.1599 & 9.7878 & 1.540 & 0.041 & -1.482 & -0.050 & -3.960 & -0.041 & 3326715714242517248 \\
    HD 47961  & B2.5V  & 100.3638 & 9.8540 & 1.248 & 0.079 & -1.976 & 0.091 & -4.034 & 0.077 & 3326737120359493504 \\
    HD 261810  & B2.5Vn  & 100.1801 & 9.7671 & 1.424 & 0.033 & -2.172 & 0.042 & -3.942 & 0.033 & 3326715439364610816 \\
    HD 261938  & B3V   & 100.2578 & 9.8800 & 1.540 & 0.059 & -1.258 & 0.056 & -3.247 & 0.041 & 3326740693772848896 \\
    EM* LkHA 25  & B4Ve  & 100.1860 & 9.8006 & 1.414 & 0.111 & -1.076 & 0.140 & -3.826 & 0.125 & 3326715851681467904 \\
    HD 261878  & B6V   & 100.2148 & 9.8637 & 1.331 & 0.037 & -1.504 & 0.041 & -3.747 & 0.030 & 3326717260430731648 \\
    CSI+09-06383  & B7    & 100.2750 & 9.8833 &... & ... & ... & ... & ... & ... & ... \\
    HD 261841  & B8IV-Ve  & 100.2037 & 9.8623 & 1.408 & 0.021 & -1.722 & 0.023 & -3.568 & 0.018 & 3326717226070995328 \\
    HD 261969  & B9IV  & 100.2933 & 9.8838 & 1.351 & 0.036 & -1.744 & 0.032 & -3.916 & 0.028 & 3326740006577519360 \\
    NGC 2264 181  & B9Vn  & 100.2968 & 9.8821 & 1.346 & 0.029 & -1.576 & -0.032 & -3.880 & 0.027 & 3326740002281694592
\enddata
\tablecomments{(1) Identifier. (2) Spectral type. (3) Right ascension (J2000). (4) Declination (J2000). (5) Parallax. (6) Uncertainty in parallax; (7) Proper motion in RA* direction. (8) Uncertainty in $\mu_{\alpha} \cos \delta$. (9) Proper motion in DEC direction. (10) Uncertainty in $\mu_{\delta}$. (11) Gaia DR3 identifier. \\
$^{\dagger}$: The astrometry parameters of 15~Mon (HIP 31978) are from \textit{Hipparcos} catalog, and those of other sources are from Gaia~DR3.}
\end{deluxetable}
\end{document}